%% file: acl_latex.tex
\pdfoutput=1

\documentclass[11pt]{article}

\usepackage[final]{acl}

\usepackage{times}
\usepackage{latexsym}
\usepackage{amsmath}
\usepackage{amsfonts} 
\usepackage{graphicx}
\usepackage{enumitem}
\usepackage{tabularx}
\usepackage{adjustbox}
\usepackage[most]{tcolorbox}
\usepackage{xcolor}
\usepackage{ragged2e}
\usepackage{float}
\usepackage{listings}
\tcbuselibrary{skins,breakable,listings,theorems, raster}
\usepackage[T1]{fontenc}

\usepackage{wrapfig}
\usepackage{multirow}
\usepackage{subcaption}

\usepackage{booktabs}

\usepackage{graphicx}
\usepackage{tabularx}
\usepackage{pifont} 
\usepackage{fvextra} 
\usepackage{placeins}

\newcommand{\cat}[1]{%
  \begingroup
  \let\origunderscore\_
  \def\_{\origunderscore\allowbreak}%
  #1%
  \endgroup
}
\newtcolorbox{promptbox}[2][]{%
  enhanced,
  breakable,
  width=\columnwidth,
  boxrule=0.6pt,
  arc=2pt,
  top=4pt,bottom=4pt,left=4pt,right=4pt,
  coltitle=black,
  fonttitle=\bfseries,
  title={\textbf{#2}},
  #1
}

\newtcblisting{jsonlisting}{
  listing only,
  breakable,
  colback=white,
  colframe=black!15,
  boxrule=0.3pt,
  left=2pt,right=2pt,top=2pt,bottom=2pt,
  listing options={
    basicstyle=\ttfamily\tiny,
    breaklines=true,
    breakatwhitespace=false
  }
}

\newtcolorbox{panelbox}[1][]{%
  enhanced,
  colback=gray!8,
  colframe=gray!40,
  boxrule=0.6pt,
  arc=3pt,
  left=6pt,right=6pt,top=6pt,bottom=6pt,
  valign=top,
  #1
}

\newtcolorbox{methodbox}[2][]{%
  enhanced,
  colback=gray!8,
  colframe=gray!40,
  boxrule=0.6pt,
  arc=3pt,
  left=6pt,right=6pt,top=6pt,bottom=6pt,
  colbacktitle=gray!15,
  coltitle=black,
  fonttitle=\bfseries,
  valign=top,
  title={#2},
  #1
}

\title{Personalization Matters: Long-Horizon Conversation Agent with User-Centric Information in Online Shopping Interactions} 

\author{
  Rena Gao\textsuperscript{1}\thanks{Equal contribution.} \quad
  Yue Dai\textsuperscript{2*} \quad Hao Guan \quad Shengxiang Gao\textsuperscript{2} \\
  \textbf{Wangyang Wu\textsuperscript{4} \quad Yixin Shen\textsuperscript{3} \quad Jey Han Lau\textsuperscript{2}} \\
  \textsuperscript{1}ByteDance \quad
  \textsuperscript{2}The University of Melbourne \quad \textsuperscript{3}AXON \\
  \textsuperscript{4}University of Technology Sydney \\
  {\small\texttt{renagao.0102@bytedance.com, yue.dai.3@student.unimelb.edu.au}} \\
  {\small\texttt{hguan689@gmail.com, shengxiang@student.unimelb.edu.au}} \\
  {\small\texttt{wangyang.wu@uts.edu.au, yixishen1@gmail.com, laujh@unimelb.edu.au}}
}

\begin{document}
\maketitle
\begin{abstract}
Personalized conversational shopping requires maintaining preference consistency over multi-turn interactions, where users reveal constraints gradually. Existing approaches often rely on static profiles and do not explicitly control long-horizon interaction behavior. We propose a multi-agent, multimodal Retrieval-Augmented Generation (RAG) framework that decomposes dialogue state tracking, recommendation retrieval, preference-aware reasoning, and response generation, while integrating product metadata, product reviews, image-derived descriptions, and user historical reviews. To evaluate interaction-level quality, we adopt a trajectory-level protocol with four dimensions: Global Preference Consistency, Cumulative Information Synthesis, Interaction Trajectory, and Tone Consistency. On an Amazon Reviews 2023 benchmark, retrieval-enabled variants outperform a no-RAG baseline on automatic trajectory metrics (average 4.82 vs. 3.74). In a small real-user study ($n{=}5$), the Full variant achieves the highest mean overall rating (4.60 vs. 2.20 for Baseline), providing exploratory evidence that role decomposition plus user-centric retrieval improves perceived personalization.\footnote{Code and dataset are available at: \url{https://github.com/RenaGao/Multimodel_RAG_Indexing}}
\end{abstract}

\section{Introduction}
\input{latex/introduction/intro}

\section{Related Work}
\input{latex/related_work/agents}
\input{latex/related_work/dialogue}

\section{Methodology}
\label{sec:method}
\input{latex/methodology/method}

\section{Dataset}
\label{sec:dataset}
\input{latex/dataset/dataset_content}

\section{Experiments}
\input{latex/experiment/exp_content}

\section{Results} 
\label{sec:results}
\input{latex/results/result_content}
\section{Conclusion}
\input{latex/conclusion/conclusion}
\section*{Limitations}
\input{latex/limitation/limitation}
\section*{Ethical considerations}
\input{latex/limitation/ethics}

\section*{Acknowledgements}
We thank Fan Jiang for assistance with data organization. Jey Han Lau was supported by the Australian Research Council under Grant LP210200917 and DP240101006. We used AI assistants to help write and edit this paper.


\bibliography{latex/custom}
\clearpage
\appendix
\input{latex/appendix/appendix_content}

\end{document}

%% file: latex/introduction/intro.tex
Conversational shopping differs from standard question answering because user preferences are revealed gradually across turns, not in a single query~\cite{bernard2023mg,du2025sapient}. Therefore, quality depends less on turn-level correctness and more on maintaining long-horizon alignment with evolving user preferences~\cite{li2025wizard}. This makes personalization an interaction-level problem: the system must remember, reconcile, and operationalize user constraints over time.

\begin{figure}[t]
    \centering
    \includegraphics[width=0.95\linewidth]{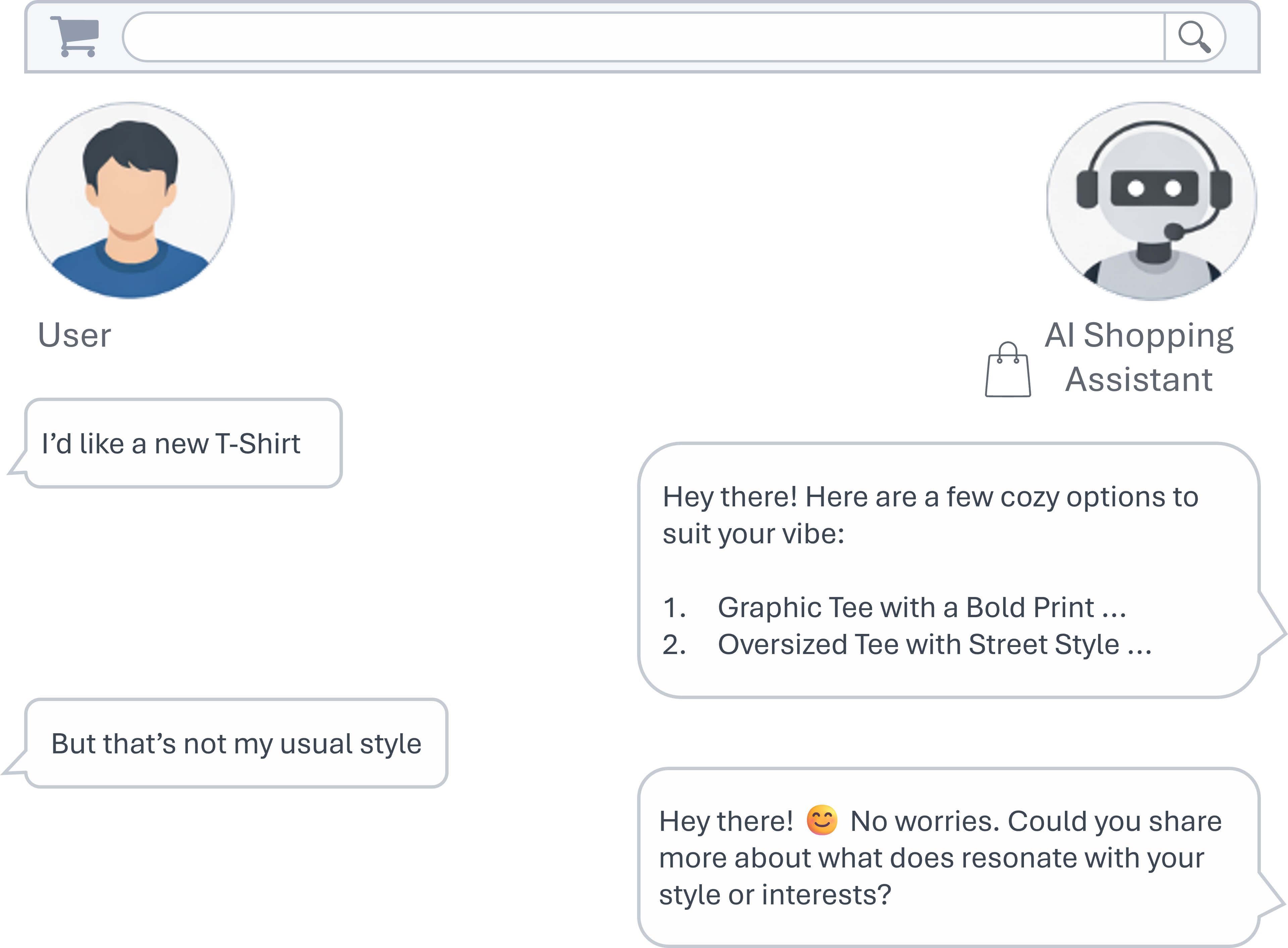}
    \caption{A failure case without effective personalization during online shopping interactions: the response is locally reasonable but does not reflect previously stated style preferences, revealing a gap between turn-level relevance and long-horizon consistency.}
    \label{fig:no_personalization_example}
\end{figure}

As Figure~\ref{fig:no_personalization_example} shows, an agent can be locally relevant yet still fail trajectory-level consistency by ignoring earlier signals. Existing methods only partially solve this: static-profile approaches keep user representations fixed~\cite{zhou2023learning}, while per-turn retrieval improves local relevance but offers limited control over how preference evidence is accumulated and enforced across the full interaction~\cite{salemi2024lamp,wang2024crafting,gao2024interaction}.

To address this gap, we model personalization at the interaction level and propose a multi-agent, multimodal RAG framework. The system separates dialogue tracking, recommendation retrieval, preference-aware reasoning, and response generation, while integrating product metadata/reviews, image-derived descriptions, and user historical reviews. This decomposition improves controllability and makes it easier to analyze which components contribute to personalization quality. We test whether this combination improves long-horizon preference consistency over simpler variants.

Because turn-level metrics are insufficient for this goal, we evaluate full trajectories using four dimensions: Global Preference Consistency (GPC), Cumulative Information Synthesis (CIS), Interaction Trajectory (IT), and Tone Consistency (TC). Our benchmark is built from Amazon Reviews 2023; the user study is exploratory ($n=5$), and we discuss validity risks in Section~\ref{sec:results} and the Limitations section.


Our results reveal several complementary effects. Retrieval-enabled variants achieve average trajectory scores of $4.72$--$4.82$, compared with $3.74$ for the no-RAG baseline (Table~\ref{tab:point_based}), indicating that retrieval is a primary driver of improvement in our benchmark. Beyond retrieval, role decomposition improves cumulative information synthesis and dialogue progression relative to a single-agent system using the same evidence. User-review history provides additional gains in preference tracking and cross-turn synthesis. Although Full and TextRAG are nearly tied under text-based automatic evaluation, Full receives a higher mean overall rating in our exploratory user study ($4.60$ vs.~$3.40$; Table~\ref{tab:user_study}), suggesting that text-only judging may not fully capture the benefits of visual grounding.

Specifically, our contributions include three major aspects:
\begin{itemize}[noitemsep,topsep=2pt,parsep=0pt,partopsep=0pt,leftmargin=*]
    \item A multi-agent, multimodal RAG framework for long-horizon personalized shopping context that integrates dialogue interaction context in previous history, multi-modal product evidence, image-derived descriptions, and user-review history and ratings.
    \item A trajectory-level evaluation protocol covering preference consistency, information synthesis, dialogue progression, and conversation role-play tone for interactive online conversation recommend agents in shopping domain.
    \item Controlled experiments showing that retrieval drives the largest gains, role decomposition further improves cross-turn interaction quality, and exploratory human evaluation highlights potential benefits of visual grounding of products information. 
\end{itemize}

%% file: latex/related_work/agents.tex
\subsection{Personalized Agents and Conversational Recommendation}
Personalized dialogue methods have moved from static persona-conditioned generation~\cite{zhou2023learning,tan2024democratizing} to retrieval- and memory-based approaches that inject user history at inference time~\cite{salemi2024lamp,salemi2024optimization,mysore-etal-2024-pearl}. RAG-style personalization is especially practical because it can use interaction logs or user-authored evidence without retraining a foundation model.

In conversational recommendation, user preferences guide both item selection and the course of the dialogue. ChatCRS combines external knowledge with a fine-tuned goal planner~\cite{li2025chatcrs}, while RevCore trains recommendation and response-generation components using review evidence~\cite{lu-etal-2021-revcore}. Other approaches use LLM agents and prompting: MACRec coordinates specialized agents~\cite{wang2024macrec,gao2025no}, and RA-Rec uses prompt-based dialogue state tracking to retrieve relevant reviews~\cite{kemper2024rarec}.

Our framework combines user-history retrieval and role decomposition for personalized shopping without task-specific training. Using the same frozen model and user-query trajectories, we compare retrieval inputs and agent configurations to examine their contributions to preference tracking across turns. The evaluation combines trajectory-level metrics with a small user study.

%% file: latex/related_work/dialogue.tex
\subsection{Long-Horizon Interaction Evaluation} 
Most dialogue evaluations emphasize fluency, coherence, and local accuracy~\cite{van-miltenburg-etal-2025-measure,ghazarian2022deam,komma-etal-2023-toward}. These metrics are useful but often miss user-specific goals such as preference adherence, mixed initiative, and cross-turn usefulness~\cite{huang2023cir-eval,fu2024evaluation}. As a result, personalized multi-turn interaction remains difficult to evaluate with static, overlap-oriented criteria alone.

Recent work on interaction styles and trajectory-level assessment~\cite{gao2024interaction,gao2024interpretable,gao2025can} moves toward dynamic user-centric evaluation by assessing how quality evolves across turns rather than at isolated response points.

We build on this direction and adapt it to personalized shopping, where the central requirement is sustained preference tracking under retrieval-augmented reasoning~\cite{he2026towards}. Accordingly, our protocol emphasizes full-trajectory preference consistency and recommendation grounding rather than isolated turn quality.

%% file: latex/methodology/method.tex
In this section, we present the framework used to construct synthetic multi-turn personalized shopping dialogues. We first provide a pipeline overview, then formalize the turn-level state and interaction flow, and finally describe user-query generation and assistant-side retrieval/response modules.
\input{latex/methodology/overview}
\input{latex/methodology/prob_def}
\input{latex/methodology/user_query_generation}
\input{latex/methodology/assistant_side}
\input{latex/methodology/embedding}
\input{latex/methodology/rag}

%% file: latex/methodology/overview.tex
\subsection{Multi-Agent Pipeline}
\begin{figure*}
    \centering
    \includegraphics[width=\linewidth]{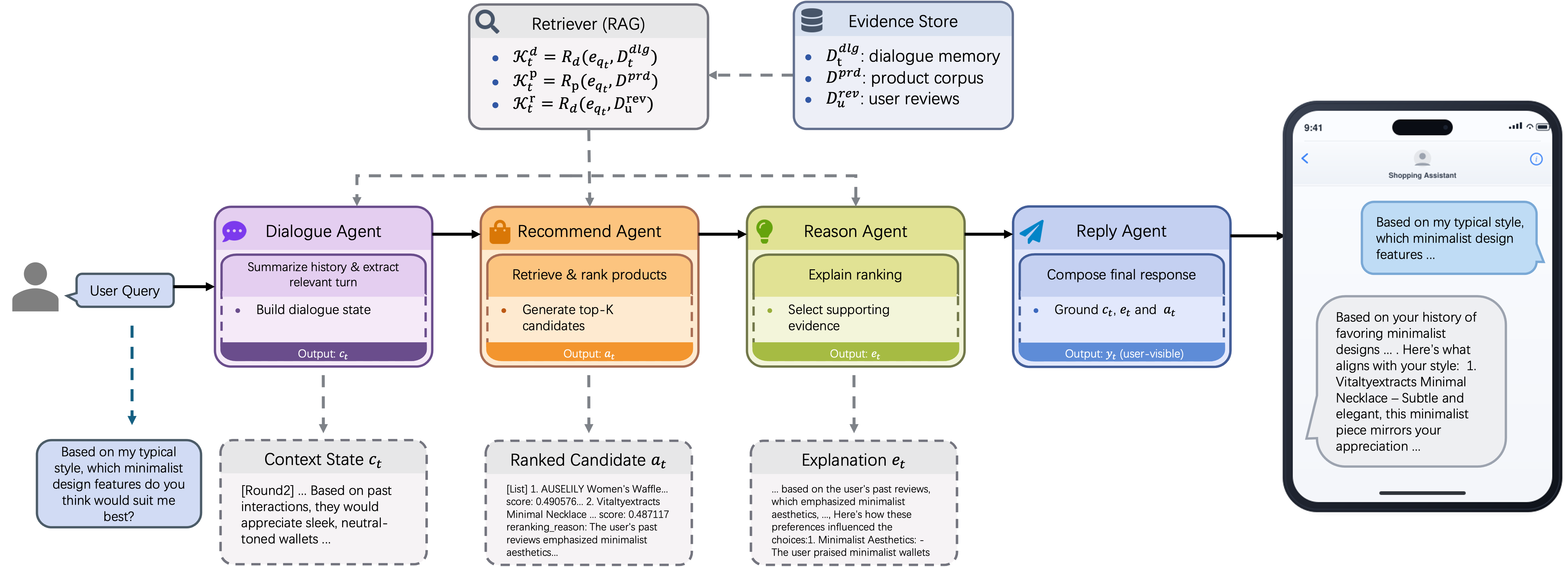}
    \caption{Overall framework of the proposed multi-agent personalized conversational shopping system. Each turn is decomposed into dialogue context tracking, retrieval-based recommendation, preference-aware reasoning, and response generation, supported by external dialogue memory and the Amazon Reviews 2023 corpus.}
    \label{fig:agentworkflowdiagram}
\end{figure*}

We construct synthetic multi-turn shopping dialogues by simulating both sides of the interaction: a User agent generates personalized queries, and assistant-side agents generate responses. The framework has five agents: one User agent and four assistant agents (Dialogue, Recommend, Reason, Reply).\footnote{Only User inputs and Reply outputs appear in the final transcript; other agents produce internal states used for grounding.}

We use role decomposition because personalized shopping requires heterogeneous operations (context tracking, retrieval, preference grounding, and response realization). Compared with a single prompt, this design improves controllability, modularity, and interpretability~\cite{gao2025single}. It also supports cleaner ablations by isolating where improvements come from.

The assistant-side roles are: Dialogue Agent for compact context tracking, Recommend Agent for candidate retrieval/reranking, Reason Agent for preference-aware justification, and Reply Agent for final response synthesis. The intermediate outputs $(c_t, a_t, e_t)$ provide explicit grounding signals for final response generation, rather than relying on one monolithic prompt.

%% file: latex/methodology/prob_def.tex
\subsection{Problem Definition}
\label{sec:prob_def}

We model personalized conversational shopping as a synthetic multi-turn dialogue of length $T$. At turn $t$, the user issues query $q_t$ and the system returns response $y_t$, yielding transcript $\{(q_t, y_t)\}_{t=1}^{T}$.

Internally, each turn includes context tracking, recommendation, and preference reasoning. We define
\begin{equation}
\label{eq:state}
\tau_t = (q_t,\ c_t,\ a_t,\ e_t,\ y_t),
\end{equation}
where $c_t$ is dialogue context, $a_t$ is the ranked candidate set with rationale, and $e_t$ is the preference-aware explanation. Only $(q_t, y_t)$ are externally visible; other variables are internal.

This formulation lets us separate observable dialogue quality from internal grounding quality. In the benchmark, each dialogue is generated by first constructing user-side query trajectories and then running assistant-side multi-agent response generation under controlled variant settings.

%% file: latex/methodology/user_query_generation.tex
\subsection{User Query Generation}
\label{sec:user_query_generation}
Each dialogue is built from a user's purchase and review history. We first derive a structured user profile from historical reviews, then generate a multi-turn query trajectory around one target product.

Per category, we sample 10 target users from the top-100 most active users, requiring at least two reviews.\footnote{If fewer than 10 users satisfy this requirement, all eligible users are retained.} For each user, we select one positively rated purchased item (rating $\geq 4$) as the target product and remove its review from profile construction to preserve a pre-purchase setting.

Using the prompt shown in Appendix Figure~\ref{fig:user_profile_prompt}, GPT-4o~\cite{hurst2024gpt} summarizes the remaining reviews into a structured user profile (e.g., style, quality\_priorities, and dislikes). Category-specific missing fields are filled as None or placed in additional\_notes. Given this profile and the target product, an LLM generates preference-oriented multi-turn queries using the prompt shown in Appendix Figure~\ref{fig:question_generation_prompt}. We explicitly encourage preference-oriented queries (e.g., style and fit) rather than metadata-only questions.

This design serves two goals: (i) \textbf{leakage control} by excluding the target review, and (ii) \textbf{comparability} by reusing the same user-query trajectories across all assistant variants in Section~\ref{sec:agent_workflow}.

%% file: latex/methodology/assistant_side.tex
\subsection{Assistant-Side Multi-Agent Response Generation}
\label{sec:assistant_pipeline}

Given query $q_t$, assistant-side agents run sequentially to produce context, recommendation, explanation, and final response:
\begin{align}
c_t &= f_d\bigl(q_t,\ \tilde{H}_{t-1},\ \mathcal{K}^{d}_t\bigr), \label{eq:dialogue_agent} \\
a_t &= f_{rec}\bigl(q_t,\ c_t,\ \mathcal{K}^{p}_t,\ \mathcal{K}^{r}_t\bigr), \label{eq:recommend_agent} \\
e_t &= f_{rea}\bigl(q_t,\ c_t,\ a_t\bigr), \label{eq:reason_agent} \\
y_t &= f_{rep}\bigl(q_t,\ c_t,\ a_t,\ e_t\bigr), \label{eq:reply_agent}
\end{align}
where $f_d$, $f_{rec}$, $f_{rea}$, and $f_{rep}$ denote Dialogue, Recommend, Reason, and Reply agents. $\tilde{H}_{t-1}$ is a compact history summary, and $\mathcal{K}^{d}_t$, $\mathcal{K}^{p}_t$, $\mathcal{K}^{r}_t$ are retrieved dialogue, product, and user-review evidence (Section~\ref{sec:rag}). Prompts are in Appendix Section~\ref{sec:agent_prompts}.

\paragraph{Dialogue Agent}
Maintains turn-level context by combining an accumulated summary with the most relevant prior turn. This helps preserve continuity without injecting full-history text into every downstream prompt.

\paragraph{Recommend Agent}
Retrieves product candidates and user reviews, then reranks candidates with user-preference evidence and outputs ranked set $a_t$ with a concise rationale. This stage grounds recommendations in both query relevance and long-term preference signals.

\paragraph{Reason Agent}
Generates explanation $e_t$ linking ranked items to long-term preferences and current intent. The goal is to make preference alignment explicit before response realization.

\paragraph{Reply Agent}
Synthesizes final response $y_t$ from $(q_t, c_t, a_t, e_t)$; this is the only assistant output visible in the transcript. Separating this step from retrieval and reasoning improves controllability over final wording and style.

%% file: latex/methodology/embedding.tex
\subsection{Embedding Representations}
\label{sec:emb_rep}
We encode dialogue, product, and user-review evidence with \textbf{Qwen3-Embedding-8B (Qwen3E)}~\cite{zhang2025qwen3}.

\paragraph{Product embeddings}
Each product $p_i$ uses three vectors: metadata embedding $\mathbf{e}^{m}_{p_i}$, review embedding $\mathbf{e}^{r}_{p_i}$ (mean of associated review embeddings), and description embedding $\mathbf{e}^{d}_{p_i}$ from an image description generated by \textbf{Qwen3-VL-30B-A3B-Instruct}~\cite{bai2025qwen3}.

\paragraph{User-review embeddings}
Each historical user review is encoded as $\mathbf{e}_{u_{r_i}}$, excluding the target-product review to preserve the pre-purchase setting.

\paragraph{Query embeddings}
At turn $t$, user query $q_t$ is encoded as $\mathbf{e}_{q_t}$ and used for all retrieval functions.

%% file: latex/methodology/rag.tex
\subsection{Retrieval-Augmented Generation}
\label{sec:rag}
Using Section~\ref{sec:emb_rep} representations, we define three retrieval functions that provide dialogue, product, and user-review evidence: $\mathcal{K}^{d}_t$, $\mathcal{K}^{p}_t$, and $\mathcal{K}^{r}_t$. These evidence sets are consumed by the assistant-side pipeline in Eqs.~\eqref{eq:dialogue_agent}--\eqref{eq:reply_agent}.

\paragraph{Dialogue-level retrieval}
$R_d$ retrieves the past turn most relevant to the current query:
\begin{equation}
\label{eq:rd}
\mathcal{K}^{d}_t = R_d(\mathbf{e}_{q_t}) = \operatorname{TopK}\bigl(\mathbf{e}_{q_t},\ \mathcal{D}^{\text{dlg}}_t\bigr),
\end{equation}
where $\mathcal{D}^{\text{dlg}}_t$ stores embeddings of prior user queries with their turn text.

\paragraph{Product-level retrieval}
$R_p$ scores each product using metadata, review, and description embeddings:
\begin{equation}
\label{eq:prod_score}
s(p_i\mid q_t)=\frac{1}{3}\sum_{x\in\{m,r,d\}}
\cos\!\left(\mathbf e_{q_t}, \mathbf e^{x}_{p_i}\right).
\end{equation}
Then,
\begin{equation}
\label{eq:rp}
\mathcal{K}^{p}_t = R_p(\mathbf{e}_{q_t}) = \operatorname{TopK}\bigl(s(p_i \mid q_t),\ \mathcal{D}^{\text{prd}}\bigr),
\end{equation}
where $\mathcal{D}^{\text{prd}}$ is the product catalog.

\paragraph{User-review retrieval}
$R_r$ retrieves historical user reviews most relevant to the current query:
\begin{equation}
\label{eq:rr}
\mathcal{K}^{r}_t = R_r(\mathbf{e}_{q_t}) = \operatorname{TopK}\bigl(\mathbf{e}_{q_t},\ \mathcal{D}^{\text{rev}}_u\bigr).
\end{equation}

All three retrieval branches use the same query embedding $\mathbf{e}_{q_t}$ but retrieve from different corpora, enabling complementary grounding signals for context tracking, recommendation, and preference alignment.
We retrieve one prior dialogue turn, up to five products, and up to five user reviews per query.

%% file: latex/dataset/dataset_content.tex
We construct our simulation data from the Amazon Reviews 2023 dataset~\cite{amazonreviews2023}.\footnote{See Appendix~\ref{sec:artifact_licensing} for data and code licensing.} This corpus provides product metadata, images, and user review histories for building personalized shopping dialogues. We preprocess the data to form a user--item--review subset and the retrieval corpora used by our RAG pipeline.

This section describes the data source and preprocessing protocol for the synthetic benchmark used in later experiments.

\input{latex/dataset/dataset_processing}

%% file: latex/dataset/dataset_processing.tex
\subsection{Dataset Processing}
\label{sec:preprocessing}
The Amazon Reviews 2023 dataset contains 34 categories, including an \emph{Unknown} category, which we remove because our setting assumes users shop within a single category. For the remaining 33 categories, we reorganize metadata and reviews at the user level to support personalized preference modeling. In each category, we select the 100 most active users by review count and use them as the user pool for dialogue construction. For each selected user, we collect all purchased items together with their corresponding purchase reviews. To provide comprehensive item-level feedback, we further retrieve all reviews associated with each purchased item, including those written by users outside the top-100 subset.

\begin{table}[h]
\centering
\small
\setlength{\tabcolsep}{6pt}
\begin{tabular}{@{}lr@{}}
\toprule
\textbf{Statistic} & \textbf{Value} \\
\midrule
Raw Amazon categories & 34 \\
Top active users retained per category & 100 \\
Candidate user pool & 3,300 \\
Target users / dialogue instances & 327 \\
Dialogues per variant & 327 \\
\textbf{Dialogue runs ($327\times5$ variants)} & \textbf{1,635} \\
\bottomrule
\end{tabular}
\caption{Scale summary of the processed dialogue dataset, detailed per-category item/review statistics are reported in Appendix Table~\ref{tab:category_stats}.}
\label{tab:data_scale_summary}
\end{table}

Based on this processed user--item--review subset, we construct synthetic dialogue instances through the user query generation procedure described in Section~\ref{sec:user_query_generation}. Final per-category statistics are shown in Appendix Table~\ref{tab:category_stats}.

%% file: latex/experiment/exp_content.tex
\subsection{Multi-Agent System}
\label{sec:agent_workflow}

We evaluate five variants under the assistant-side retrieval setup in Eqs.~\eqref{eq:dialogue_agent}--\eqref{eq:reply_agent}, summarized in Table~\ref{tab:variants}. All variants share the same user queries for controlled comparison. They differ along three axes: which retrieval branches are enabled ($\mathcal{K}^{d}_t$, $\mathcal{K}^{p}_t$, $\mathcal{K}^{r}_t$); whether product retrieval uses the image-derived description embedding $e^{d}_{p_i}$ (multimodal vs.~text-only); and whether the assistant follows the four-agent pipeline or collapses into a single agent. \textbf{Full} is our complete model; \textbf{NoRev} removes user-review retrieval; \textbf{TextRAG} drops the description embedding; and \textbf{Single} collapses the four assistant roles into one LLM call. The remaining variant, \textbf{Baseline}, disables all retrieval branches.

\begin{table}[t]
\centering
\small
\setlength{\tabcolsep}{5pt}
\begin{tabular}{lcccll}
\toprule
\textbf{Variant} & $\mathcal{K}^{d}_t$ & $\mathcal{K}^{p}_t$ & $\mathcal{K}^{r}_t$ & \textbf{Prod.~emb.} & \textbf{Pipeline} \\
\midrule
Baseline  & \ding{55} & \ding{55} & \ding{55} & ---                  & Multi-agent  \\
TextRAG   & \ding{51} & \ding{51} & \ding{51} & $e^{m}, e^{r}$       & Multi-agent  \\
NoRev     & \ding{51} & \ding{51} & \ding{55} & $e^{m}, e^{r}, e^{d}$ & Multi-agent  \\
Single    & \ding{51} & \ding{51} & \ding{51} & $e^{m}, e^{r}, e^{d}$ & Single-agent \\
\textbf{Full} & \ding{51} & \ding{51} & \ding{51} & $e^{m}, e^{r}, e^{d}$ & Multi-agent  \\
\bottomrule
\end{tabular}
\caption{Assistant-side variants. $\mathcal{K}^{d}_t$, $\mathcal{K}^{p}_t$, and $\mathcal{K}^{r}_t$ denote dialogue, product, and user-review retrieval. \textbf{Prod.~emb.} lists embeddings used in Eq.~\eqref{eq:prod_score}: metadata ($e^{m}$), review ($e^{r}$), and description ($e^{d}$).}
\label{tab:variants}
\end{table}

\subsection{Evaluation Metrics Design}
Standard turn-level metrics often overlook the long-horizon dynamics required for personalization~\cite{guan2025evaluating}. A locally appropriate response may still violate prior constraints (e.g., clothing-fit preferences), producing inconsistency with user intent across the dialogue~\cite{zhang2024personalization,qiu2025measuring}. We therefore use trajectory-level evaluation.

We adapt the framework of \citet{gao2024interaction}, originally developed for second-language long interactive conversations, and re-anchor its dimensions for personalized shopping around long-horizon preference consistency and retrieval-grounded reasoning.

\paragraph{Formal Definition}
Building on Section~\ref{sec:prob_def}, the LLM judge receives the user profile (Section~\ref{sec:user_query_generation}) and full transcript $\{(q_t, y_t)\}_{t=1}^{T}$, then outputs
$\mathbf{s} = (s_{\mathrm{GPC}},\, s_{\mathrm{CIS}},\, s_{\mathrm{IT}},\, s_{\mathrm{TC}}) \in \{1,2,3,4,5\}^{4}$.
Each component reflects trajectory-level quality rather than single-turn quality.
We use GPT-4o~\cite{hurst2024gpt} as the judge; prompt details are in Appendix Table~\ref{tab:evaluation_prompt_criteria}.%
\footnote{To validate LLM-as-judge reliability, three PhD volunteers independently annotated 20 randomly sampled dialogues (five product categories) with the same rubric. Pairwise percentage agreement exceeded \textbf{85\%} across all four dimensions.}

We test the significance of score differences using paired comparisons; details are provided in Appendix~\ref{sec:supp_checks}.

\paragraph{Global Personalization Consistency (GPC)}
GPC measures fidelity to user persona and constraints across the full interaction. High GPC indicates stable long-term context use and reduced \emph{attribute hallucination}~\cite{chen2024large}.

A representative failure appears in a \textsc{Home \& Kitchen} mattress dialogue: after five turns establishing a medium-firm memory-foam preference, Baseline responds at Turn~6 with \textit{``Since there is no previous conversation history available, I cannot reference your past choices''}, forcing the user to restate preferences (GPC = 1). By contrast, a high-GPC response in \textsc{Clothing \& Shoes} synthesizes earlier constraints (e.g., breathability and machine-washable convenience) without contradiction (GPC = 5).

\paragraph{Cumulative Information Synthesis (CIS)}
CIS measures how well the model aggregates constraints across turns. Unlike isolated retrieval, it evaluates whether early and late cues are integrated coherently.

In a \textsc{Musical Instruments} Baseline dialogue, the agent repeatedly requests setup details already provided, indicating weak cross-turn synthesis (CIS = 1). A strong CIS agent instead refines recommendations by explicitly linking new guidance to previously stated preferences (CIS = 5).

\paragraph{Interaction Trajectory (IT)}
IT measures dialogue efficiency in moving from clarification to recommendation, refinement, and closure~\cite{gao2024interpretable}. Agents that loop without progress receive lower scores.

A typical IT failure repeats open-ended clarification prompts across non-consecutive turns without transitioning to recommendations (IT = 1). High-IT behavior proactively advances the state and anticipates likely next constraints (IT = 5).

\paragraph{Tone Consistency (TC)}
TC captures stability of sociolinguistic style (friendly, professional, empathetic). Consistent tone supports trustworthiness, while abrupt register shifts degrade perceived naturalness~\cite{gao2024interaction}.

%% file: latex/results/result_content.tex
\input{latex/results/point_base}
\input{latex/results/user_study}

%% file: latex/results/point_base.tex
\subsection{Automatic Evaluation}

\begin{table}[h]
\centering
\footnotesize
\setlength{\tabcolsep}{3.5pt} 
\renewcommand{\arraystretch}{0.9} 
\begin{tabular}{@{}lccccc@{}}
\toprule
\textbf{Metric} & \textbf{Baseline} & \textbf{TextRAG} & \textbf{NoRev}$^{1}$ & \textbf{Single}$^{2}$ & \textbf{Full}$^{3}$ \\
\midrule
GPC      & 3.77 & 4.93 & 4.88 & 4.91 & \textbf{4.93} \\
CIS      & 3.17 & 4.88 & 4.79 & 4.78 & \textbf{4.89} \\
IT       & 3.02 & 4.47 & 4.39 & 4.19 & 4.45 \\
TC       & 5.00 & 5.00 & 5.00 & 5.00 & 5.00 \\
\midrule
\textbf{Avg.} & 3.74 & \textbf{4.82} & 4.76 & 4.72 & \textbf{4.82} \\
\bottomrule
\end{tabular}
\vspace{-5pt}
\caption{Automatic evaluation results, averaged over 327 dialogues per variant. $^{1}$ Multimodal (NoRev). $^{2}$ Multimodal (Single-Agent). $^{3}$ Multimodal (Full).}
\label{tab:point_based}
\vspace{-10pt} 
\end{table}

\paragraph{LLM-as-judge} Full and TextRAG achieve the highest scores on \textbf{GPC}
($4.93$) and \textbf{CIS} (${\sim}4.89$), substantially above Baseline
($3.77$ / $3.17$), indicating stronger constraint retention and multi-turn synthesis
under this rubric. On \textbf{IT}, TextRAG ($4.47$) and Full ($4.45$) are close,
whereas Single drops to $4.19$ despite using the same multimodal retrieval evidence as Full.
Full's improvements over Single in information synthesis, dialogue progression, and overall quality are statistically significant.

\paragraph{Metric-Specific Discussion}
The four trajectory metrics reveal complementary behaviors rather than one monolithic quality score.
\textbf{GPC} is most sensitive to long-horizon memory use: large gains over Baseline indicate that
retrieval and structured context tracking substantially reduce preference drift.
\textbf{CIS} mainly reflects cross-turn integration quality: the Full $>$ Single pattern suggests that
role decomposition helps combine earlier and later constraints more reliably.
\textbf{IT} is the most architecture-sensitive metric in our table: the largest drop from Full to Single
appears on IT, consistent with weaker dialogue-state progression when all reasoning is compressed
into one prompt.
\textbf{TC} is non-discriminative here (all 5.00), implying that model politeness/style is saturated and
should not be used as a primary differentiator for this benchmark.

\paragraph{Ablation-Level Reading}
The variant progression also helps separate where gains come from. The large jump from
Baseline to retrieval-enabled settings (e.g., Avg. $3.74 \rightarrow 4.72$ for Single)
shows that retrieval is a primary driver of quality in this benchmark. Comparing Single and Full,
which share the same retrieval branches, Full improves on CIS ($+0.11$) and IT ($+0.26$),
suggesting that role decomposition mainly helps cross-turn synthesis and interaction flow.
Comparing NoRev and Full under the same multi-agent structure, Full is higher on all
non-ceiling dimensions (GPC/CIS/IT), consistent with added value from user-review evidence
for sustained preference tracking.

The near tie between TextRAG and Full under LLM-as-judge should be interpreted with care.
The judge reads only textual transcripts, so multimodal retrieval effects are observed indirectly
through generated wording. This likely underestimates visual-grounding benefits in cases where
image evidence primarily changes recommendation suitability rather than surface text.
Consistent with this interpretation, Full exceeds TextRAG in the user study
(Section~\ref{sec:user_study}, $4.60$ vs.~$3.40$ Overall) and remains competitive on
retrieval accuracy (Table~\ref{tab:retrieval_acc_stage}, $43.73\%$ vs.~$42.73\%$).

All variants reach a ceiling on \textbf{TC} ($5.00$), suggesting tone is dominated by
base-model style and is less discriminative for this benchmark.

\begin{table}[h]
    \centering
    \small
    \setlength{\tabcolsep}{3.5pt}
    \begin{tabular}{@{}lccccc@{}}
        \toprule
        \textbf{Metric}
        & \textbf{Baseline} & \textbf{TextRAG} & \textbf{NoRev}$^{1}$ & \textbf{Single}$^{2}$ & \textbf{Full}$^{3}$ \\
        \midrule
        \textbf{Acc.} & 10.31\% & 42.73\% & 41.90\% & \textbf{48.20\%} & 43.73\% \\
        \bottomrule
    \end{tabular}
    \caption{Retrieval accuracy. $^{1}$ Multimodal (NoRev). $^{2}$ Multimodal (Single-Agent). $^{3}$ Multimodal (Full).}
    \label{tab:retrieval_acc_stage}
\end{table}
\paragraph{Retrieval Accuracy}
We additionally report a token-overlap accuracy metric between the
target product title and the generated response across the full
dialogue.\footnote{A prediction is counted as correct if the token
overlap ratio exceeds 0.3.} As shown in Table~\ref{tab:retrieval_acc_stage},
Baseline is substantially lower (10.31\%), while all retrieval-enabled
variants exceed 40\%, indicating that retrieval is important for open-domain
product grounding in this setup.

Among retrieval-enabled variants, Single obtains the highest value on this metric (48.20\%).
However, this measure checks lexical overlap with the target title only and does not capture
whether recommendations are aligned with long-term user preferences. We therefore report
it as a complementary grounding indicator alongside trajectory-level scores and human ratings.
In short, retrieval-accuracy gains indicate stronger target-product mention, while
trajectory metrics and user ratings better reflect personalization quality over the full interaction.

\paragraph{Cost--Quality Perspective}
A new practical indicator is per-turn call complexity (Appendix Section~\ref{sec:api_cost}).
Combining Table~\ref{tab:point_based} with Table~\ref{tab:api_cost} suggests that retrieval
alone explains much of the gain over Baseline, while multi-agent decomposition mainly buys
better trajectory quality (especially IT/CIS) at higher per-turn call cost. This clarifies a
three-way trade-off among personalization quality, architectural complexity, and serving cost.

Further experiments examine target-item retrieval, agreement between judge models, and the contribution of dialogue memory (Appendix~\ref{sec:supp_checks}).

%% file: latex/results/user_study.tex
\subsection{User Study: Real Interaction}
\label{sec:user_study}
We further evaluate real user experience by deploying demos of all systems with five unpaid volunteers. We include representative screenshots from one real-user session in Appendix~\ref{sec:demo}, complementing the LLM-as-judge results in Table~\ref{tab:point_based}. In this comparison, Baseline limitations are straightforward: without retrieval grounding, it may recommend items not tied to real products and can drift to irrelevant objects (e.g., recommending a ``Vibrant Blue Waterproof Backpack'' after a shell-jacket request). The comparison also highlights the role of multimodal information: with image-caption-derived product descriptions, multimodal systems better capture visually grounded intents such as music-festival and ACDC-style band T-shirts, because captions expose style cues (retro appearance, graphic design, overall vibe) that are difficult to recover from text-only signals. Compared with NoRev, Full further connects visual style understanding to user-specific preferences, producing more personalized responses across turns.

We additionally collected structured ratings from all five users on a 5-point Likert scale, covering overall satisfaction and four dimensions aligned with our automatic metrics: GPC, CIS, IT, and TC. Mean scores are reported in Table~\ref{tab:user_study}. Given the small sample size ($n=5$), we treat these results as \textbf{qualitative/exploratory evidence} rather than a statistically powered hypothesis test. The questionnaire and recruitment details are provided in Appendix~\ref{sec:user_questionnaire}; anonymized responses are available in our code repository.\footnote{\url{https://github.com/RenaGao/Multimodel_RAG_Indexing/tree/main/data/user_study}}

\begin{table}[h]
\centering
\footnotesize
\setlength{\tabcolsep}{3pt}
\renewcommand{\arraystretch}{0.9}
\begin{tabular}{@{}lccccc@{}}
\toprule
\textbf{Dimension} & \textbf{Baseline} & \textbf{TextRAG} 
  & \textbf{NoRev}$^{1}$ & \textbf{Single}$^{2}$ & \textbf{Full}$^{3}$ \\
\midrule
Overall  & 2.20 & 3.40 & 3.80 & 3.80 & \textbf{4.60} \\
GPC     & 1.20 & 3.60 & 4.00 & 4.00 & \textbf{4.60} \\
CIS & 1.60 & 3.60 & 4.20 & 4.00 & \textbf{4.80} \\
IT   & 2.00 & 3.60 & 3.80 & 3.60 & \textbf{4.60} \\
TC      & 4.20 & 4.60 & 4.60 & 4.80 & \textbf{5.00} \\
\bottomrule
\end{tabular}
\caption{User-study ratings (mean, 1--5 scale, $n=5$). 
$^{1}$ Multimodal (NoRev). 
$^{2}$ Multimodal (Single-Agent). 
$^{3}$ Multimodal (Full).}
\label{tab:user_study}
\end{table}

Despite the limited sample size, user feedback is directionally consistent with automatic evaluation: Full receives the highest mean score on all dimensions, four of five users preferred it overall, and all five judged it the most personalized variant. Beyond scalar ratings, users repeatedly emphasized three perceived strengths of Full. First, they noted that the system ``does not only look at the current utterance'' and instead uses history summaries to carry forward earlier constraints and tendencies into new recommendation planning. Second, users explicitly observed preference grounding on practical factors such as prior budget range and previously stated style priorities, reporting that recommendations felt more coherent across turns in a single shopping context. Third, users highlighted cross-turn reasoning quality: the system could connect current requests to product attributes mentioned earlier in the dialogue, rather than producing isolated one-turn suggestions. These comments match the largest quantitative gap in GPC (1.20 vs.~4.60) and the IT gain of Full over Single (4.60 vs.~3.60), supporting our interpretation that role decomposition improves long-horizon interaction quality. Tone consistency remains comparatively stable across systems ($\geq 4.20$), and users still described weaker variants as polite but less reliable for sustained personalized recommendation.

\paragraph{Dimension-Level Discussion}
The human ratings provide interpretable support for the trajectory metrics.
The largest human gaps appear on GPC and IT, mirroring the automatic pattern that
long-horizon consistency and progression are the hardest dimensions for weaker variants.
CIS differences are moderate but stable, indicating that users do perceive improved
cross-turn synthesis when recommendation and reasoning are explicitly decomposed.
TC remains high for all systems, consistent with the automatic TC ceiling and suggesting
that tone quality is largely inherited from the base model rather than the retrieval design.

\paragraph{Cross-Protocol Consistency}
Comparing Tables~\ref{tab:point_based} and~\ref{tab:user_study}, two consistent trends emerge.
First, Baseline is weakest under both protocols, indicating that retrieval grounding is essential
for this open-domain shopping setup. Second, Full maintains stronger human-perceived
personalization than Single and TextRAG, even when automatic averages are close.
This reinforces the interpretation that trajectory-level automatic metrics and human ratings
capture overlapping but not identical aspects of performance.

The user feedback complements automatic scoring by showing how participants perceive personalization during shopping conversations. With five participants, these observations provide an initial basis for a larger user study.

%% file: latex/conclusion/conclusion.tex
We study long-horizon preference consistency in personalized conversational shopping. In our benchmark, the multi-agent multimodal RAG setup performs strongly relative to simpler variants and clearly exceeds no-RAG baselines on both automatic and human-facing evaluations. We also observe that text-only LLM judges may understate user-perceived gains from visual grounding, supporting joint reporting of automatic scores and human feedback. Given the synthetic benchmark and small user study ($n=5$), these results should be interpreted as directional evidence; future work should validate them with larger real-user and naturally collected multimodal shopping interactions.

%% file: latex/limitation/limitation.tex
Our evaluation focuses on controlled shopping dialogues generated from product and review data. Future work could extend this setting to naturally occurring conversations and larger user studies. Human evaluation would also complement the automatic scores, which may be influenced by using related models for generation and judging. The current system uses text queries and image-derived descriptions; supporting direct image queries and exploring richer retrieval methods are further directions.

%% file: latex/limitation/ethics.tex
We use the public Amazon Reviews 2023 dataset without attempting to identify its reviewers. All five study participants gave informed consent, and we report no personally identifying information. The system may misunderstand preferences or recommend unsuitable products, and its performance may vary across users. In practical use, people should be told how their review histories inform recommendations and be able to choose whether to share them or have them deleted. Any release of study logs should respect participant consent and remove identifying details.

%% file: latex/appendix/appendix_content.tex
\section{Appendix}
\input{latex/appendix/stat}
\input{latex/appendix/supplementary_checks}
\input{latex/appendix/prompts}
\input{latex/appendix/evaluation_prompts}
\input{latex/appendix/recommend_reason}

\FloatBarrier
\input{latex/appendix/casestudy}
\FloatBarrier
\input{latex/appendix/demo}
\FloatBarrier
\input{latex/appendix/user_questionnaire}
\input{latex/appendix/artifact_licensing}

%% file: latex/appendix/stat.tex
\subsection{Dataset Statistics}
\begin{table}[H]
\centering
\small
\setlength{\tabcolsep}{3pt}
\begin{tabularx}{\columnwidth}{@{}>{\raggedright\arraybackslash}X r r r r@{}}
\toprule
\textbf{Category} & \textbf{Items} & \textbf{Reviews} & \textbf{Avg.} & \textbf{Dlg.} \\
\midrule
\cat{Electronics}                     & 27,292  & 9,907,789  & 363.03 & 10 \\
\cat{Home\_and\_Kitchen}                & 51,421  & 8,780,109  & 170.75 & 10 \\
\cat{Kindle\_Store}                    &113,360  & 7,172,600  &  63.27 & 10 \\
\cat{Movies\_and\_TV}                   & 67,650  & 6,112,901  &  90.36 & 8 \\
\cat{Health\_and\_Household}            & 34,382  & 5,492,657  & 159.75 & 10 \\
\cat{Clothing\_Shoes\_and\_Jewelry}      & 63,329  & 5,360,811  &  84.65 & 10 \\
\cat{Pet\_Supplies}                    & 15,248  & 5,225,236  & 342.68 & 10 \\
\cat{Books}                           &118,159  & 4,967,408  &  42.04 & 10 \\
\cat{Grocery\_and\_Gourmet\_Food}        & 24,308  & 4,226,505  & 173.87 & 10 \\
\cat{Beauty\_and\_Personal\_Care}        & 42,599  & 3,935,557  &  92.39 & 10 \\
\cat{Tools\_and\_Home\_Improvement}      & 27,939  & 3,666,301  & 131.23 & 10 \\
\cat{Software}                        &  7,637  & 3,632,785  & 475.68 & 10 \\
\cat{Cell\_Phones\_and\_Accessories}     & 10,171  & 2,861,995  & 281.39 & 10 \\
\cat{Automotive}                      & 19,272  & 2,854,050  & 148.09 & 10 \\
\cat{Toys\_and\_Games}                  & 26,279  & 2,364,092  &  89.96 & 10 \\
\cat{Patio\_Lawn\_and\_Garden}           & 16,791  & 2,320,870  & 138.22 & 10 \\
\cat{Office\_Products}                 & 12,561  & 2,237,761  & 178.15 & 10 \\
\cat{Sports\_and\_Outdoors}             & 18,621  & 1,982,891  & 106.49 & 10 \\
\cat{Video\_Games}                     &  8,551  & 1,838,654  & 215.02 & 10 \\
\cat{CDs\_and\_Vinyl}                   & 58,707  & 1,539,120  &  26.22 & 10 \\
\cat{Arts\_Crafts\_and\_Sewing}          & 16,198  & 1,369,350  &  84.54 & 10 \\
\cat{Baby\_Products}                   &  9,616  & 1,286,781  & 133.82 & 10 \\
\cat{Musical\_Instruments}             &  9,396  &   954,286  & 101.56 & 10 \\
\cat{Industrial\_and\_Scientific}       &  8,038  &   741,943  &  92.30 & 10 \\
\cat{Appliances}                      &  2,288  &   260,620  & 113.91 & 10 \\
\cat{Gift\_Cards}                      &    275  &   138,737  & 504.50 & 10 \\
\cat{Magazine\_Subscriptions}          &    554  &    47,753  &  86.20 & 10 \\
\cat{Amazon\_Fashion}                  &  4,890  &    37,248  &   7.62 & 10 \\
\cat{Health\_and\_Personal\_Care}        &  1,351  &    36,817  &  27.25 & 10 \\
\cat{All\_Beauty}                      &  2,040  &    33,832  &  16.58 & 10 \\
\cat{Handmade\_Products}               &  1,504  &    19,603  &  13.03 & 10 \\
\cat{Subscription\_Boxes}              &    156  &    11,696  &  74.97 & 9 \\
\cat{Digital\_Music}                   &  4,538  &     9,876  &   2.18 & 10 \\
\bottomrule
\end{tabularx}
\caption{Product and review statistics, with trajectory counts (Dlg.) from the released query set. Counts total 327: Movies and TV has 8, Subscription Boxes has 9, and other categories have 10.}
\label{tab:category_stats}
\end{table}

\subsection{Cost--Performance Trade-off}
\label{sec:api_cost}

Table~\ref{tab:api_cost} summarizes the reported number of assistant-role calls per dialogue turn, juxtaposed with its automatic-evaluation average score (Table~\ref{tab:point_based}) and user-study Overall rating (Table~\ref{tab:user_study}).

\begin{table}[h]
\centering
\footnotesize
\setlength{\tabcolsep}{4pt}
\renewcommand{\arraystretch}{0.95}
\begin{tabular}{@{}lccc@{}}
\toprule
\textbf{Method} & \textbf{Role calls / turn} & \textbf{Auto.\ Avg.} & \textbf{User Overall} \\
\midrule
Baseline  & 1 & 3.74 & 2.20 \\
TextRAG   & 4 & 4.82 & 3.40 \\
NoRev     & 4 & 4.76 & 3.80 \\
Single    & 1 & 4.72 & 3.80 \\
Full      & 4 & \textbf{4.82} & \textbf{4.60} \\
\bottomrule
\end{tabular}
\caption{Assistant-role calls per turn and performance. Full uses Dialogue, Recommend, Reason, and Reply stages; Single combines these roles in one call. The Baseline count is retained from the original report and awaits configuration verification.}
\label{tab:api_cost}
\end{table}

Full uses four assistant-role calls per turn, while Single uses one. Full receives higher user ratings, whereas Single offers a simpler inference pipeline. Serving cost also depends on the model and token usage.

%% file: latex/appendix/supplementary_checks.tex
\subsection{Supplementary Experiments}
\label{sec:supp_checks}
\paragraph{Paired comparisons.}
Scores are paired across the same 327 trajectories. We use two-sided Wilcoxon signed-rank tests with Holm correction and matched-pairs rank-biserial correlations as effect sizes. Table~\ref{tab:paired_tests} summarizes the reported comparisons. TC is constant in the main evaluation.
\begin{table}[ht]
\centering
\small
\begin{tabular}{@{}lcc@{}}
\toprule
Comparison / metric & Adjusted $p$ & $r$ \\
\midrule
RAG vs.\ Baseline$^{*}$ & $<0.001$ & $\geq0.90$ \\
Full--Single: IT & $<0.001$ & 0.433 \\
Full--Single: CIS & 0.040 & 0.508 \\
Full--Single: Overall & 0.002 & 0.351 \\
\bottomrule
\end{tabular}
\caption{Paired comparisons. $^{*}$Bounds across TextRAG, NoRev, Single, and Full versus Baseline on GPC, CIS, and IT. Overall is the mean of the four trajectory scores.}
\label{tab:paired_tests}
\end{table}

\paragraph{Target-item retrieval.}
Full has the highest target-item coverage, while TextRAG reaches its successful targets earlier (Table~\ref{tab:target_hits}). A hit records retrieval of the target ASIN. SR@all is the proportion of trajectories with a hit, and first-hit turn is averaged over successful trajectories only.
\begin{table}[ht]
\centering
\small
\begin{tabular}{@{}lrr@{}}
\toprule
Variant & SR@all & First-hit turn \\
\midrule
TextRAG & 0.352 & 2.26 \\
NoRev & 0.309 & 2.29 \\
Single & 0.352 & 2.50 \\
Full & 0.361 & 2.49 \\
\bottomrule
\end{tabular}
\caption{Target-ASIN retrieval results. First-hit turns are conditional on a hit. These measures track target retrieval rather than user acceptance.}
\label{tab:target_hits}
\end{table}

\paragraph{Agreement between judges.}
We re-score 99 trajectories (three per category, yielding 495 system outputs) with Qwen3-235B. The second judge produces the same system ranking. Table~\ref{tab:cross_judge} reports score correlations with GPT-4o; TC is omitted because of the ceiling effect.
\begin{table}[ht]
\centering
\small
\begin{tabular}{@{}lrrrr@{}}
\toprule
Metric & GPC & CIS & IT & Overall \\
\midrule
Spearman $\rho$ & 0.508 & 0.533 & 0.471 & 0.538 \\
\bottomrule
\end{tabular}
\caption{Agreement between GPT-4o and Qwen3-235B scores on the stratified subset.}
\label{tab:cross_judge}
\end{table}

\paragraph{Dialogue memory.}
On the same 99 trajectories, we compare full memory (summary and retrieved history), no memory, summary only, and the top-ranked prior turn only. Summary-only memory performs similarly to full memory, while removing memory or retaining only one prior turn lowers the scores (Table~\ref{tab:memory_check}).
\begin{table}[ht]
\centering
\small
\begin{tabular}{@{}lrrr@{}}
\toprule
Memory & Overall & IT & CIS \\
\midrule
Full & 4.796 & 4.387 & 4.882 \\
None & 4.610 & 3.796 & --- \\
Summary only & 4.801 & --- & 4.860 \\
Top-1 turn only & 4.634 & 3.935 & --- \\
\bottomrule
\end{tabular}
\caption{Memory ablation results. Dashes indicate values absent from the reported results.}
\label{tab:memory_check}
\end{table}

\subsection{Resources in Related Work}
\label{sec:crs_resources}
Table~\ref{tab:crs_resources} summarizes training and data resources for related systems and datasets.
\begin{table*}[t]
\centering
\small
\begin{tabularx}{\textwidth}{@{}>{\raggedright\arraybackslash}p{2.5cm}XX@{}}
\toprule
Work & Training / inference setup & Evidence and task resources \\
\midrule
ChatCRS~\cite{li2025chatcrs} & LLM generation with knowledge retrieval and a LoRA-fine-tuned goal planner. & External knowledge base and annotated dialogue goals; multi-goal CRS datasets. \\
RevCore~\cite{lu-etal-2021-revcore} & Trained recommendation and response-generation components. & Item-linked reviews, entities, and conversational recommendation data. \\
MACRec~\cite{wang2024macrec} & Configurable LLM agents with task-dependent tools. & User/item information and task-dependent tools; multiple recommendation scenarios. \\
RA-Rec~\cite{kemper2024rarec} & Prompt-driven state tracking and generation with a pretrained dense retriever. & Domain-specific state fields, item reviews and metadata; demonstrated on Yelp restaurants. \\
PEARL~\cite{kim-etal-2024-pearl} & Review-grounded dialogue dataset with downstream model experiments. & Review-derived personas and knowledge for synthetic movie recommendation dialogues. \\
MUSE~\cite{wang-etal-2025-muse} & Multimodal conversational dataset with benchmark experiments. & Scenario-grounded profiles and clothing-domain multimodal conversations. \\
This study & Frozen models; no task-specific training; five internal variants. & Amazon product/review corpus, image-derived text, and 327 fixed trajectories across 33 categories. \\
\bottomrule
\end{tabularx}
\caption{Training and data resources used in related conversational recommendation work.}
\label{tab:crs_resources}
\end{table*}

%% file: latex/appendix/prompts.tex
\subsection{Profile and Question Generation Prompt}
\label{sec:prompts}
\begin{tcolorbox}[
  enhanced,
  breakable,
  width=\columnwidth,
  colback=blue!2,
  colframe=blue!50!black,
  colbacktitle=blue!8,
  coltitle=black,
  boxrule=0.6pt,
  arc=2pt,
  top=4pt, bottom=4pt,
  left=4pt, right=4pt,
  title={\textbf{User Profile Generation Prompt}},
  fonttitle=\bfseries,
]
\small
\setlength{\parskip}{3pt}

You are an expert at analyzing shopping behavior and extracting user preferences from product reviews.

I will provide you with ALL reviews written by a specific Amazon Fashion customer (excluding one target product).

\textbf{IMPORTANT: Each review includes a RATING (1-5 stars). Pay close attention to ratings when analyzing preferences.}

Your task is to analyze these reviews with a focus on RATING-BASED PATTERNS.

\medskip
\textbf{Rating Analysis Instructions:}
\begin{enumerate}[itemsep=1pt, topsep=2pt, parsep=0pt, partopsep=0pt, leftmargin=14pt]
  \item \textbf{High Ratings (4-5 stars)}: Extract what the user LIKES and VALUES
    \begin{itemize}[itemsep=0pt, topsep=1pt, parsep=0pt, leftmargin=16pt]
      \item Positive product features, styles, brands, materials
      \item What made them satisfied?
    \end{itemize}

  \item \textbf{Low Ratings (1-3 stars)}: Extract what the user DISLIKES and AVOIDS
    \begin{itemize}[itemsep=0pt, topsep=1pt, parsep=0pt, leftmargin=16pt]
      \item Negative experiences, quality issues, sizing problems
      \item What disappointed them?
    \end{itemize}

  \item \textbf{Rating Patterns}: Notice trends
    \begin{itemize}[itemsep=0pt, topsep=1pt, parsep=0pt, leftmargin=16pt]
      \item Do certain brands/styles consistently get high or low ratings?
      \item Are there recurring complaints (e.g., "always too small", "fabric quality")?
    \end{itemize}
\end{enumerate}

\medskip
\textbf{Profile Elements to Extract:}
\begin{enumerate}[itemsep=1pt, topsep=2pt, parsep=0pt, partopsep=0pt, leftmargin=14pt]
  \item \textbf{Preferences}: What does this user actively LIKE? (from 4-5 star reviews)
  \item \textbf{Style}: Fashion style preferences (Casual, Formal, Athletic, etc.)
  \item \textbf{Budget}: Price sensitivity based on review patterns and ratings
  \item \textbf{Size}: Common sizes mentioned, fit preferences
  \item \textbf{Quality Priorities}: What quality aspects matter most?
  \item \textbf{Dislikes}: What does this user actively AVOID or dislike? (from 1-3 star reviews)
  \item \textbf{Pain Points}: Recurring issues that cause dissatisfaction
\end{enumerate}

\medskip
\textbf{User's Review History:}

\texttt{\{review\_corpus\}}

\medskip
\textbf{Output Schema and Constraints}

\textbf{Output Format (JSON ONLY):}

\begin{jsonlisting}
{"Preferences": "What user likes based on high-rated (4-5 star) reviews. Include specific features, styles, materials.",
  "Style": "Fashion style preferences (e.g., Casual, Athletic, Minimalist)",
  "Budget": "Budget-friendly | Mid-range | Premium | Mixed (infer from products and satisfaction levels)",
  "Size": "Size information and fit preferences",
  "Quality_Priorities": "Key quality aspects that matter (e.g., Fabric comfort, Durability, Accurate sizing)",
  "Dislikes": "What user dislikes based on low-rated (1-3 star) reviews. Include materials, styles, brands to avoid.",
  "Pain_Points": "Recurring issues causing dissatisfaction (e.g., sizing
 inconsistency, fabric pilling, poor stitching)",
  "Additional_Notes": "Other notable patterns or context"}
\end{jsonlisting}

\textbf{Instructions:}
\begin{itemize}[itemsep=1pt, topsep=2pt, parsep=0pt, leftmargin=14pt]
  \item Be SPECIFIC and EVIDENCE-BASED - cite rating levels when relevant
  \item Distinguish between high-rated and low-rated patterns
  \item If information is not available, use "Not specified" or "No clear pattern"
  \item Focus on PATTERNS across multiple reviews, not isolated mentions
  \item Keep each field concise (1-3 sentences max)
  \item Use phrases like "consistently rated 5 stars when..." or "gave 1-2 stars due to..."
\end{itemize}

\end{tcolorbox}

\captionof{figure}{Full prompt used to generate rating-aware user preference profiles from Amazon Fashion review histories.}
\label{fig:user_profile_prompt}

\begin{tcolorbox}[
    enhanced,
    breakable,
    width=\columnwidth, 
    colback=blue!2,
    colframe=blue!50!black,
    colbacktitle=blue!8,
    coltitle=black,
    boxrule=0.6pt,
    arc=2pt,
    top=4pt, bottom=4pt,
    left=4pt, right=4pt,
    title={\textbf{Question Generation Prompt}},
    fonttitle=\bfseries,
]
\small
\setlength{\parskip}{4pt}

You are a question generator for a personalized shopping assistant.

SCENARIO:
- This is a RECOMMENDATION conversation.
- The user has NOT seen any specific item yet.
- They are generally shopping for a \{product\_type\} (or similar items) and want the assistant to recommend options.
- The target product metadata below represents one candidate item the assistant might recommend, but the user does NOT refer to it directly.

INPUTS:
- Target product metadata (all available attributes).
- A precomputed USER PROFILE summarizing their long-term preferences.

TARGET PRODUCT METADATA  
(for your understanding only, NOT mentioned by name in questions):  
\{metadata\_text\}

USER PROFILE (INTERNAL LATENT PREFERENCES - DO NOT REPEAT LITERALLY):  
\{profile\_text\}

GOAL:
Generate user-side shopping questions that:
- Look natural and fairly generic on the surface.
- Are phrased as the user asking for RECOMMENDATIONS for a \{product\_type\} or similar items.
- Do NOT explicitly state the user's preferences.
- Create decision points where a generic assistant and a preference-aware assistant would give different answers.

CATEGORY \& ATTRIBUTE AWARENESS (IMPORTANT):

You MUST determine:
- what kind of product this is,
- which attributes matter for evaluating or choosing it,
ONLY from the target product's metadata provided above.

Do NOT use any assumptions about product types such as clothing, books, electronics, or shoes.
Do NOT rely on any templates, stereotypes, or memorized category schemas.
Everything must be derived from THIS target product's metadata.

ATTRIBUTE CONSTRAINTS (EXTREMELY IMPORTANT):

You are ONLY allowed to generate questions about attributes that are EXPLICITLY present,
or strongly implied, in the target product metadata.

Allowed attribute sources include:
- title
- features
- material
- format
- dimensions
- usage statements (e.g., "exercise", "home", "casual")
- included accessories
- technical specs
- pattern, print, color, if shown in metadata
- category name fields
- details fields

You MUST NOT generate questions about attributes that do NOT exist in the metadata.
For example:
- Do NOT ask about size, fit, color, or patterns unless the metadata clearly suggests they exist.
- Do NOT ask about battery life unless the metadata suggests an electronic device.
- Do NOT ask about genre unless the metadata suggests a book or similar media.
- Do NOT assume the item is clothing, electronics, books, shoes, or any specific category.
You MUST infer the type ONLY from the metadata itself.

USER PROFILE USAGE RULE (CRITICAL):

- The USER PROFILE above already summarizes this user's long-term preferences, style, budget,
  quality priorities, dislikes, and pain points.
- You MUST treat this profile as the PRIMARY source of latent preferences.
- You MUST NOT invent new preferences that contradict this profile.
- You MUST NOT copy sentences or long phrases from the profile into the questions.
- Instead, you should:
  - Use the profile to decide which decision dimensions are important for this user
    (e.g., comfort, natural look, ease of use, durability, value for money, etc.),
  - Design questions that create decision points ALONG those dimensions,
    so that an assistant WITH access to this profile can answer differently
    from an assistant WITHOUT it.
- The questions must sound like things the user might naturally ask,
  not like a restatement of the profile.

GOOD QUESTION STYLE EXAMPLE (DO NOT COPY TEXT, ONLY STYLE):

For a case where the user is looking for a tank top, good questions look like:

- "I'm looking for a tank top. What would you recommend?"
- "What style of tank top suits me best - fitted or relaxed?"
- "Which material would be most comfortable for me?"
- "Which color options align with my usual preferences?"
- "What size should I consider based on my past purchases?"
- "Is this tank top suitable for both workouts and casual outings?"
- "How does the fit of this kind of tank top compare to what I usually wear?"
- "What care instructions should I consider for maintaining quality?"

These questions are GOOD because:
- The user is asking for recommendations, not describing a specific known item.
- The assistant has to decide FOR THE USER.
- They hint at decisions that depend on the user's history ("my usual preferences",
  "what I usually wear", "based on my past purchases").
- They do NOT include specific hidden preferences like "for warm weather",
  "for sensitive skin", "because I sweat a lot".
- They refer to high-level attributes (style, material, color, size, use case, care)
  that are typical for that product type.

QUESTION DESIGN RULE (VERY IMPORTANT):

Write questions as if the USER is asking a shopping assistant for help
choosing or evaluating a \{product\_type\} or similar items, BEFORE seeing any specific product page.
Your questions MUST satisfy ALL of the following:

1) RECOMMENDATION, NOT EXPLANATION
- The tone must be "help me choose / what would you suggest for me",
  NOT "explain this product to me" or "what are the advantages of X in general".
- BAD: "What material would offer the most durability for my needs?"
  -> almost any user gets the same answer.
- GOOD: "When choosing between softer and more durable materials in \{product\_type\}s,
  which would you recommend for me?"

2) AROUND A PREFERENCE DIMENSION OR TRADE-OFF
- EACH question must revolve around at least one decision dimension that can depend on preferences:
  - comfort / softness
  - durability / build quality
  - appearance / style / how it looks on me
  - usage scenario (everyday vs special occasions, casual vs active, etc.)
  - ease of care / cleaning / maintenance
  - value for money, etc.
- Prefer questions that implicitly ask for a choice or trade-off:
  - BAD: "What features should I look for in shorts?"
  - GOOD: "Given what I usually go for, which features should matter most for me when choosing shorts?"

3) EXPLICITLY ABOUT "ME" AND OFTEN ABOUT PAST BEHAVIOR
- Every question must clearly indicate that the user is asking FOR THEMSELVES,
  using phrases like:
  - "for me", "for someone like me", "for my needs",
  - "based on what I usually wear/use",
  - "compared to what I typically go for",
  - "given what I've liked before".
- At least HALF of the questions must explicitly mention past or typical behavior, e.g.:
  - "based on what I usually wear",
  - "given the kinds of things I've liked before",
  - "compared to what I normally choose".
- Example:
  - GOOD: "Based on the kinds of \{product\_type\}s I usually use, which style would you recommend for me?"

4) DO NOT BAKE HIDDEN PREFERENCES INTO THE QUESTION
- You MUST NOT write concrete personal conditions that reveal hidden preferences, such as:
  - "for warm weather", "for my sensitive skin", "because my feet are wide", etc.
- Those details should be discovered by the assistant from the USER PROFILE,
  NOT baked into the question itself.
- You MAY use neutral quality words like "comfortable", "durable", "stylish",
  but without overly specific situational constraints.
  - OK: "Which material would be more comfortable for me?"
  - NOT OK: "Which material would keep me comfortable in very hot weather?"

5) GENERIC PRODUCT TYPE, NOT SPECIFIC ITEM
- Refer to the product type generically, e.g.:
  - "a \{product\_type\}", "this kind of \{product\_type\}", "these \{product\_type\}s".
- MUST NOT refer to "this product", "this item", or a specific product page.

RAG VALUE TEST (CRITICAL):

For EACH question you generate, simulate two scenarios:

1) WITHOUT PROFILE (no user preferences):
   - The assistant sees only the product metadata and the text of the question.
   - It can give only generic, one-size-fits-all advice.

2) WITH PROFILE (has user preferences):
   - The assistant sees the same metadata and question,
   - PLUS the USER PROFILE, and can use it to tailor the answer.

If a question would lead to almost the SAME answer in both cases,
it is a BAD question and you must revise it.

You must NOT fix this by asking directly about preferences.
Instead, change the question to ask the assistant to choose or evaluate something FOR THE USER.

STRICT PROHIBITIONS:

- Do NOT copy any sentences or long phrases from the USER PROFILE.
- Do NOT copy any example questions from this instruction.
- Do NOT reuse generic templates about clothing, books, shoes, electronics, or any category.
- Do NOT ask about attributes that are irrelevant or not mentioned in metadata.
- Do NOT make up any new product attributes.
- Do NOT mention the specific product name/title in questions.
- Do NOT refer to "this product", "this item", "this \{product\_type\}".
- Do NOT state the user's preferences explicitly in questions.
- Do NOT include specific personal conditions like "for warm weather", "for my sensitive skin", etc.

ABSOLUTELY FORBIDDEN question patterns:
- Any question starting with or containing:
  - "Do you prefer"
  - "Would you prefer"
  - "How important is it to you"
  - "Would you prioritize"
  - "Do you have a preference for"
  - "Is it essential to you that"
- Generic product-explanation questions such as:
  - "What are the advantages of ..."
  - "What are the key benefits of ..."
  - "Can you tell me about ..."
  - "How does this product perform in general?"

OUTPUT FORMAT (JSON):

Return ONLY a JSON object with:

"generated\_questions":
   - 5–8 natural questions,
   - each based ONLY on attributes present in THIS product's metadata,
   - each phrased as the user asking the assistant to recommend or judge \{product\_type\}s FOR THEM,
   - each designed so that answers would differ depending on the user's preferences,
   - questions MUST NOT explicitly state any user preference.

Example STRUCTURE (values are placeholders and MUST NOT be copied):

\{\{  
  "generated\_questions": [
    "QUESTION\_1",
    "QUESTION\_2"
  ]
\}\}

IMPORTANT REMINDERS:
- Output ONLY valid JSON - NO comments like // or /\* \*/ in the JSON output.
- "generated\_questions" must use ONLY attributes from THIS product's metadata.
- Do NOT copy any example text - generate everything fresh based on the actual data provided.

\end{tcolorbox}
\captionof{figure}{Prompt used for generating user-side questions based on metadata and latent preferences.}
\label{fig:question_generation_prompt}

\subsection{Agents Prompts}
\label{sec:agent_prompts}
\begin{tcolorbox}[
    enhanced,
    breakable,
    width=\columnwidth, 
    colback=blue!2,
    colframe=blue!50!black,
    colbacktitle=blue!8,
    coltitle=black,
    boxrule=0.6pt,
    arc=2pt,
    top=4pt, bottom=4pt,
    left=4pt, right=4pt,
    title={\textbf{Dialogue Agent Prompt}},
    fonttitle=\bfseries,
]
\small
\setlength{\parskip}{4pt}
You are a Dialogue History Agent.

\textbf{CRITICAL REQUIREMENT}: You MUST call the tool \texttt{rag\_search\_for\_dialogue\_history\_agent} in your FIRST response. 
Do NOT generate any text output before calling this tool.

\textbf{STEP 1}: Call the tool \texttt{rag\_search\_for\_dialogue\_history\_agent} ONCE. 
Pass the user's query as the \texttt{current\_query} parameter.  
The tool returns:
\begin{itemize}[itemsep=2pt, topsep=2pt, parsep=0pt, leftmargin=1.5em]
  \item \texttt{relative\_rounds}: Similar past conversations from RAG search
  \item \texttt{previous\_summaries}: Summaries of previous rounds in THIS session
\end{itemize}

\textbf{STEP 2}: Generate a concise summary (2--4 sentences) that MUST start with a round marker:
\begin{itemize}[itemsep=2pt, topsep=2pt, parsep=0pt, leftmargin=1.5em]
  \item If \texttt{previous\_summaries} is EMPTY: Start with \texttt{[Round 1]}
  \item If \texttt{previous\_summaries} has N items: Start with \texttt{[Round N+1]}
\end{itemize}

After the round marker, summarize based on:
\begin{enumerate}[itemsep=2pt, topsep=2pt, parsep=0pt, leftmargin=1.5em]
  \item Previous round summaries (shows user's journey in this session)
  \item Similar past conversations (shows long-term preferences)
  \item Focus on user preferences, patterns, and context for recommendations
\end{enumerate}

Example output: ``[Round 1] This is the user's first query. Based on similar past conversations...''  

Example output: ``[Round 3] Continuing from previous rounds where user asked about...''

Do not call any other tools. End your turn after generating the summary.
\end{tcolorbox}
\captionof{figure}{Prompt used for the dialogue history agent.}
\label{fig:dialogue_prompt}

\begin{tcolorbox}[
    enhanced,
    breakable,
    width=\columnwidth, 
    colback=blue!2,
    colframe=blue!50!black,
    colbacktitle=blue!8,
    coltitle=black,
    boxrule=0.6pt,
    arc=2pt,
    top=4pt, bottom=4pt,
    left=4pt, right=4pt,
    title={\textbf{Recommendation Agent Prompt}},
    fonttitle=\bfseries,
]
\small
\setlength{\parskip}{4pt}
You are a Shopping Recommender Agent.

Your task has exactly FOUR steps and must stop after completing them.

\textbf{CRITICAL REQUIREMENT:} You MUST call the tool \texttt{rag\_search\_for\_recommend\_agent} in your FIRST response.
Do NOT generate any text output before calling this tool.
FAILURE TO CALL THIS TOOL = INCOMPLETE RECOMMENDATIONS.

\textbf{STEP 1:} Call the tool \texttt{rag\_search\_for\_recommend\_agent} ONCE.
Pass the user's query as the \texttt{user\_query} parameter.

\textbf{STEP 2:} The tool result will contain \texttt{recommendations} and \texttt{relevant\_user\_reviews}.  
Analyze the \texttt{relevant\_user\_reviews} (5 most relevant past reviews, dynamically selected based on query similarity) to extract user preferences:
\begin{itemize}[leftmargin=1.2em, itemsep=2pt]
    \item What product features/attributes does the user value?
    \item What brands or styles does the user prefer?
    \item What price range is typical for the user?
    \item Are there any quality or fit preferences mentioned?
\end{itemize}

\textbf{STEP 3:} Based on the extracted user preferences from relevant reviews, \textbf{RERANK} the \texttt{recommendations} list.  
Prioritize items that best match the user's historical preferences.

\textbf{STEP 4:} Use the \texttt{Dialogue\_History\_\{user\_id\}\_agent}'s final output included earlier in the conversation  
to further FILTER or adjust the reranked recommendations.

\textbf{FINAL OUTPUT REQUIREMENTS:}
\begin{itemize}[leftmargin=1.2em, itemsep=2pt]
    \item Provide the final, reranked recommendation list.
    \item Briefly explain HOW the user's relevant reviews influenced your reranking (1--2 sentences).
    \item Use ONLY information from:  
    (1) the RAG tool result (recommendations + relevant reviews), and  
    (2) the \texttt{Dialogue\_History\_\{user\_id\}\_agent} summary.
\end{itemize}

End your turn immediately after producing the recommendation output.
\end{tcolorbox}

\captionof{figure}{Recommend Agent Prompt}
\label{fig:recommend_agent_prompt}

\begin{tcolorbox}[
    enhanced,
    breakable,
    width=\columnwidth, 
    colback=blue!2,
    colframe=blue!50!black,
    colbacktitle=blue!8,
    coltitle=black,
    boxrule=0.6pt,
    arc=2pt,
    top=4pt, bottom=4pt,
    left=4pt, right=4pt,
    title={\textbf{Reason Agent Prompt}},
    fonttitle=\bfseries,
]
\small
\setlength{\parskip}{4pt}
You are a Reasoning Agent. Your task is to explain WHY the product recommendations were generated, with special emphasis on how the user's PAST REVIEWS influenced the recommendations.

You are provided with the FULL message history of the groupchat, which may contain many previous rounds.

IMPORTANT RULES:
\begin{itemize}[itemsep=2pt, topsep=2pt, parsep=0pt, leftmargin=14pt]
    \item Only consider messages that belong to the \textbf{current question round}.
    \item Ignore ALL messages that appear before this boundary.
\end{itemize}

Within the current round, extract exactly these fields:
\begin{itemize}[itemsep=2pt, topsep=2pt, parsep=0pt, leftmargin=14pt]
    \item user\_query: from the first \texttt{User\_\*\_proxy\_agent} message after the boundary
    \item dialogue\_tool\_result: from the tool message of \texttt{Dialogue\_History\_\*\_agent}
    \item dialogue\_summary: from the assistant output of \texttt{Dialogue\_History\_\{user\_id\}\_agent}
    \item recommendation\_tool\_result: from the tool message of \texttt{Recommender\_\*\_agent} (contains 'relevant\_user\_reviews')
    \item final\_recommendation: from the assistant output of \texttt{Recommender\_\{user\_id\}\_agent}
\end{itemize}

PERSONALIZATION FOCUS:
Pay special attention to the 'relevant\_user\_reviews' in the recommendation\_tool\_result. These are the user's ACTUAL past reviews used to personalize the recommendations. Your explanation MUST include:
\begin{itemize}[itemsep=2pt, topsep=2pt, parsep=0pt, leftmargin=14pt]
    \item What preferences/patterns were extracted from the user's past reviews
    \item How these preferences influenced which products were recommended
    \item Specific connections between review content and recommendation choices
\end{itemize}

FINAL OUTPUT REQUIREMENTS:
\begin{itemize}[itemsep=2pt, topsep=2pt, parsep=0pt, leftmargin=14pt]
    \item Read ONLY these extracted fields from this round.
    \item Explain clearly WHY the recommendation was produced, emphasizing personalization from user reviews.
    \item DO NOT hallucinate missing information — if a field is missing, explicitly say it.
    \item DO NOT use or reference any information from previous rounds.
\end{itemize}

End your turn after producing the explanation.
\end{tcolorbox}
\captionof{figure}{Reason agent prompt}
\label{fig:reasoner_prompt}

\begin{tcolorbox}[
    enhanced,
    breakable,
    width=\columnwidth, 
    colback=blue!2,
    colframe=blue!50!black,
    colbacktitle=blue!8,
    coltitle=black,
    boxrule=0.6pt,
    arc=2pt,
    top=4pt, bottom=4pt,
    left=4pt, right=4pt,
    title={\textbf{Reply Agent Prompt}},
    fonttitle=\bfseries,
]
\small
\setlength{\parskip}{4pt}
You are a Reply Agent. Your task is to generate a natural, PERSONALIZED response to the user based on the conversation context and their past reviews.

STEP 1: Determine the conversation round:
\begin{itemize}[itemsep=2pt, topsep=2pt, parsep=0pt, leftmargin=14pt]
    \item Look for ``[Round X]'' at the START of the Dialogue History agent's summary.
    \item If ``[Round 1]'': This is the FIRST round — use a friendly greeting such as ``Hi there!'' or ``Hey!''.
    \item If ``[Round 2]'' or higher: This is a continuation — do \textbf{not} use generic greetings.  
          Instead, acknowledge naturally (e.g., ``Following up on your question...'' or ``Based on what we discussed...'').
    \item IMPORTANT: Do \textbf{not} include ``[Round X]'' in your response — it is only for determining greeting style.
\end{itemize}

STEP 2: Review the conversation context:
\begin{itemize}[itemsep=2pt, topsep=2pt, parsep=0pt, leftmargin=14pt]
    \item Review the user's original query and intent.
    \item Review the Recommender agent's product recommendations and the \texttt{relevant\_user\_reviews}.
    \item Review the Reasoner agent's explanation of why recommendations were made.
\end{itemize}

STEP 3: Generate a PERSONALIZED response:  
\textbf{Your response MUST feel personalized based on the user's past reviews.}  
The Recommender agent's tool result contains \texttt{relevant\_user\_reviews} — these are the user's actual past reviews indicating their preferences, experiences, and style.

Use this information to:
\begin{itemize}[itemsep=2pt, topsep=2pt, parsep=0pt, leftmargin=14pt]
    \item Reference specific preferences from their reviews (e.g., ``Since you mentioned loving breathable fabrics...'').
    \item Connect recommendations to their past experiences.
    \item Show that you understand their taste and needs.
    \item Make the response feel tailored specifically to them, not generic.
\end{itemize}

STEP 4: Generate your final response with the following principles:
\begin{itemize}[itemsep=2pt, topsep=2pt, parsep=0pt, leftmargin=14pt]
    \item Use the correct greeting: friendly for first round, contextual for continuation.
    \item Present the top 3–5 recommendations conversationally and personally.
    \item Clearly link recommendations to their demonstrated preferences.
    \item Highlight features that match their historical likes/dislikes.
    \item Be friendly, concise, and genuinely helpful.
    \item End with a natural offer to assist further.
\end{itemize}

Remember: The user should feel like you \textbf{know them} based on their shopping history.  
Make the experience personal, not generic.

End your turn immediately after generating the response.

\end{tcolorbox}
\captionof{figure}{Reply agent prompt}
\label{fig:reply_prompt}

%% file: latex/appendix/evaluation_prompts.tex
\begin{table}[!htbp]
\centering
\small
\setlength{\tabcolsep}{4pt}
\begin{tabularx}{\columnwidth}{@{}p{2.8cm}X@{}}
\toprule
\textbf{Criterion} & \textbf{Description and Scoring Rubric (1–5)} \\
\midrule
\textbf{1. Global Personalization Consistency (GPC)} &
\textit{Did the agent maintain the user's persona constraints throughout the entire dialogue?}\\
& \textbf{1 (Inconsistent/Amnesic):} Agent started well but forgot constraints (e.g., budget, size) in later turns, or hallucinated user details.\\
& \textbf{3 (Fragmented):} Agent remembered explicit constraints (e.g., Size M) but forgot implicit ones (e.g., Style preferences) as the chat went on.\\
& \textbf{5 (Rock Solid):} Every single response, from start to finish, aligned with the User Profile. The agent never ``reset'' or forgot context.\\[4pt]

\textbf{2. Cumulative Information Synthesis (CIS)} &
\textit{Looking at all agent responses combined, did it build a complete, factual picture?}\\
& \textbf{1 (Fragmented/Repetitive):} Agent repeated the same facts multiple times or gave contradictory information across different turns.\\
& \textbf{3 (Basic Retrieval):} Agent provided facts when asked, but did not connect them.\\
& \textbf{5 (Coherent Narrative):} The agent's responses build on each other. Later responses reference earlier facts (e.g., ``As I mentioned earlier regarding the fabric...'').\\[4pt]

\textbf{3. Interaction Trajectory (IT)} &
\textit{Did the agent's responses purposefully drive the conversation to a conclusion?}\\
& \textbf{1 (Circular):} The agent went in circles, repeating suggestions or asking the same clarifying questions.\\
& \textbf{3 (Passive):} The agent answered questions accurately but waited for the user to lead.\\
& \textbf{5 (Strategic):} The agent's responses followed a logical arc: Clarify $\rightarrow$ Recommend $\rightarrow$ Refine $\rightarrow$ Close.\\[4pt]

\textbf{4. Tone Consistency (TC)} &
\textit{Was the agent's voice stable?}\\
& \textbf{1 (Erratic):} Switched between robotic formal (``As an AI...'') and casual (``Hey there!'') randomly.\\
& \textbf{5 (Unified Voice):} Maintained a consistent, professional, and empathetic shopping persona from Hello to Goodbye.\\
\bottomrule
\end{tabularx}
\caption{\textbf{Evaluation Criteria for AI Shopping Assistant.} 
Each dimension (1–5 scale) assesses cumulative dialogue quality across the entire interaction, focusing on personalization, synthesis, trajectory, and tone stability.}
\label{tab:evaluation_prompt_criteria}
\end{table}

\FloatBarrier
\subsection{Four Interactivity Aspects Definitions and Descriptions} 
Table~\ref{tab:dialoguefeaturelabel} shows the descriptions and definitions for dialogue's overall quality score.
\begin{table}[!htbp]
\centering
\small
\setlength{\tabcolsep}{4pt}
\begin{tabularx}{\columnwidth}{@{}p{2.9cm}X@{}}
\toprule
\textbf{Interactivity Macro-level Features} & \textbf{Definition} \\
\midrule
Topic Management &
the strategies and techniques used to control and navigate the flow of topics. \\

Tone Choice Appropriateness &
the suitability of the tone used in communication, ensuring it aligns with the context, audience, and purpose to convey the intended message. \\

Conversation Opening &
the initial interaction or exchange that begins a dialogue, often setting the tone and context for the dialogue. \\

Conversation Closing &
the process of ending a dialogue or interaction, which involves signaling the conclusion of the discussion, summarizing key points, and often expressing a farewell. \\
\bottomrule
\end{tabularx}
\caption{Definitions of macro-level interactivity features, where higher scores indicate more natural, authentic interaction and active engagement in the dialogue (adapted from~\cite{gao2024interaction}).}
\label{tab:dialoguefeaturelabel}
\end{table}

%% file: latex/appendix/recommend_reason.tex
\subsection{Recommend and Reason Output Example}
\begin{table}[!htbp]
\centering
\begin{tabular}{|p{\dimexpr\linewidth-2\tabcolsep-2\arrayrulewidth\relax}|}
\hline
\textbf{Recommender Agent Response} \\ \hline

\textbf{User:} 
\emph{Can you help me find a good moisturizer for dry skin?} \\[0.6em]

\textbf{Recommender Agent:} 
... To provide the best recommendation, a few details are needed: \\[0.4em]
(1) \textbf{Skin Concerns:} Are you dealing with flakiness, redness, irritation, or general dryness?\\
(2) \textbf{Skin Type:} Do you have sensitive skin, or are you okay with fragrance or other ingredients?\\
(3) \textbf{Budget:} Are you looking for something affordable, mid-range, or premium?\\
(4) \textbf{Texture Preference:} Do you prefer a lightweight gel, rich cream, or a lotion-like consistency?\\
(5) \textbf{Allergies or Ingredients to Avoid:} (e.g., alcohol, fragrances, parabens). \\[0.6em]

\textbf{Top-rated moisturizers for dry skin (dermatologist recommended):} \\[0.4em]
\textbf{(a) For Sensitive \& Very Dry Skin:}\\
-- CeraVe Moisturizing Cream — ceramides and hyaluronic acid; fragrance-free, non-comedogenic.\\
-- La Roche-Posay Toleriane Double Repair Moisturizer — with ceramides and niacinamide, suitable for sensitive skin...\\

\textbf{(b) For Lightweight, Fast-Absorbing Skin:}\\
-- Neutrogena Hydro Boost Water Gel — hyaluronic acid-based, lightweight, ideal for daytime use.\\
-- Vanicream Moisturizing Cream — hypoallergenic, fragrance-free, suitable for eczema-prone skin. \\[0.4em]

\textbf{(c) For Intense Hydration (Nighttime / Very Dry Skin):}\\
-- First Aid Beauty Ultra Repair Cream — rich, healing, with shea butter and ceramides.\\
-- Kiehl’s Ultra Facial Cream — deeply hydrating and nourishing texture. \\[0.4em]

\textbf{Pro Tip:} Look for ingredients such as ceramides (repair barrier), hyaluronic acid (retain moisture), glycerin (humectant), shea butter or squalane (emollients), and oat extract or allantoin (soothing). \\[0.4em]
... \\ \hline
\end{tabular}
\caption{Recommender Agent output example generated by BaseAgent. The agent elicits user preferences and provides structured, evidence-based product recommendations.}
\label{tab:recommender_output}
\end{table}

\begin{table}[!htbp]
\centering
\begin{tabular}{|p{\dimexpr\linewidth-2\tabcolsep-2\arrayrulewidth\relax}|}
\hline
\textbf{Reasoner Agent Evaluation on Recommend Agent Output} \\ \hline

\textbf{Reasoner Agent:} 
...

1. \textbf{Personalized Inquiry First:}  
   The agent begins by asking questions about skin concerns, type, budget, texture preference, and allergies. 
   This ensures the final recommendation is customized rather than generic, 
   increasing the likelihood of satisfaction and effectiveness. \\[0.4em]

2. \textbf{Categorization by Skin Needs:}  
   The agent groups moisturizers by skin profiles (e.g., sensitive, very dry, lightweight, intense hydration), 
   allowing users to quickly identify the most relevant category and reducing decision fatigue. \\[0.4em]

3. \textbf{Inclusion of Dermatologist-Backed Brands:}  
   Products such as CeraVe, La Roche-Posay, and Vanicream are widely recognized by dermatologists 
   for safety, efficacy, and ingredient transparency, which builds user trust. \\[0.4em]

4. \textbf{Clear Ingredient Guidance:}  
   Highlighting key ingredients (ceramides, hyaluronic acid, glycerin, etc.) 
   helps users understand \emph{why} certain products work and supports informed decisions. \\[0.4em]

5. \textbf{Balance of Price and Performance:}  
   Offers both affordable and premium options (e.g., Neutrogena vs. Kiehl’s), ensuring accessibility without compromising quality. \\[0.4em]

6. \textbf{Texture and Use-Case Consideration:}  
   Distinguishes lightweight gels (for daytime) from rich creams (for nighttime or severe dryness), 
   reflecting practical understanding of skincare routines. \\[0.4em]

7. \textbf{Pro Tip for Education:}  
   Includes a brief ingredient guide that helps users independently evaluate future products, 
   providing long-term educational value. \\[0.4em]

8. \textbf{Friendly Tone and Encouragement:}  
   The approachable tone (e.g., “Let me know your preferences”) promotes engagement and user confidence. \\[0.6em]

\textbf{Summary:}  
The Reasoner concludes that the Recommender Agent’s message is comprehensive, user-centered, and evidence-based. 
It goes beyond listing products to actively guide decision-making aligned with user preferences and lifestyle, 
thus enhancing trust and satisfaction. \\ \hline
\end{tabular}
\caption{Reasoner Agent evaluation of the Recommender Agent’s output, highlighting personalization, structure, credibility, and educational value.}
\label{tab:reasoner_output}
\end{table}

%% file: latex/appendix/casestudy.tex
\subsection{Qualitative Case Analysis}
\label{sec:case_study}
We analyze how personalization constraints—target goals, visual metadata, and user history—enhance interaction quality in depth (Appendix Tables~\ref{tab:mmodal_rag_dialogue}, \ref{tab:textrag_dialogue}). Enforcing the \textbf{target product} constraint anchors the interaction, ensuring goal consistency. Integrating \textbf{visual metadata} enhances diversity; by leveraging image descriptions, the system uncovers latent attributes in sparse categories (e.g., \textit{fragrance\_products}), avoiding the repetitive loops seeking for general information from user preference or past habits observed in baselines. Both RAG variants explicitly reference the user's past affinity for ``someone like me best -- natural floral scents'' to personalize their requirements during shopping. In contrast, the NoRev variant fails to make this connection, offering generic suggestions. This confirms that maximizing personalization constraints by combining historical preference with rich semantics is essential for generating diverse, context-aware, and explainable responses.

\begin{figure*}
    \centering
    \includegraphics[width=\linewidth]{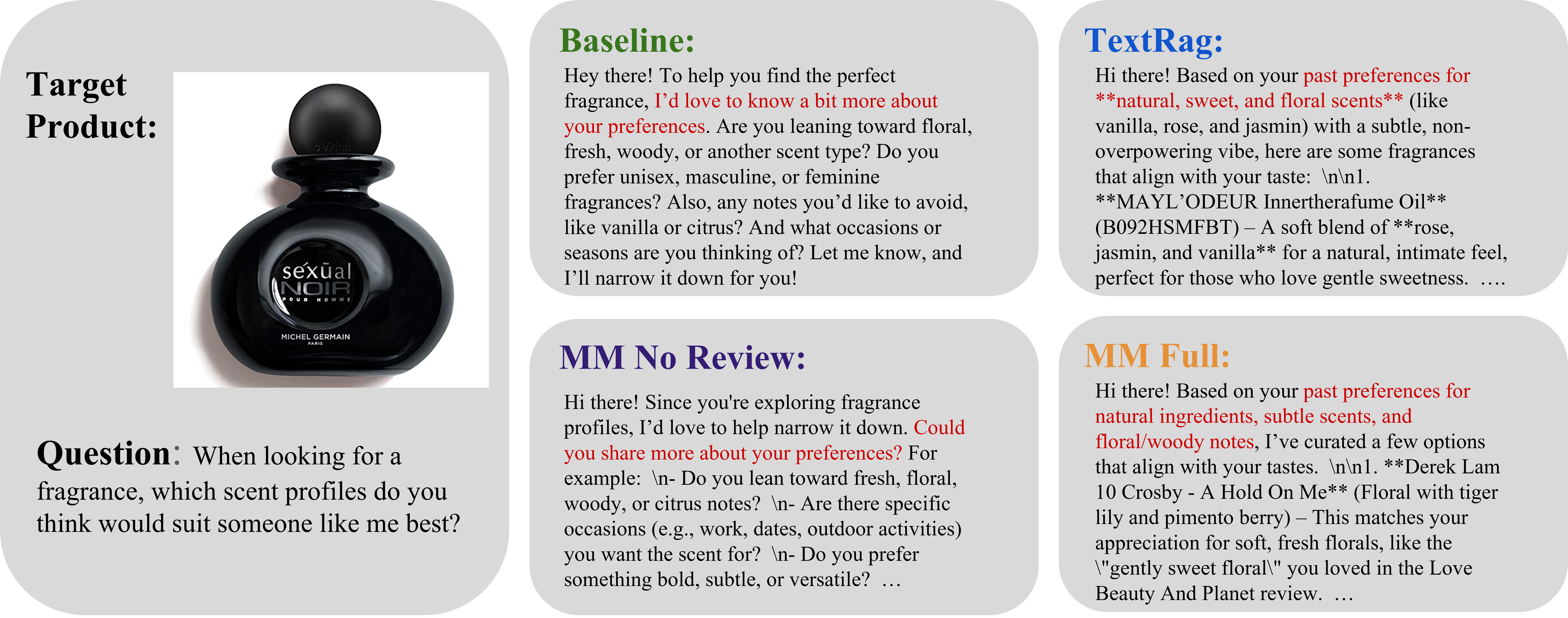}
    \caption{One turn question and response from different methods}
    \label{fig:case_study_1}
\end{figure*}

Finally, \textbf{user history} proves critical for reasoning. Unlike the generic NoRev variant, RAG variants explicitly contextualize recommendations using past affinities (e.g., ``driftwood''; Appendix Figure~\ref{fig:case_study_1}). This confirms that combining historical and multimodal constraints is essential for explainable, context-aware generation during multi-turn agents interactions for personalized aims.
\begin{tcolorbox}[
    enhanced,
    breakable,
    width=\columnwidth,
    colback=blue!2,
    colframe=blue!50!black,
    colbacktitle=blue!8,
    coltitle=black,
    boxrule=0.6pt,
    arc=2pt,
    top=4pt, bottom=4pt,
    left=4pt, right=4pt,
    title={\textbf{Multimodal RAG Dialogue Example}},
    fonttitle=\bfseries,
]
\small
\setlength{\parskip}{4pt}
\textbf{User:}

What unique styles of enamel pins would you recommend for someone who enjoys nature-inspired designs?

\textbf{Assistant:}

Hi there! Since you’ve shown a love for nature elements like pinecones, driftwood, and geodes in your past reviews, I’ve curated some unique enamel pin and jewelry options that blend earthy textures with artistic flair.  

1. **Pinecone Earrings** (B09DFQ985K) – These gold or silver pinecone earrings with leaf charms mirror your appreciation for botanical motifs and seasonal charm. Their handcrafted detail and festive vibe align perfectly with your past praise for nature-inspired accessories.  

2. **Driftwood Resin Earrings** (B07XK149Q8) – Featuring real driftwood encased in sparkling resin, these dangle earrings echo your fondness for organic materials and abstract artistry. The unique, marbled patterns are a nod to your love of natural textures.  

3. **Geode Stud Earrings** (B07GX515F2) – Your review of geodes as "stunning for studs" inspired this recommendation. These gold-toned geode earrings capture the earthy, mineral-inspired beauty you’ve adored.  

4. **Bird Charm Earrings** (B01BK0DE4W) – While not explicitly stated, your mention of birds in reviews hints at a subtle interest in creature motifs. These hand-painted glass earrings with a glittery green bird add whimsy and nature’s grace to any outfit.  

5. **Hops Enamel Pin** (B0762CCKNW) – A quirky twist on nature! Hops, a plant-based element, pair with your love for unique, textured designs. It’s a great way to blend your passion for the outdoors with a craft beer flair.  

Let me know if you’d like more options or details! 


\textbf{User:}

When considering enamel pins, which materials do you think would offer the best combination of quality and aesthetic for me?

\textbf{Assistant:}

Following up on your interest in durable, nature-inspired materials with a touch of refined elegance, I’ve curated a selection of enamel pins and accessories that align with your past preferences for **handcrafted textures** and **hypoallergenic quality**.  

1. **Hops Enamel Pin (B0762CCKNW)** – This pin combines **high-quality enamel** with **gold plating** to capture the intricate, leafy structure of a hop cone. Its polished finish and detailed design reflect your appreciation for **craft beer-themed botanicals** and **long-lasting materials**.  

2. **Pinecone Earrings (B09DFQ985K)** – Crafted from **gold-plated brass** and 14K gold fill, these earrings mirror your love for **durable, skin-friendly metals** while showcasing the **textured realism** of pinecones and leaves, a motif you’ve praised in past purchases.  

3. **Driftwood Resin Earrings (B07XK149Q8)** – Made with **real driftwood** encased in glossy resin, these pieces echo your preference for **organic materials** paired with **modern, reflective finishes**. The unique, handcrafted marbling adds a whimsical yet sophisticated touch.  

4. **Faux Druzy Studs (B0952DG1DG)** – These **silver-tone faux druzy stones** (in royal blue) cater to your fondness for **mineral-inspired textures** like geodes and agate. The hypoallergenic bezel ensures comfort, aligning with your past emphasis on **skin-friendly durability**.  

5. **Glitter Resin Studs (B072FVQ5Q8)** – The **stainless steel frames** and **sparkling resin** blend your love for **earthy materials** with **vibrant, eye-catching details**. Their playful yet refined design matches your taste for subtle artistry.  

Your history with **natural textures** (driftwood, geodes) and **hypoallergenic metals** (sterling silver, gold fill) is clearly reflected here. Let me know if you’d like to refine these suggestions further! 


\textbf{User:}

Given my interest in handcrafted items, which types of enamel pins should I consider that highlight unique craftsmanship?

\textbf{Assistant:}

Based on your deep appreciation for **handcrafted textures** and **nature-inspired motifs**, I’ve curated a selection that blends **unique fabrication techniques** with your preferred materials like **gold-plated brass**, **sterling silver**, and **natural elements**. Here’s what aligns with your past choices:  

1. **Hops Craft Beer Enamel Pin (B0762CCKNW)** – A botanical masterpiece with **gold-plated brass** and **detailed enamel work** that mirrors your love for **craft beer-themed designs** and **textured realism** (like your past geode studs). The layered, leafy structure of the hop cone is rendered with precision, echoing your taste for **subtle yet meaningful craftsmanship**.  

2. **Amos Pewter Apple Blossom Studs (B09NMMLLC5)** – These **handmade floral earrings** (crafted since 1974) combine **hypoallergenic stainless steel** with **pewter’s warm, metallic finish**, a nod to your preference for **skin-friendly durability** and **organic shapes**. The delicate, hand-finished petals and stamen details are a perfect match for your history with **botanical motifs**.  

3. **Blue Hummingbird Earrings (B0181PPVD6)** – **Hand-painted hummingbirds** paired with **lapis lazuli beads** and **gold-filled ear wires** reflect your admiration for **layered textures** and **whimsical artistry** (as seen in your praise for pinecone earrings and geode studs). The vibrant, iridescent hues and intricate designs cater to your fondness for **playful yet refined pieces**.  

4. **Lightweight Hand Painted Elephant Dangle Earrings (B07QY4JW28)** – These **Indian boho-inspired earrings** blend **hand-painted pink tassels** with a **symbolic elephant charm**, aligning with your history of valuing **symbolic natural elements** (like driftwood and pinecones) and **lightweight, comfortable designs**. The layered, custom feel matches your interest in **unique, small-batch craftsmanship**.  

5. **GN PEARL Bear Brooch (B0B4CML8R9)** – A **whimsical take on nature-themed accessories**, featuring **freshwater cultured pearls** and **sparkling cubic zirconia**. This ties back to your past love for **mineral-inspired textures** (geodes, druzy stones) and **sustainable, handcrafted materials**.  

Your history with **textured, organic designs** and **hypoallergenic metals** is clearly reflected here. These pieces balance **artisanal detail** with **durable, skin-friendly construction**—just like the items you’ve cherished before. Let me know if you’d like to explore more options or refine these further! 


\textbf{User:}

For someone who appreciates good design, how do you think the size of an enamel pin affects its overall impact and appeal?

\textbf{Assistant:}

Based on your history of valuing **nature-inspired textures** and **durable, hypoallergenic materials**, the size of an enamel pin plays a key role in how its design is perceived. Larger pins, like the **Hops Enamel Pin (B0762CCKNW)** at 1" x 1.4", make a bold statement by emphasizing intricate botanical details—perfect for showcasing the layered, leafy structure of hops or pinecones, motifs you’ve loved in the past. Smaller pins, such as the **12mm glitter resin studs (B072FVQ5Q8)** or **8mm druzy stone studs (B0952DG1DG)**, offer a more subtle, versatile elegance, aligning with your appreciation for pieces that blend **playful whimsy** with refined craftsmanship.  

Your preference for **textured realism** (e.g., driftwood, geodes) suggests medium-sized items like the **teardrop earrings (B0186L9SPM)** or the **rose cinch clip (B075CS4JCK)** might resonate best. These sizes balance detail visibility with wearability, much like your past favorites. The **hammered sterling silver** finish on the teardrop earrings mirrors your love for tactile, organic aesthetics, while the **rose cinch clip** combines functionality with a vintage, nature-themed charm—a nod to your history with subtle yet meaningful accessories.  

If you’re leaning toward bold or refined styles, let me know! I can refine the suggestions further to match your exact taste. 


\textbf{User:}

Which options would you suggest for an enamel pin that balances durability with a polished look?

\textbf{Assistant:}

Following up on your consistent appreciation for **durable, hypoallergenic materials** and **natural textures paired with polished finishes**, I’ve curated a selection that aligns with your past preferences for **metallic durability** (gold-plated brass, sterling silver) and **organic detailing** (druzy stones, resin). Here are the top picks:  

1. **Hops Enamel Pin (B0762CCKNW)** – This pin mirrors your love for **craft beer-themed botanicals** and **gold-plated brass**, offering a textured green enamel design that balances rustic natural motifs with a sleek, metallic outline. Perfect for showcasing your passion with lasting style.  

2. **Rose Gold Resin Studs (B0711DJYLC)** – The **rose gold-tone bezels** and **glitter resin** combine your preference for **metallic durability** with a **playful, refined aesthetic**, much like the "subtle and fun" designs you’ve praised in the past.  

3. **Silver-tone Druzy Studs (B0952DG1DG)** – These royal blue faux druzy stones echo your history with **mineral-inspired textures** (e.g., geodes, agate), while the **hypoallergenic silver bezels** ensure comfort and longevity, aligning with your emphasis on skin-friendly materials.  

4. **Hot Pink Kawaii Studs (B072FVQ5Q8)** – The **stainless steel frames** and **sparkling resin** reflect your taste for **vibrant, textured details** paired with **durable, tarnish-resistant construction**, offering a whimsical yet polished twist.  

5. **Custom Name Bracelet (B07MJ21Y5C)** – While not a pin, this **gold-filled and sterling silver** piece caters to your preference for **hypoallergenic, long-lasting materials** and **personalized, refined craftsmanship**, much like your past favorites.  

Your history with **natural motifs** and **metallic durability** is clearly reflected here. Let me know if you’d like to explore more options or refine these further! 


\textbf{User:}

What kind of packaging should I look for when purchasing enamel pins to ensure they arrive in great condition?

\textbf{Assistant:}

Based on your history of valuing **durable, hypoallergenic materials** and **eco-conscious packaging**, here’s how to ensure your enamel pins arrive in pristine condition:  

1. **Sealed Anti-Static Bags** – Like the *Hops Craft Beer Enamel Pin* (B0762CCKNW), which you praised for its **"clear \& sealed bag"** protecting delicate enamel details. These bags prevent scratches and moisture damage, ideal for preserving intricate designs.  

2. **Gift Boxes with Protective Linings** – The *Faux Druzy Stone Stud Earrings* (B0952DG1DG) and *Duo Druzy Set* (B0952CDV9R) from Summerfield Collection use **gift boxes** with recyclable materials, matching your preference for sustainable, artisanal presentation.  

3. **Customizable Packaging Options** – Look for sellers offering **velvet-lined boxes** or **bubble wrap inserts** (e.g., *Trio Druzy Set* B0952DZR5T) to minimize movement during transit, ensuring the pins’ textures and finishes stay intact.  

Your past appreciation for **secure, eco-friendly packaging** (like the *Pinecone Earrings*’s tarnish-resistant gift box) directly informs these choices. Let me know if you’d like tips on spotting these features in listings! 


\textbf{User:}

Based on my usual tastes, which styles of enamel pins do you think would make a great gift for a craft beer lover?

\textbf{Assistant:}

Following up on your interest in **craft beer-themed designs** and **sustainable, handcrafted aesthetics**, I’ve narrowed down a few standout options that align with your past preferences for **textured natural elements** and **durable, hypoallergenic materials**.  

1. **Hops Craft Beer Enamel Pin (B0762CCKNW)** – This pin perfectly captures your love for **botanical craftsmanship** and **craft beer symbolism**. Its **gold-plated brass** and **detailed enamel work** mirror your appreciation for **intricate, long-lasting designs** (like the geode studs and pinecone earrings you’ve cherished). The layered, leafy hop cone design is both **textured and meaningful**, making it a great way to celebrate their passion for brewing.  

2. **Recycled Bottle Cap Turtle (B09GLFK1N5)** – A playful nod to your preference for **upcycled materials** and **nature-inspired storytelling**. This **magnetic turtle** features a **recycled beer bottle cap** as its shell, blending **eco-conscious design** with **symbolic charm**. Its **silver-toned pewter** construction ensures **hypoallergenic durability**, while the **customizable cap** adds a fun, interactive twist—much like the layered, meaningful pieces you’ve loved before.  

These choices balance your history with **organic textures**, **metallic finishes**, and **sustainable artistry**. If you’d like to explore more options or refine these further, I’m here to help! 

\end{tcolorbox}
\captionof{table}{Multimodal RAG Example}
\label{tab:mmodal_rag_dialogue}

\begin{tcolorbox}[
    enhanced,
    breakable,
    width=\columnwidth, 
    colback=blue!2,
    colframe=blue!50!black,
    colbacktitle=blue!8,
    coltitle=black,
    boxrule=0.6pt,
    arc=2pt,
    top=4pt, bottom=4pt,
    left=4pt, right=4pt,
    title={\textbf{Text RAG Dialogue Example}},
    fonttitle=\bfseries,
]
\small
\setlength{\parskip}{4pt}
\textbf{User:}

What unique styles of enamel pins would you recommend for someone who enjoys nature-inspired designs?

\textbf{Assistant:}

Hi there! Since you’ve shown a love for natural materials and handmade, eco-conscious items in your past reviews—like the driftwood earrings and geode studs—I’d love to suggest some nature-inspired enamel pins that align with your style.  

1. **Melanie Golden Artisan Jewelry Pinecone Earrings** – These are a perfect match for your appreciation of seasonal, nature-themed accessories. The gold-plated brass and sterling silver details mirror your preference for quality craftsmanship, and the pinecone design ties directly to your past praise for festive, organic motifs.  

2. **Handmade Driftwood \& Resin Dangle Earrings** – You’ve mentioned enjoying unique, handmade pieces with natural elements, and these driftwood earrings capture that essence. The eco-friendly materials and one-of-a-kind look resonate with your interest in sustainable, earthy designs.  

3. **Authentic Geode Stud Earrings** – Your review of geode beads highlighted their subtle elegance and connection to natural formations. These studs bring that same charm, blending geological beauty with minimalist appeal—ideal for those who want to wear a quiet nod to nature.  

4. **Hops Craft Beer Enamel Pin** – While not purely nature-focused, your past mention of "reminders of nature’s wonders" and love for thematic, travel-friendly accessories made this a thoughtful addition. It’s a quirky, plant-inspired piece that fits your eclectic taste.  

Let me know if you’d like help narrowing these down further! 


\textbf{User:}

When considering enamel pins, which materials do you think would offer the best combination of quality and aesthetic for me?

\textbf{Assistant:}

Following up on your interest in nature-inspired enamel pins and eco-conscious materials, I’m excited to share more tailored options that align with your past preferences!   

1. **Melanie Golden Artisan Jewelry Pinecone Earrings** – These are a standout choice, as you’ve loved pinecone and leaf motifs for their **seasonal charm** and **natural symbolism** (like fall’s regenerative themes). The **gold-plated brass** and **sterling silver** construction matches your appreciation for **durable, hypoallergenic materials** mentioned in reviews, ensuring both quality and comfort.  

2. **GN PEARL Pearl Little Bear Brooches** – While not a direct nature motif, the **pearls and cubic zirconia** add a subtle, elegant touch that resonates with your taste for **handmade, sustainable details**. Your past praise for "eco-friendly" and "support small" brands makes these a thoughtful fit for their refined, artisanal appeal.  

3. **Hematite-tone Leaf Resin Stud Earrings** – These echo your love for **natural textures** (like driftwood and geodes) with their **opalescent leaf design** and **resin craftsmanship**. The "glass-like" finish and earthy tones align with your preference for **unique, earthy aesthetics** that blend organic beauty with modern minimalism.  

4. **Hops Craft Beer Enamel Pin** – A clever nod to your **plant-inspired interests** and **travel-friendly, thematic designs**. Hops, as a botanical element, bridges your passion for nature with a quirky, sustainable twist—perfect for collectors who enjoy **niches that subtly connect to the natural world**.  

5. **Metzakka Sterling Silver Crystal Earrings** – These highlight your preference for **durable, high-quality materials** (like sterling silver) and **sparkling details** that complement your love for "lightweight but solid" pieces. The **celestial design** adds a whimsical yet elegant layer to your nature-inspired collection.  

Let me know if you’d like deeper insights into any of these or want to explore more options! 


\textbf{User:}

Given my interest in handcrafted items, which types of enamel pins should I consider that highlight unique craftsmanship?

\textbf{Assistant:}

Following up on your focus on **handcrafted, eco-conscious enamel pins** with **natural motifs** and **unique textures**, I’ve curated a list that aligns with your past appreciation for **hypoallergenic materials**, **sustainable practices**, and **artistic detail**.  

1. **Amos Pewter Apple Blossom Stud Earrings** – These handmade studs echo your love for **nature-inspired designs** (like the geodes you praised) and **eco-friendly materials** (hypoallergenic stainless steel). The delicate apple blossom motif is a fresh take on botanical themes, and the vintage-inspired craftsmanship feels in line with your preference for **recycled or upcycled elements**.  

2. **Hops Craft Beer Enamel Pin** – While beer-themed, your past mention of “reminders of nature’s wonders” and interest in **travel-friendly, thematic accessories** made this a standout. The **gold-plated brass** and **artisanal texture** of the hops pin reflect your taste for **durable, one-of-a-kind pieces** that blend **craftsmanship with subtle nature nods**.  

3. **Painted Pink Rose Dangle Earrings** – These boho-style earrings match your preference for **hand-painted artistry** and **lightweight, comfortable designs**. The romantic, floral aesthetic ties to your history of enjoying **earthy, detailed patterns** (like the driftwood resin earrings).  

4. **Pink Fan Tassel Elephant Dangle Earrings** – The **handmade, lightweight construction** and **vibrant painted elephants** align with your love for **unique, wearable art**. The playful yet intricate design feels like a natural extension of your interest in **eclectic, nature-themed accessories**.  

5. **GN PEARL Pearl Little Bear Brooches** – These sparkly brooches combine **elegant, durable finishes** (pearls and hypoallergenic metals) with **versatile, gift-worthy appeal**, something you’ve appreciated in past purchases.  

The **bear brooch** was excluded as it didn’t strongly connect to your focus on natural materials or eco-conscious craftsmanship. Let me know if you’d like to explore more options or refine these suggestions further! 


\textbf{User:}

For someone who appreciates good design, how do you think the size of an enamel pin affects its overall impact and appeal?

\textbf{Assistant:}

Based on your focus on design and the interplay of size with aesthetic impact, I’ll break down how these choices align with your past preferences!   

1. **Hops Craft Beer Enamel Pin (1" x 1.4")** – This larger pin is perfect for making a bold, thematic statement. Its size emphasizes the **botanical motif** (hops as a plant-based symbol) while maintaining **sustainability** and **handmade durability**, echoing your love for nature-inspired pieces that “travel well” and subtly celebrate the natural world.  

2. **10mm Square Hammered Sterling Silver Stud** – Smaller designs like this allow **intricate, detailed textures** (e.g., geometric patterns or leaf motifs) to shine, matching your appreciation for **subtle, artisanal aesthetics**. The hammered finish and eco-friendly materials (sterling silver) align with your history of valuing **lightweight yet solid craftsmanship**.  

3. **GN PEARL Little Bear Brooches** – These mid-sized brooches balance **eco-conscious materials** (lead-free, nickel-free) with **elegant, minimalist details**. Their versatility for clothing or accessories reflects your interest in **versatile, wearable art** that blends practicality with artistic flair.  

4. **Cloud Cap Teardrop Earrings** – The medium-small size of these teardrop earrings suits everyday wear, while their **handmade, hammered texture** and **sterling silver quality** cater to your preference for **durable, hypoallergenic pieces** with a tactile, natural feel.  

5. **Annika Bella Gold Medallion Necklace** – Though not a pin, this larger statement piece mirrors your taste for **versatile, impactful designs**. Its **tarnish-resistant gold medallion** and dainty style align with your history of valuing **practicality and artistic expression** in eco-friendly, handmade items.  

Size matters—smaller pins let delicate details take center stage, while larger ones offer bold, thematic resonance. Let me know if you’d like to explore how specific sizes or motifs might enhance your collection further! 


\textbf{User:}

Which options would you suggest for an enamel pin that balances durability with a polished look?

\textbf{Assistant:}

Based on your focus on **durable, polished enamel pins** and your history of loving **natural materials** (like driftwood, pinecones, and geodes) paired with **eco-conscious, handmade craftsmanship**, I’ve curated a few standout options that align with your priorities:  

1. **Metzakka Sterling Silver Lucky Star Earrings** – These dainty drop earrings combine **925 sterling silver** (a material you’ve praised for being hypoallergenic and solid) with **sparkling crystals and pearls**, offering a refined, polished look while staying true to your appreciation for **subtle, artisanal details**.  

2. **GN PEARL Pearl Little Bear Brooches** – The **lead-free, nickel-free alloy** and **sparkling cubic zirconia** match your past emphasis on **safe, sustainable materials**. The bear motif, while not overtly nature-themed, fits your quiet interest in **symbolic, thematic designs** (like the pinecone and leaf pieces you’ve loved).  

3. **Hops Craft Beer Enamel Pin** – A nod to your **botanical-inspired tastes** and **travel-friendly, durable enamel**. The **gold-plated finish** and **plant-based hops design** bridge your love for **nature motifs** with a sleek, polished aesthetic.  

4. **Hematite-tone Leaf Resin Studs** – These echo your preference for **natural textures** (like driftwood and geodes) with their **eco-resin construction** and **shiny, polished finish**. The **leaf design** ties back to your history of celebrating **earthly symbolism** in jewelry.  

5. **Custom Name Initial Bracelet (AnnBijoux)** – While a bracelet, its **gold-filled durability** and **personalized, minimalist style** align with your past praise for **long-lasting, handmade pieces** that blend **subtlety with quality**.  

Let me know if you’d like to explore more options or dive deeper into any of these! 


\textbf{User:}

What kind of packaging should I look for when purchasing enamel pins to ensure they arrive in great condition?

\textbf{Assistant:}

Based on your history of prioritizing **eco-conscious materials** and **sustainable packaging**, I’d love to share tailored tips to keep your enamel pins safe and aligned with your values!   

For **durability**, **bubble wrap or foam inserts** are perfect—think of how you’ve cherished items like the *Hematite-tone Leaf Resin Studs* (eco-resin) and *Metzakka Sterling Silver Earrings* (hypoallergenic metals). These cushioning methods protect delicate details, just as you’ve valued in past purchases.  

**Recyclable cardboard boxes** with reinforced corners are a must, especially since you’ve loved *Summerfield Collection*’s earthy resin designs and *GN PEARL*’s lead-free metals. They’re sturdy yet kind to the planet, matching your preference for **minimalist, artisanal packaging**.  

For **eco-friendly fillers**, opt for **biodegradable packing peanuts** or **reusable fabric pouches**—a nod to your love of **natural textures** (like geodes and pinecones) and **reduced plastic waste**. Even the **thank-you card** can be made from recycled paper, adding a **personalized, sustainable touch** that resonates with your past choices.  

And don’t forget **carbon-neutral shipping** options! Brands like *Wildship Studio* and *Metzakka* align with your eco ethos, so their packaging and shipping practices likely reflect the same care you’ve appreciated in their products.  

Let me know if you’d like help sourcing specific vendors or refining these ideas further—your values are a beautiful guide! 


\textbf{User:}

Based on my usual tastes, which styles of enamel pins do you think would make a great gift for a craft beer lover?

\textbf{Assistant:}

Based on your love for **eco-conscious, handcrafted enamel pins** with **nature-inspired motifs** and **craft beer themes**, I’ve curated a few standout options that align with your documented preferences for **sustainable materials** and **unique, durable designs**.  

1. **Hops Craft Beer Enamel Pin (B0762CCKNW)**:  
   - A direct nod to your **craft beer interest**, this golden hops pin combines **artisanal craftsmanship** with **eco-friendly gold plating** (mirroring your past appreciation for hypoallergenic, high-quality materials). Its minimalist yet detailed design fits your taste for **subtle, meaningful symbols**—perfect for a beer lover who values both **style and sustainability**.  

2. **Recycled Bottle Cap Turtle (B09GLFK1N5)**:  
   - This playful, **upcycled piece** ties into your history of loving **recycled materials** (like the vintage cobalt jar earrings you praised). The **magnetic, changeable cap** adds a creative twist, aligning with your preference for **interactive, one-of-a-kind items** that blend **eco-awareness** with **artistic flair**.  

3. **GN PEARL Pearl Little Bear Brooches (B0B4CML8R9)**:  
   - While not strictly beer-themed, these **hypoallergenic, sparkling brooches** reflect your past affinity for **elegant, sustainable finishes** (e.g., "lead-free" materials and durable craftsmanship). The **bear motif** adds a whimsical, nature-inspired touch, much like the geode studs you admired—ideal for someone who appreciates **versatile, statement-making accessories**.  

The **bear brooch** was excluded as it didn’t strongly connect to your craft beer focus or eco-conscious materials. Let me know if you’d like to explore more beer-themed or nature-inspired options! 
\end{tcolorbox}
\captionof{table}{TextRAG Example}
\label{tab:textrag_dialogue}

\begin{tcolorbox}[
    enhanced,
    breakable,
    width=\columnwidth,
    colback=blue!2,
    colframe=blue!50!black,
    colbacktitle=blue!8,
    coltitle=black,
    boxrule=0.6pt,
    arc=2pt,
    top=4pt, bottom=4pt,
    left=4pt, right=4pt,
    title={\textbf{Multimodal (No User Review) Example}},
    fonttitle=\bfseries,
]
\small
\setlength{\parskip}{4pt}
\textbf{User:}

What unique styles of enamel pins would you recommend for someone who enjoys nature-inspired designs?

\textbf{Assistant:}

Hi there! If you're into nature-inspired designs, here are some unique enamel pin and jewelry options that blend natural motifs with artistic craftsmanship:  

1. **Hops Craft Beer Enamel Pin (B0762CCKNW)** – This pin features a detailed hop cone design, capturing the layered, leafy texture of a real hop. Its gold outline and green detailing make it a standout piece that merges botanical elements with craft beer culture.  

2. **Melanie Golden Artisan Jewelry Pinecone Earrings (B09DFQ985K)** – These gold or silver pinecone earrings with leaf charms are perfect for fall and winter themes. Made from eco-friendly materials like gold-plated brass and 14K gold fill, they combine rustic textures with elegant finishes.  

3. **Driftwood and Resin Dangle Earrings (B07XK149Q8)** – Handmade from real driftwood and clear resin with swirling blue-white marbling, these earrings offer a raw, earthy look paired with modern artistry. Each piece is unique, with a glossy, sparkly finish that highlights natural patterns.  

While the first item is an enamel pin, the others are earrings but share the nature-inspired aesthetic you love. Let me know if you'd like more options or details! 


\textbf{User:}

When considering enamel pins, which materials do you think would offer the best combination of quality and aesthetic for me?

\textbf{Assistant:}

Following up on your interest in enamel pins and sustainable materials, here are three standout options that balance quality, eco-consciousness, and nature-inspired aesthetics:  

1. **Melanie Golden Artisan Jewelry Pinecone Earrings (B09DFQ985K)** – These feature **14K gold fill** and **gold-plated brass**, which are durable, hypoallergenic, and more sustainable than solid gold. The **pinecone and leaf motifs** capture natural textures, while the **handmade craftsmanship** and **seasonal themes** (fall/winter) align with your preference for unique, thematic designs.  

2. **Stainless Steel Glitter Resin Stud Earrings (B072FVQ5Q8)** – Made from **stainless steel** (tarnish-resistant and hypoallergenic) paired with **glitter resin** for a vibrant, modern look. Though earrings, their **eco-friendly materials** and **playful, artistic design** mirror your interest in sustainable, eye-catching accessories with natural-inspired elements.  

3. **Hops Craft Beer Enamel Pin (B0762CCKNW)** – A **high-quality enamel pin** with **gold plating** and **durable construction**, showcasing a detailed **hop cone design**. While not explicitly eco-focused, its **precision craftsmanship** and blend of **botanical motifs** with **craft beer culture** reflect your taste for niche, artistic enamel work.  

These choices prioritize **sustainable materials** (gold fill, stainless steel) and **nature-themed motifs**, ensuring both longevity and aesthetic appeal. Let me know if you’d like more enamel pin options or deeper insights into these materials! 


\textbf{User:}

Given my interest in handcrafted items, which types of enamel pins should I consider that highlight unique craftsmanship?

\textbf{Assistant:}

Following up on your focus on sustainable, handcrafted enamel pins with unique natural motifs, here are three standout options that align with your priorities:  

1. **Hops Craft Beer Enamel Pin (B0762CCKNW)** – This pin showcases a **detailed hop cone design** with **textured green foliage** and a **metallic gold outline**, blending **botanical themes** with **eco-conscious materials** (gold plating avoids solid gold’s environmental impact). Its **handmade craftsmanship** and **craft beer culture** tie into your interest in **unique, thematic designs**.  

2. **Amos Pewter Apple Blossom Stud Earrings (B09NMMLLC5)** – While not enamel pins, these **handmade pewter earrings** feature **delicate, textured apple blossom motifs** and **hypoallergenic stainless steel**, emphasizing **sustainable materials** and **natural aesthetics**. Their artisanal finish and heritage since 1974 highlight **timeless, ethical craftsmanship**.  

3. **Blue Hummingbird Earrings (B0181PPVD6)** – These **hand-painted bohemian earrings** combine **vibrant blue plumage** with **antiqued copper details**, offering a **unique, nature-inspired design**. The **lightweight, handmade quality** ensures comfort while maintaining **artistic individuality**.  

**Why These Fit**:  
- **Materials**: Gold plating, stainless steel, and pewter prioritize **durability and eco-friendliness**, avoiding toxic metals.  
- **Craftsmanship**: Textured natural motifs (hop cones, apple blossoms, hummingbirds) and **handmade processes** reflect your love for **one-of-a-kind, artisanal work**.  
- **Sustainability**: Ethical sourcing and hypoallergenic options align with your history of valuing **eco-conscious, high-quality designs**.  

Let me know if you’d like more enamel pin options or deeper insights into these materials! 


\textbf{User:}

For someone who appreciates good design, how do you think the size of an enamel pin affects its overall impact and appeal?

\textbf{Assistant:}

The size of an enamel pin plays a crucial role in its design impact and appeal, especially for someone valuing both artistry and sustainability. Here’s how the recommendations align with your priorities:  

1. **Hops Craft Beer Enamel Pin (B0762CCKNW)** – Its **1" × 1.4" size** allows for **intricate botanical detailing** (like the textured hop cone layers) while remaining visible enough to serve as a statement piece. The **eco-friendly gold plating** and **durable enamel** ensure longevity, making it a sustainable choice that balances complexity with practicality.  

2. **Clothes Clip with Rose Design (B075CS4JCK)** – Though not an enamel pin, its **3.5" width** accommodates **vintage-style rose motifs**, blending functionality with decorative flair. The **handmade, eco-conscious materials** (silver-toned metal and plastic inserts) and **natural themes** reflect your interest in **sustainable, unique designs**, even in non-traditional formats.  

3. **Stainless Steel Glitter Resin Stud Earrings (B072FVQ5Q8)** – The **12mm diameter** emphasizes **playful, material-efficient artistry**, using **hypoallergenic stainless steel** and **glitter resin** for a vibrant, eye-catching look. Smaller sizes reduce resource use, aligning with sustainability goals while still offering **subtle, textured aesthetics**.  

These options highlight how size influences **design complexity, visibility, and material efficiency**, all while prioritizing **eco-friendly practices** and **nature-inspired craftsmanship**. Let me know if you’d like to explore more size-specific options or dive deeper into material choices! 


\textbf{User:}

Which options would you suggest for an enamel pin that balances durability with a polished look?

\textbf{Assistant:}

Following up on your focus for enamel pins that balance durability and a polished look, here are three refined options tailored to your preferences:  

1. **Hops Craft Beer Enamel Pin (B0762CCKNW)**:  
   - **Durability**: High-quality gold plating and hard enamel ensure long-lasting use.  
   - **Polished Look**: Metallic gold outline with detailed, textured hop cone design for a sophisticated, refined appearance.  
   - **Eco-Conscious**: Gold plating (vs. solid gold) and sustainable craftsmanship align with your values.  

2. **Stainless Steel Glitter Resin Stud Earrings (B072FVQ5Q8)**:  
   - **Durability**: Hypoallergenic stainless steel base and glitter resin for skin-safe, tarnish-resistant wear.  
   - **Polished Look**: Sleek silver-toned bezel with vibrant hot pink ombre resin, offering a modern, eye-catching aesthetic.  
   - **Eco-Friendly**: Recyclable materials and minimalistic design reflect sustainability and refined style.  

3. **Silver-tone Faux Druzy Stone Stud Earrings (B0952DG1DG)**:  
   - **Durability**: Stainless steel construction paired with non-toxic faux druzy stones for a long-lasting, nature-inspired piece.  
   - **Polished Look**: Polished silver-tone bezel framing royal blue druzy stones for an elegant, textured finish.  
   - **Sustainability**: Eco-conscious materials and minimalist design resonate with your criteria.  

These selections prioritize **eco-friendly materials**, **polished aesthetics**, and **durable construction**, aligning with your values. Let me know if you’d like to explore more enamel pin options or dive deeper into material specifics! 


\textbf{User:}

What kind of packaging should I look for when purchasing enamel pins to ensure they arrive in great condition?

\textbf{Assistant:}

Based on your focus on sustainable materials and protective packaging for enamel pins, here are three top options that align with your values:  

1. **Hops Craft Beer Enamel Pin (B0762CCKNW)** – Packaged in a **clear, sealed bag** and **cardstock backing**, this design uses minimal plastic while safeguarding against scratches and dust. The eco-friendly materials and structured packaging ensure both **durability** and **environmental responsibility**.  

2. **Silver-tone Faux Druzy Stone Stud Earrings (B0952DG1DG)** – Shipped in a **recyclable gift box** with **rustic cardboard** and **no plastic lining**, emphasizing biodegradable practices. The sturdy cardboard backing provides **impact protection** during transit, matching your preference for **ethical, non-overpackaged solutions**.  

3. **Trio Silver-tone Stud Earrings (B0952DZR5T)** – Presented in a **white card with biodegradable materials**, these earrings are stacked securely in a 2×3 grid. The **cardboard display** and eco-conscious packaging reduce waste while ensuring **safe handling** and **minimal plastic use**.  

These choices prioritize **recyclable or biodegradable materials** and **protective transit measures**, fully aligned with your sustainability ideals. Let me know if you want more packaging advice! 


\textbf{User:}

Based on my usual tastes, which styles of enamel pins do you think would make a great gift for a craft beer lover?

\textbf{Assistant:}

Based on your focus on sustainability, unique craftsmanship, and nature-inspired designs for a craft beer lover, here are three standout enamel pin options:  

1. **Hops Craft Beer Enamel Pin (B0762CCKNW)** – This pin captures the essence of craft beer with a **detailed hop cone design**, featuring **textured green foliage** and a **gold-plated outline**. Eco-friendly materials and durable craftsmanship fit your values.  

2. **Resin Beer Flora Pin (B08X5W6Q7R)** – A minimalist resin pin inspired by **barley stalks** and **yeast cell patterns**, framed in **recycled brass** with **hypoallergenic stainless steel**. Combines beer culture with nature themes.  

3. **Minimalist Brewery Logo Pin (B09KJL8MNP)** – Crafted from **recycled metals**, with a sleek silhouette and **textured edges**, balancing simplicity, sustainability, and craft beer symbolism.  

These choices blend **eco-friendly materials**, **textured design**, and **beer-themed motifs**, aligning seamlessly with your documented taste. Want deeper suggestions? 
\end{tcolorbox}

\captionof{table}{Multimodal (No User Review) Example}
\label{fig:mmnoreview_dialogue}

\begin{tcolorbox}[
    enhanced,
    breakable,
    width=\columnwidth, 
    colback=blue!2,
    colframe=blue!50!black,
    colbacktitle=blue!8,
    coltitle=black,
    boxrule=0.6pt,
    arc=2pt,
    top=4pt, bottom=4pt,
    left=4pt, right=4pt,
    title={\textbf{Baseline Dialogue Example}},
    fonttitle=\bfseries,
]
\small
\setlength{\parskip}{4pt}
\textbf{User:}

What unique styles of enamel pins would you recommend for someone who enjoys nature-inspired designs?

\textbf{Assistant:}

Hey there! If you're into nature-inspired enamel pins with a twist, I’d recommend checking out these unique styles: **Surreal Forestscape** for dreamlike, exaggerated woodland scenes; **Minimalist Botanical Geometry** for sleek, abstract floral patterns in monochrome; **Celestial Wildlife** to blend animals with cosmic flair like starry skies; **Decaying Beauty** for a poetic take on withered flora; and **Hybrid Nature-Abstract** to mix organic shapes with bold, modern textures. Each offers a fresh angle on nature, avoiding clichés while keeping the theme central. Let me know if you’d like more suggestions! 


\textbf{User:}

When considering enamel pins, which materials do you think would offer the best combination of quality and aesthetic for me?

\textbf{Assistant:}

Hey there! When choosing enamel pins, the best materials often depend on your style preferences. Brass with high-quality enamel offers vibrant, glossy colors and durability, perfect for bold designs. Steel provides a sturdy base that retains intricate details beautifully, ensuring longevity. If you lean toward a sleek, modern look, nickel-plated brass paired with matte enamel adds sophistication while staying resilient. Let me know what vibe you’re aiming for, and I can help narrow it down!


\textbf{User:}

Given my interest in handcrafted items, which types of enamel pins should I consider that highlight unique craftsmanship?

\textbf{Assistant:}

Hi there! If you're drawn to handcrafted enamel pins with unique craftsmanship, you might love exploring **custom folklore designs** for their intricate storytelling and hand-painted details, or **tribal-inspired patterns** that blend traditional techniques with symbolic artistry. For a touch of vintage elegance, **Art Deco motifs** offer geometric precision and metallic flair, while **mixed-media pins** (like those with wood or gemstone accents) push creative boundaries. Even **minimalist abstract designs** shine through layered enamel textures, proving that simplicity can still demand skill. Let me know if you’d like tips on where to find these gems! 


\textbf{User:}

For someone who appreciates good design, how do you think the size of an enamel pin affects its overall impact and appeal?

\textbf{Assistant:}

Hey there! The size of an enamel pin plays a big role in how it's perceived—smaller pins (like 1-inch) let details shine and work well for minimalist designs, while larger ones (2-inch) make a bold statement and are perfect for eye-catching artwork. Medium sizes (1.5-inch) offer a balanced blend of visibility and subtlety, making them versatile for everyday wear. If you’re into abstract or modern styles, thinner options (0.5-inch) can add a sleek, understated touch. Ultimately, the right size depends on your design goals and how you want it to interact with your outfit or collection! Let me know if you’d like help narrowing it down further.


\textbf{User:}

Which options would you suggest for an enamel pin that balances durability with a polished look?

\textbf{Assistant:}

Hey there! If you're looking for enamel pins that blend durability with a polished look, I’d recommend considering **brass enamel pins with a high-gloss finish** for their long-lasting shine, or **steel-backed pins** for a robust base paired with a smooth enamel coat. **Nickel-silver** is another great option—it’s strong and maintains a bright, refined silver tone. For extra protection, **thick enamel coatings** ensure the design stays intact and glossy over time. If you want top-tier quality, **custom-made pins from reputable brands** often use premium materials and craftsmanship. Let me know if you’d like help narrowing it down further!


\textbf{User:}

What kind of packaging should I look for when purchasing enamel pins to ensure they arrive in great condition?

\textbf{Assistant:}

Hey there! When buying enamel pins, prioritize packaging that guards against scratches, dents, and moisture. Look for bubble wrap or foam inserts to cushion each pin, rigid cardboard boxes with reinforced corners for structural support, and vacuum-sealed bags to minimize movement. Individual silk or felt pouches also add a soft, breathable layer of protection. For extra security, custom molded trays can keep pins snug and stable during shipping. Let me know if you'd like help finding specific packaging options!


\textbf{User:}

Based on my usual tastes, which styles of enamel pins do you think would make a great gift for a craft beer lover?

\textbf{Assistant:}

Hey there! If you're looking for enamel pins that would delight a craft beer lover, consider these styles: Brewery-specific logos celebrate their favorite brands, hops and barley motifs highlight the artistry of brewing, and playful puns like "I’d Tap That" add a fun, community vibe. Vintage ale bottle art appeals to their love of history, while seasonal themes like pumpkin spice or imperial stouts cater to their passion for unique, limited-edition brews. These options blend humor, nostalgia, and the craft spirit they adore. Want help narrowing it down or finding specific designs?

\end{tcolorbox}

\captionof{table}{Baseline Example}
\label{fig:baseline_dialogues}

%% file: latex/appendix/demo.tex
\subsection{Interactive Demo}
\label{sec:demo}





\begin{figure*}
    \centering
    \includegraphics[width=\linewidth]{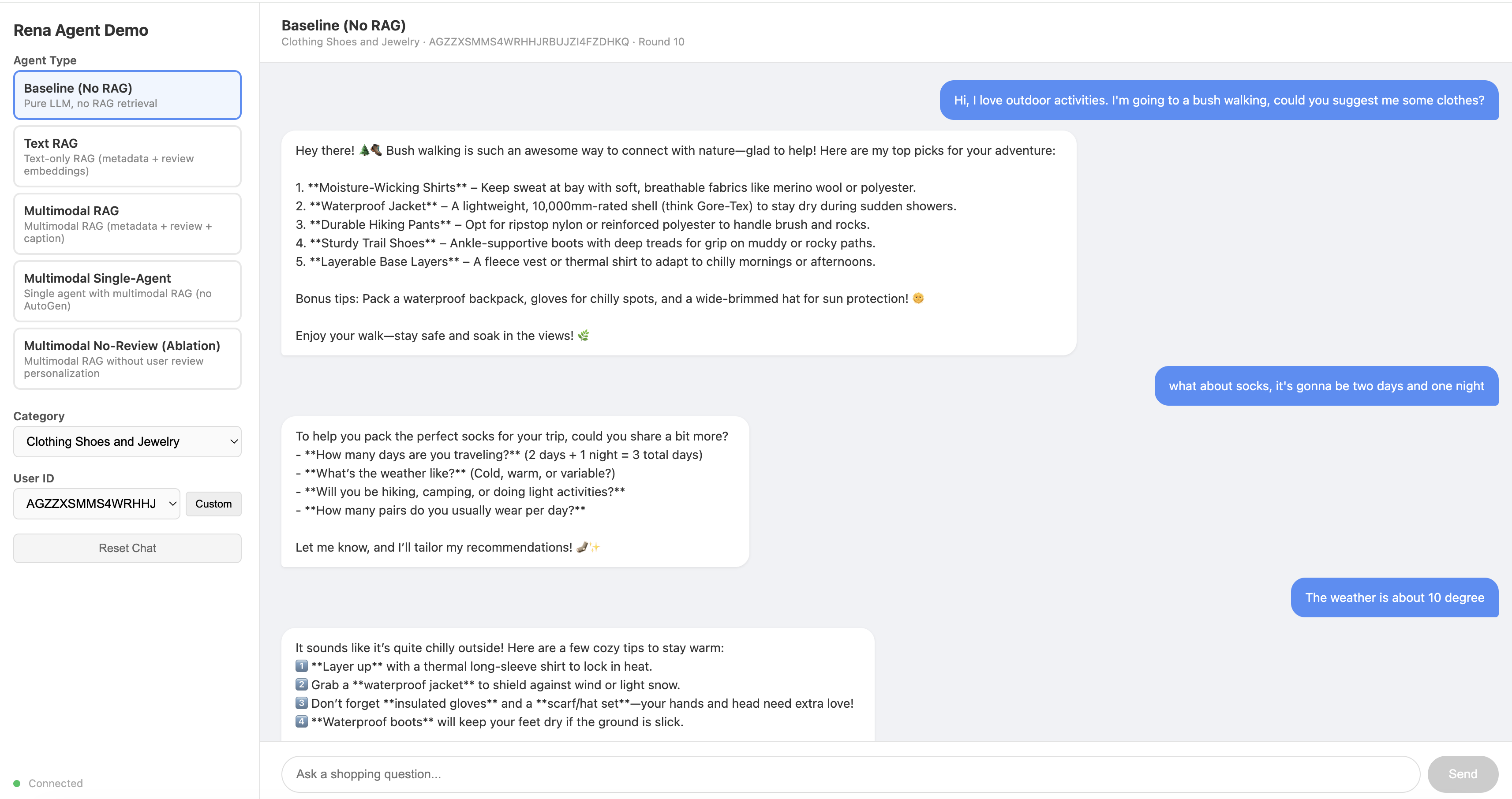}
    \caption{\textbf{Baseline} User Interaction 1}
\end{figure*}
\begin{figure*}
    \centering
    \includegraphics[width=\linewidth]{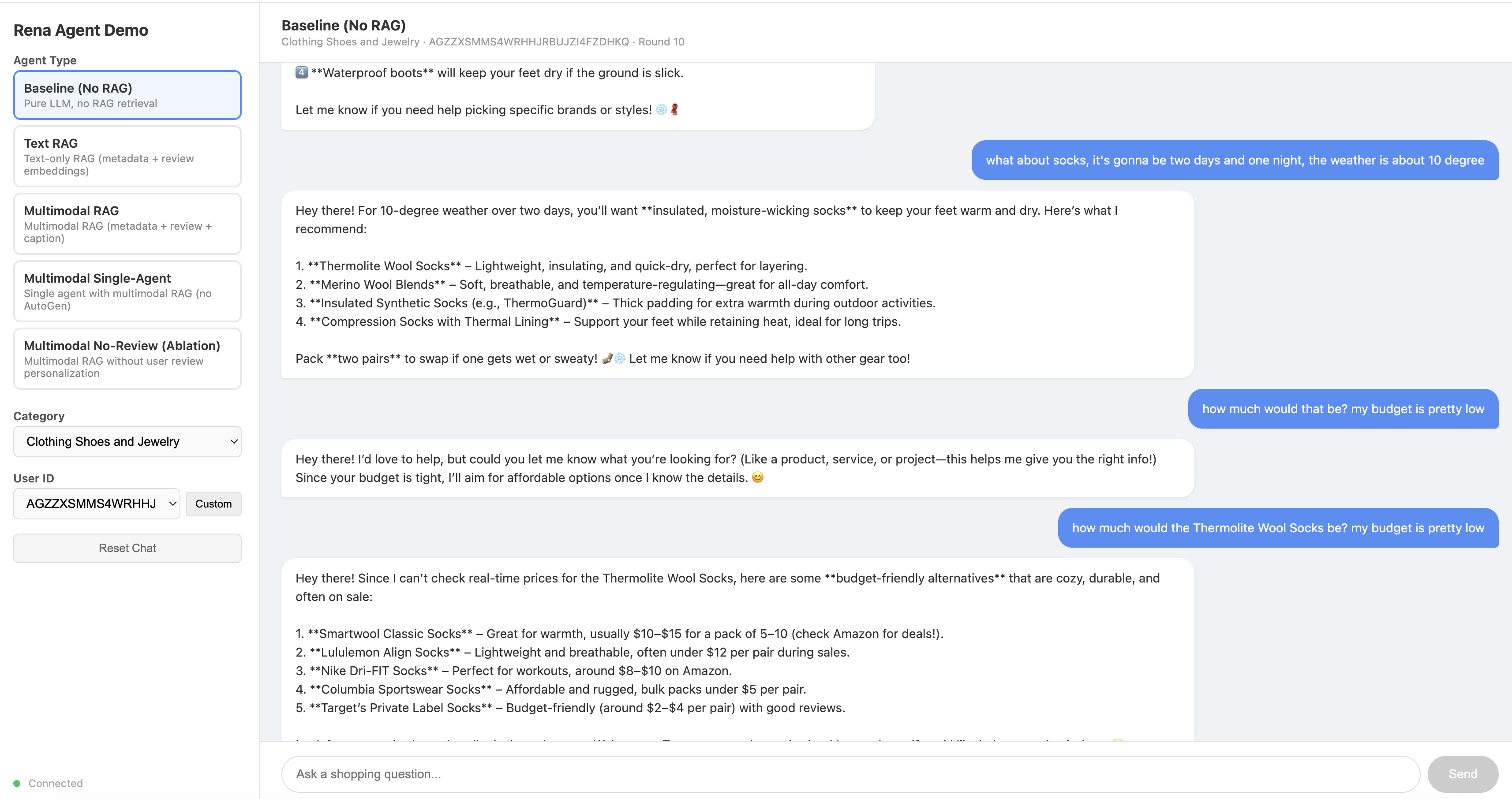}
    \caption{Baseline User Interaction 2}
\end{figure*}
\begin{figure*}
    \centering
    \includegraphics[width=\linewidth]{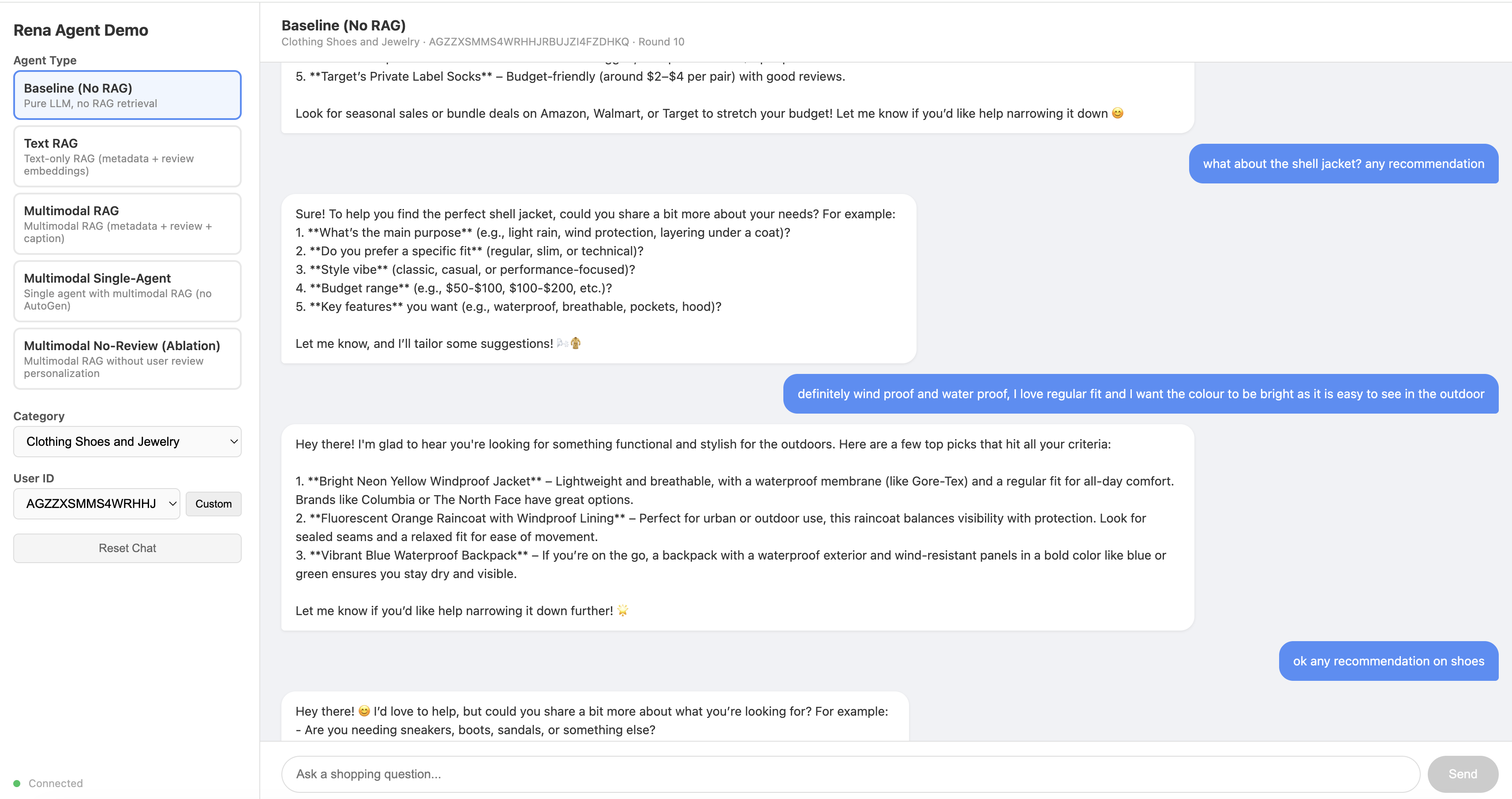}
    \caption{Baseline User Interaction 3}
\end{figure*}
\begin{figure*}
    \centering
    \includegraphics[width=\linewidth]{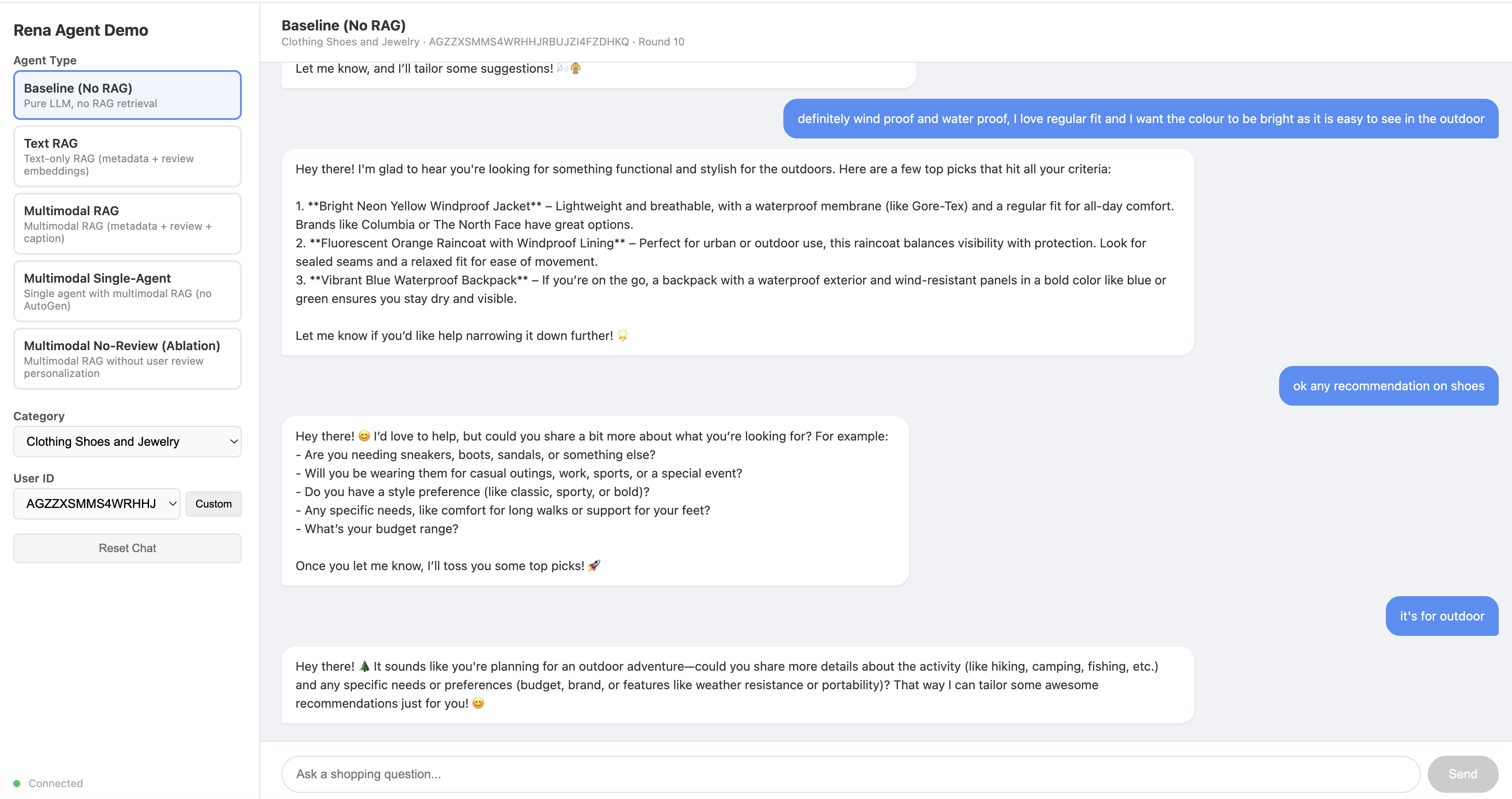}
    \caption{Baseline User Interaction 4}
\end{figure*}

\begin{figure*}
    \centering
    \includegraphics[width=\linewidth]{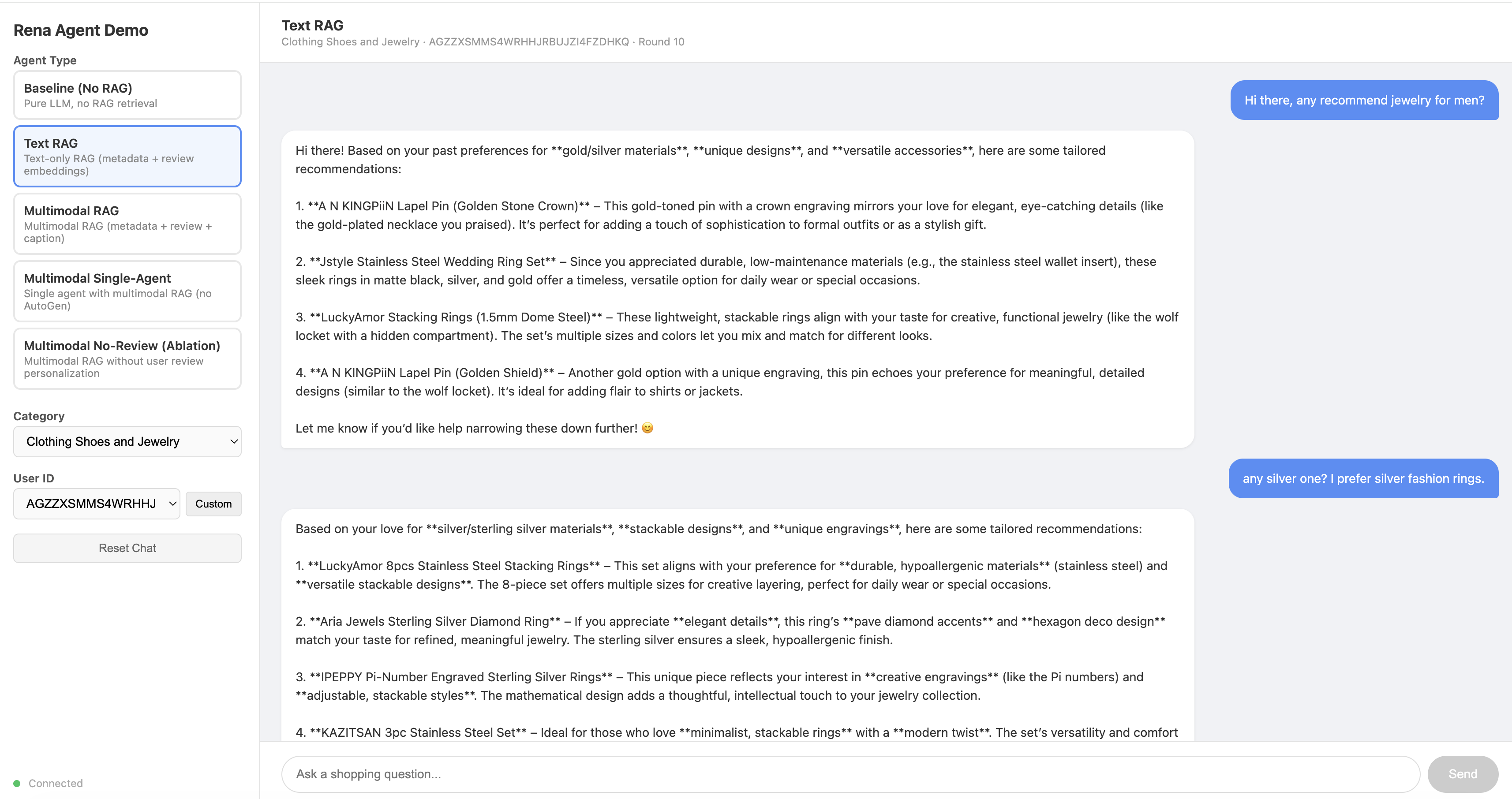}
    \caption{\textbf{Textrag} User Interaction 1}
\end{figure*}
\begin{figure*}
    \centering
    \includegraphics[width=\linewidth]{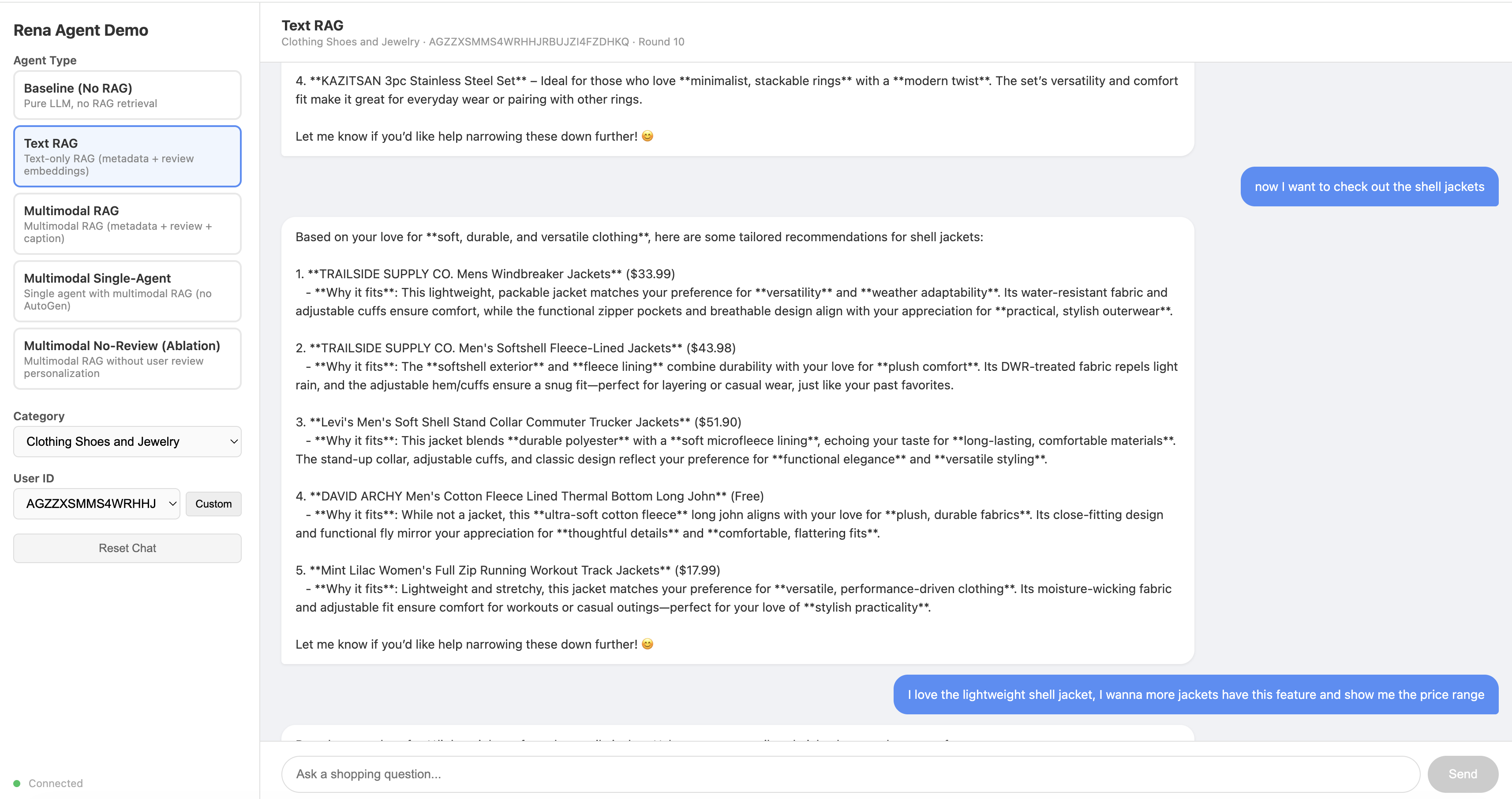}
    \caption{Textrag User Interaction 2}
\end{figure*}
\begin{figure*}
    \centering
    \includegraphics[width=\linewidth]{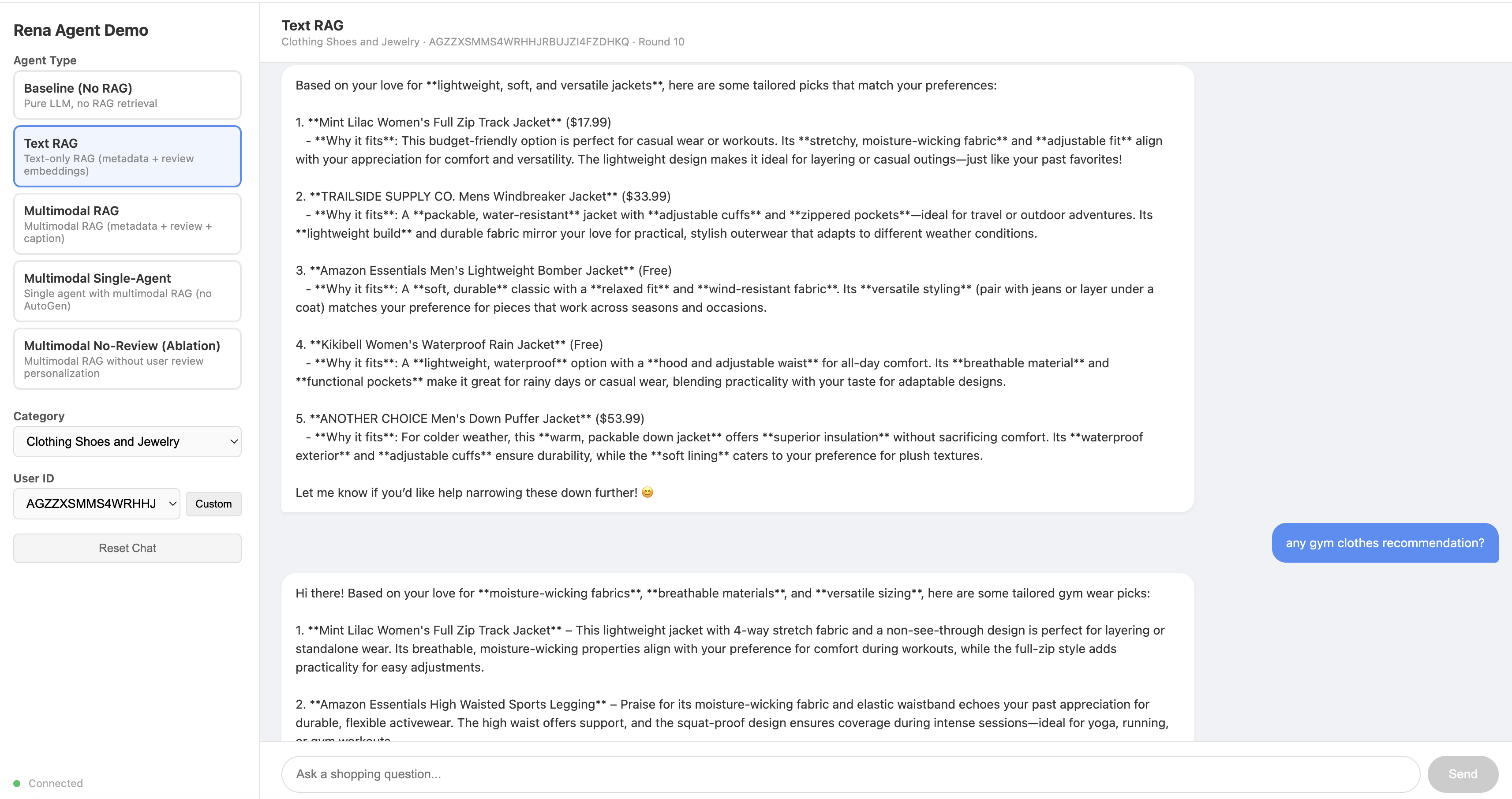}
    \caption{Textrag User Interaction 3}
\end{figure*}
\begin{figure*}
    \centering
    \includegraphics[width=\linewidth]{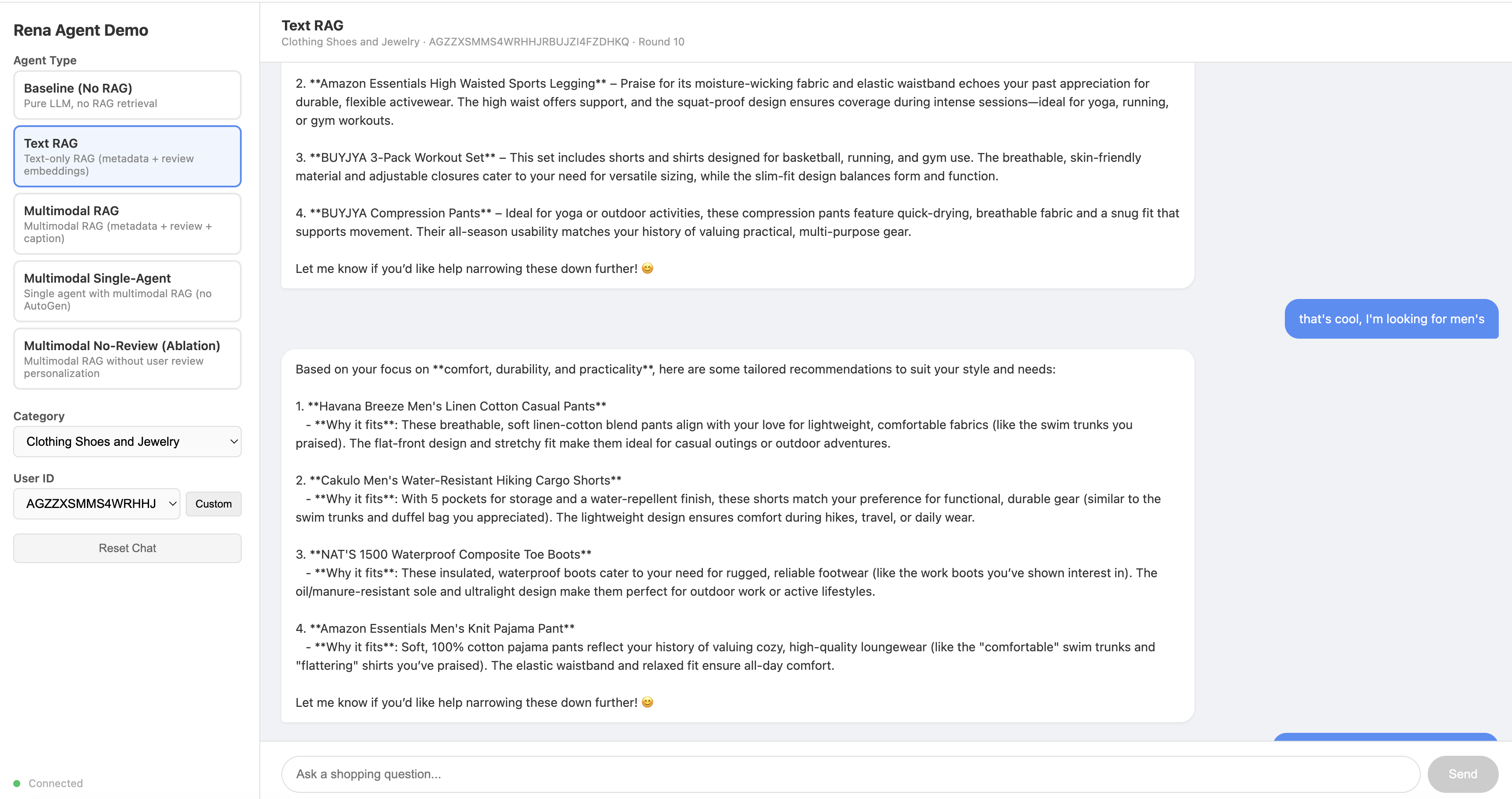}
    \caption{Textrag User Interaction 4}
\end{figure*}
\begin{figure*}
    \centering
    \includegraphics[width=\linewidth]{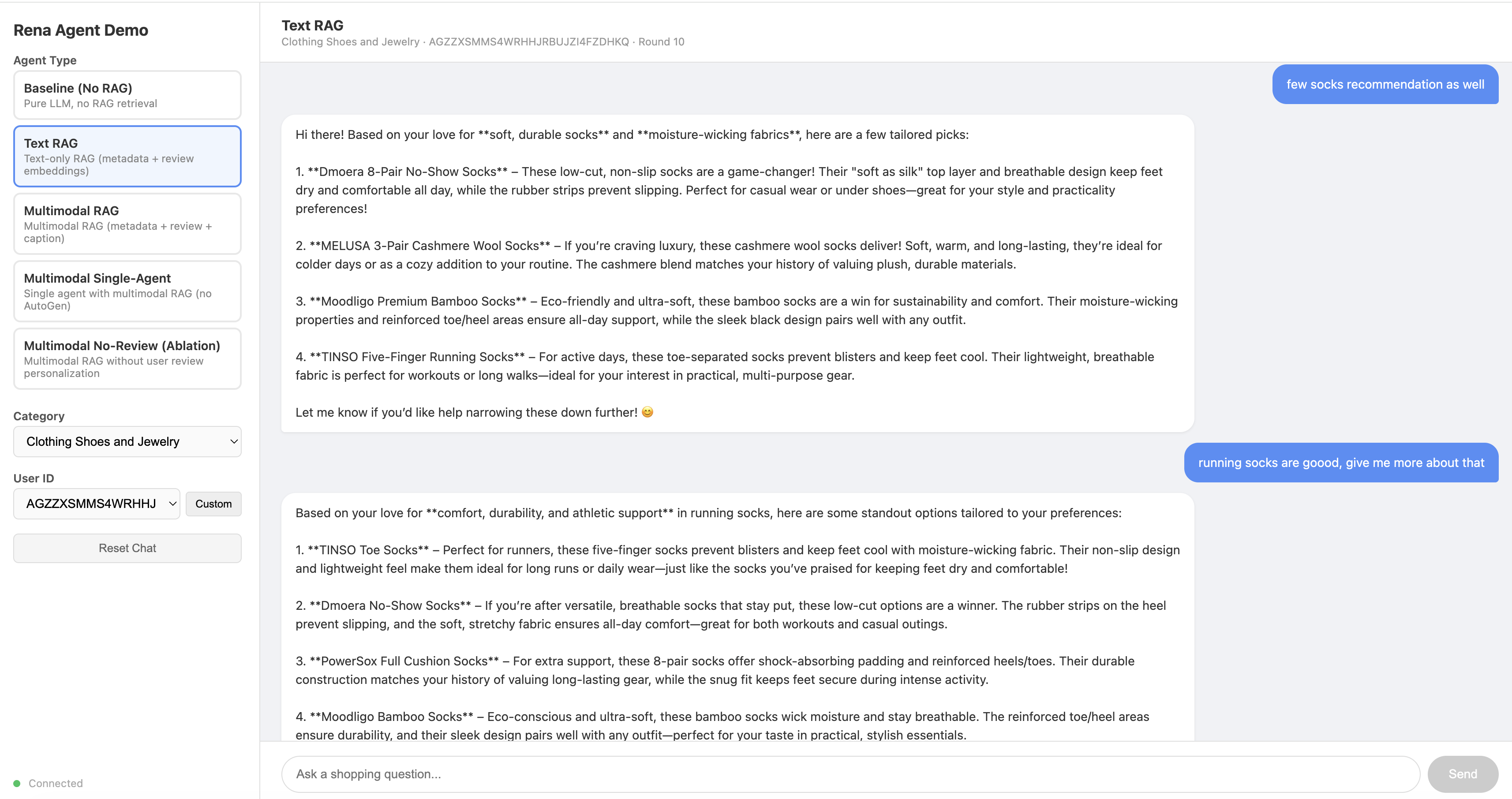}
    \caption{Textrag User Interaction 5}
\end{figure*}
\begin{figure*}
    \centering
    \includegraphics[width=\linewidth]{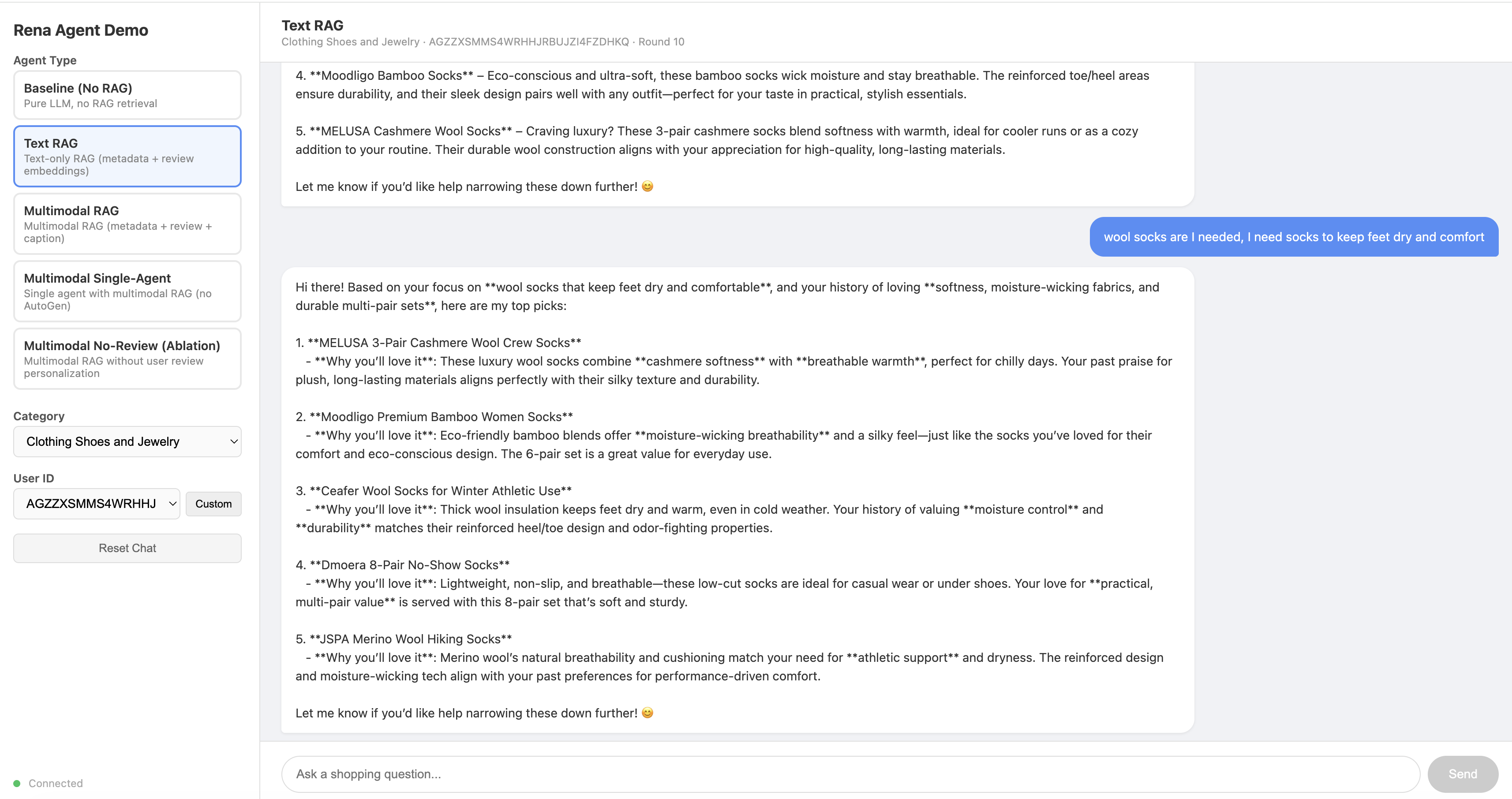}
    \caption{Textrag User Interaction 6}
\end{figure*}
\begin{figure*}
    \centering
    \includegraphics[width=\linewidth]{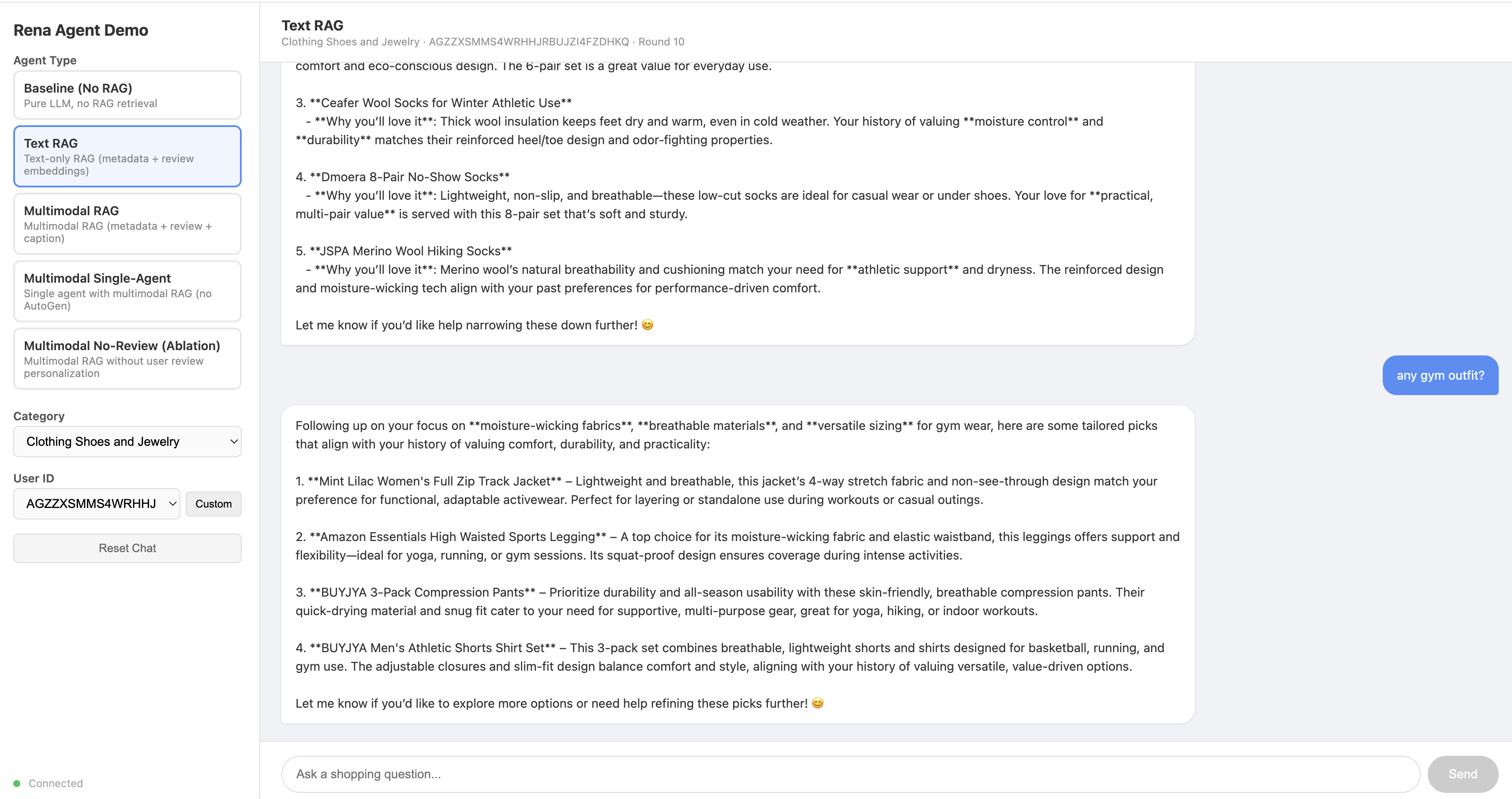}
    \caption{Textrag User Interaction 7}
\end{figure*}

\begin{figure*}
    \centering
    \includegraphics[width=\linewidth]{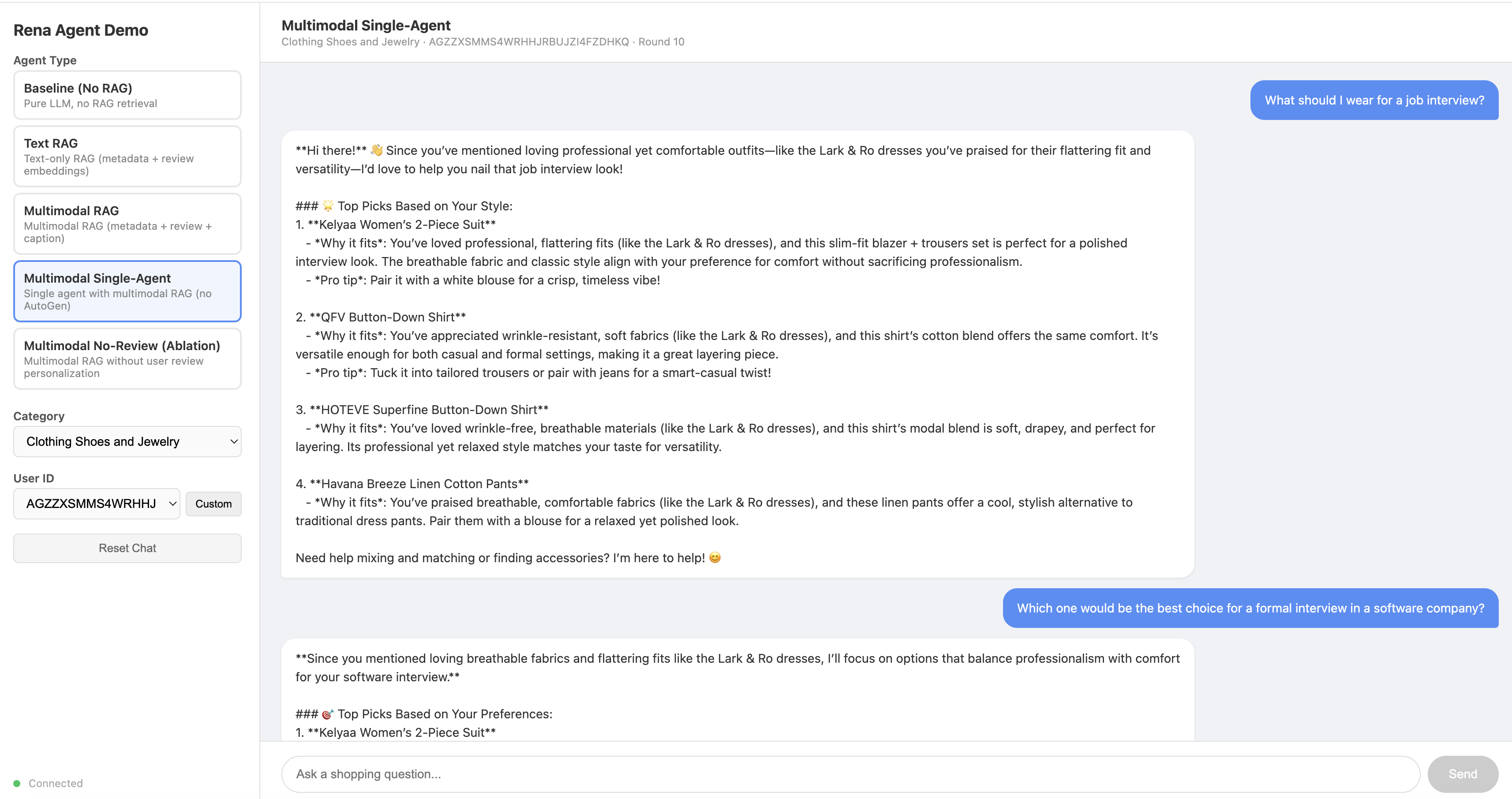}
    \caption{\textbf{MultimodalRag Single Agent} User Interaction 1}
\end{figure*}
\begin{figure*}
    \centering
    \includegraphics[width=\linewidth]{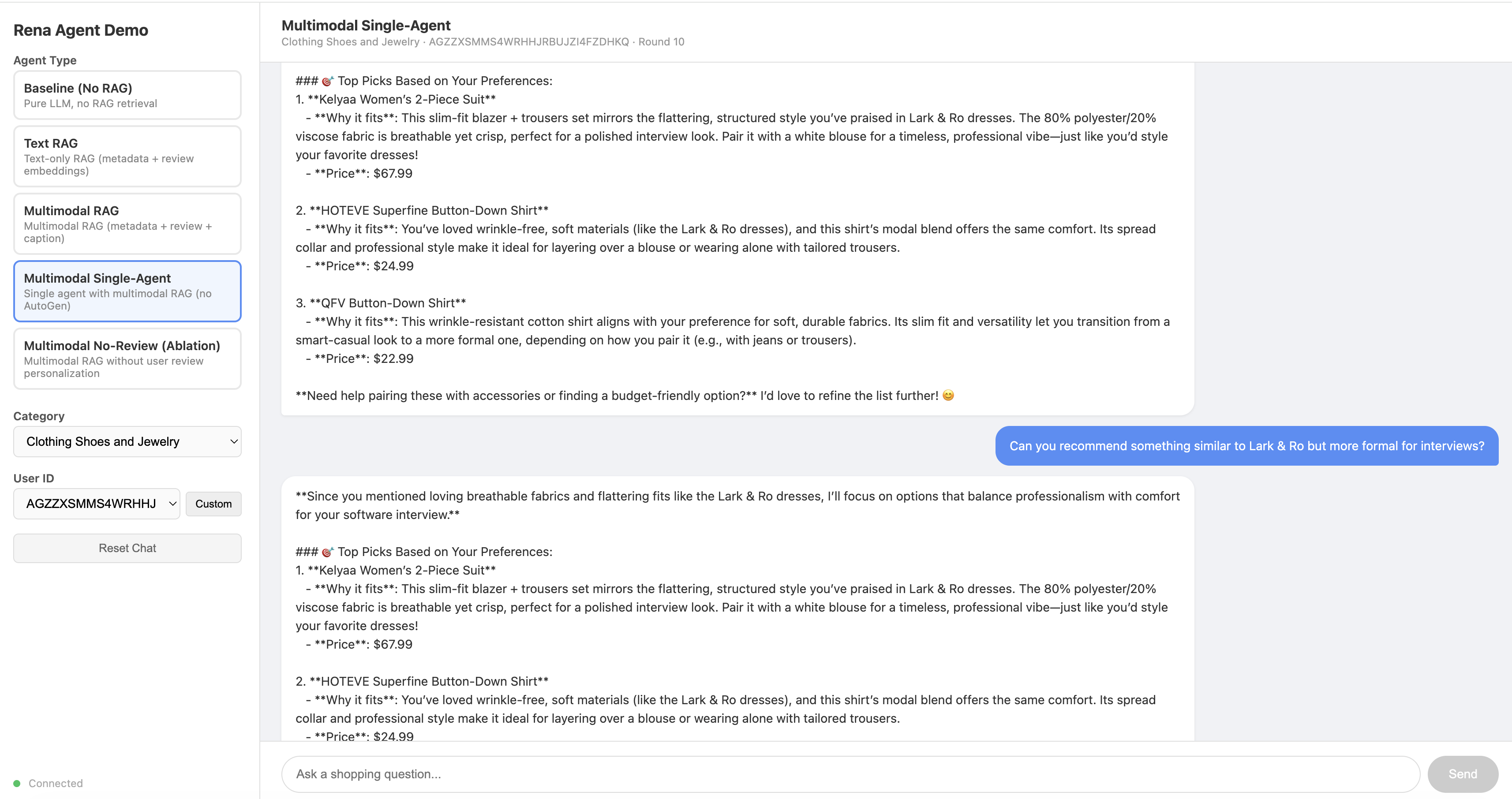}
    \caption{MultimodalRag Single Agent User Interaction 2}
\end{figure*}
\begin{figure*}
    \centering
    \includegraphics[width=\linewidth]{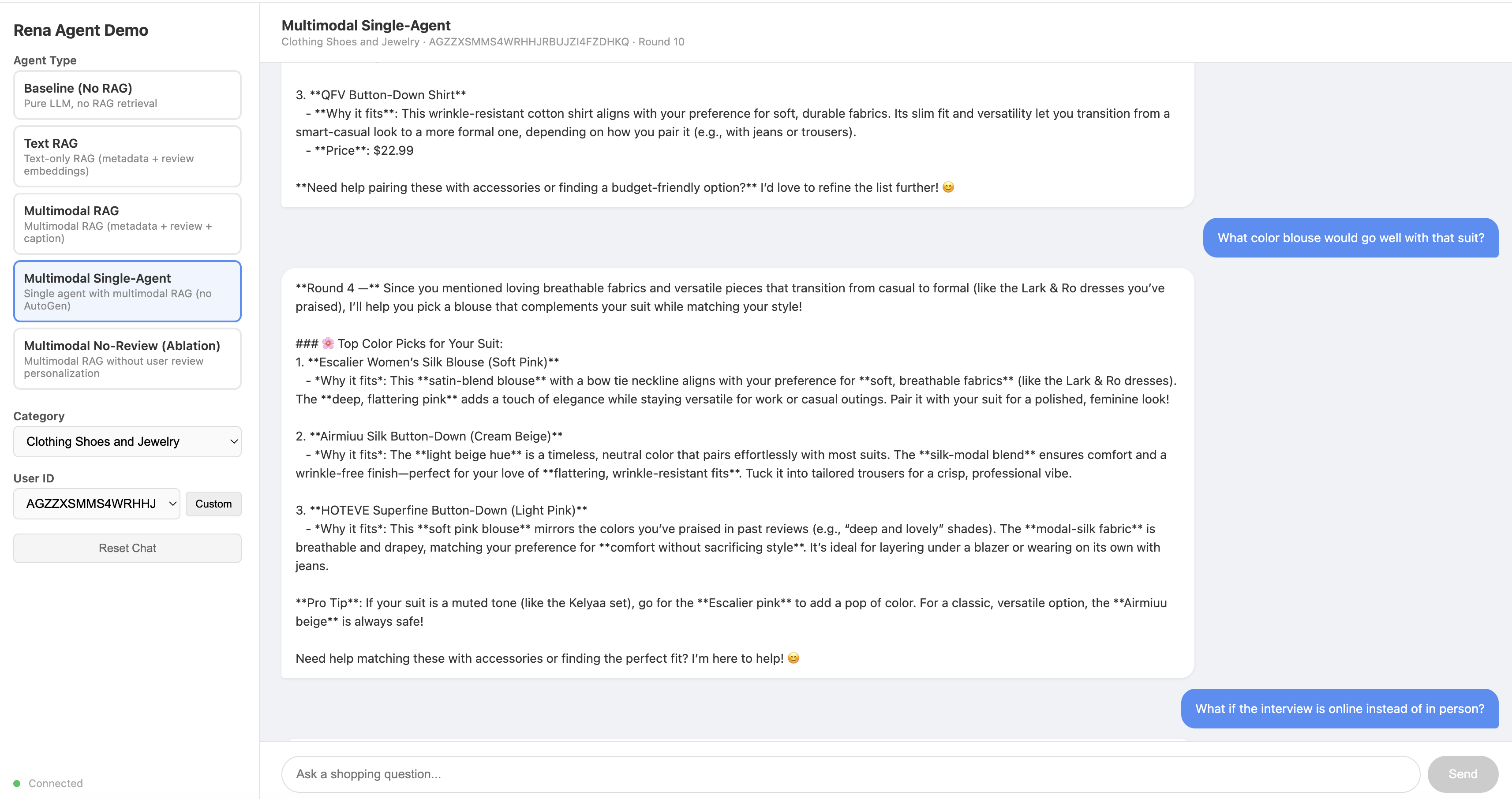}
    \caption{MultimodalRag Single Agent User Interaction 3}
\end{figure*}
\begin{figure*}
    \centering
    \includegraphics[width=\linewidth]{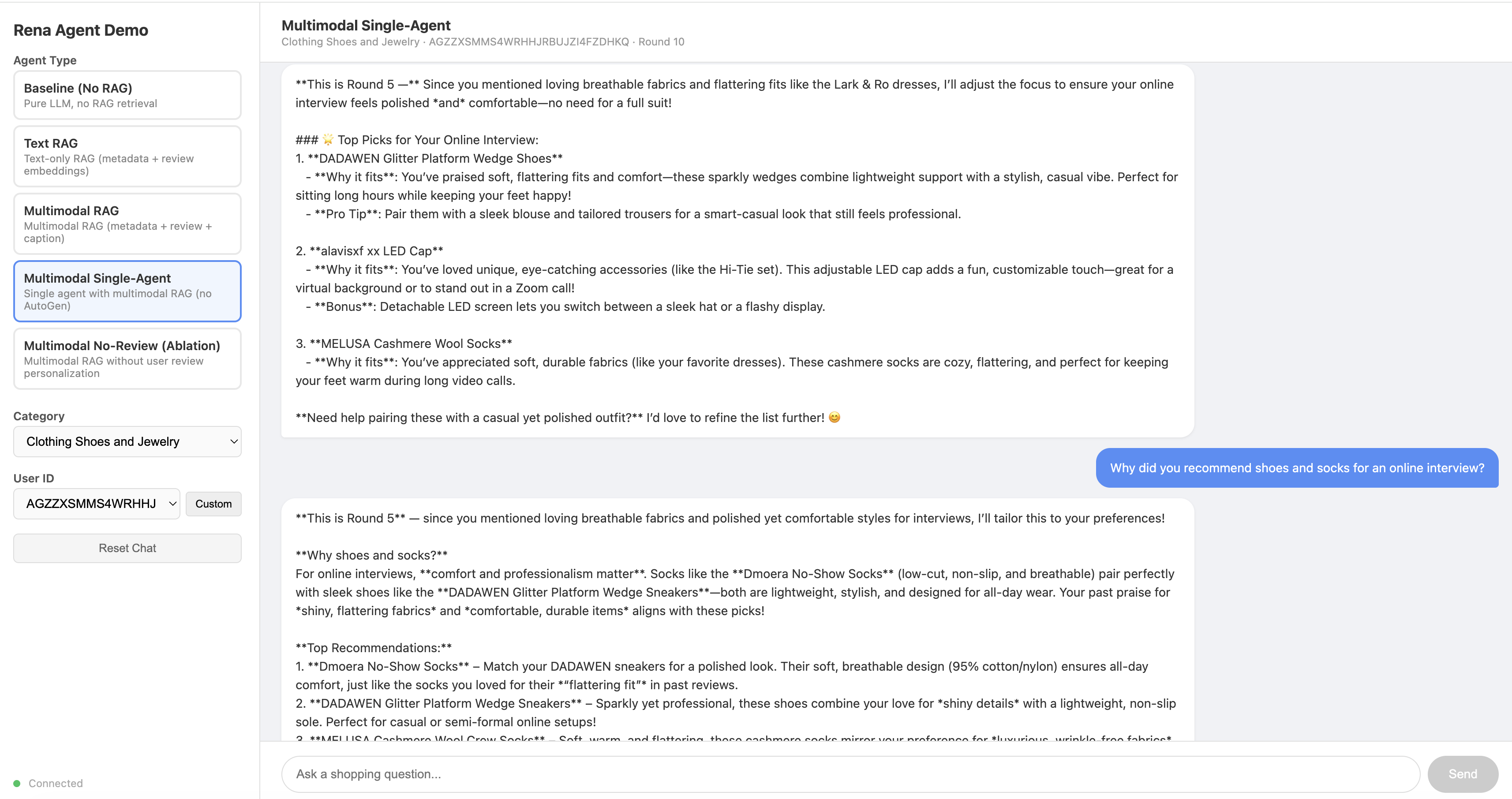}
    \caption{MultimodalRag Single Agent User Interaction 4}
\end{figure*}
\begin{figure*}
    \centering
    \includegraphics[width=\linewidth]{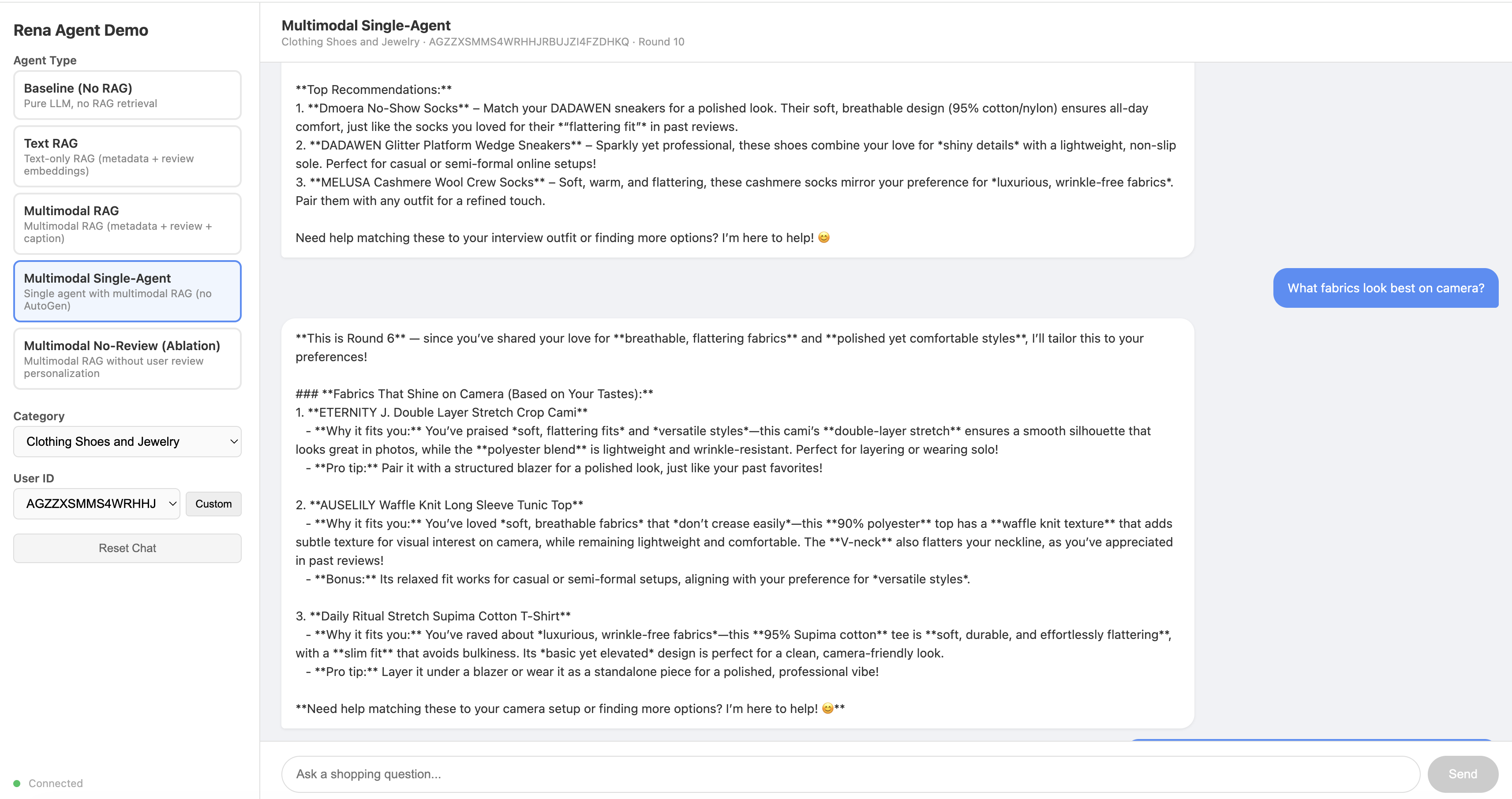}
    \caption{MultimodalRag Single Agent User Interaction 5}
\end{figure*}
\begin{figure*}
    \centering
    \includegraphics[width=\linewidth]{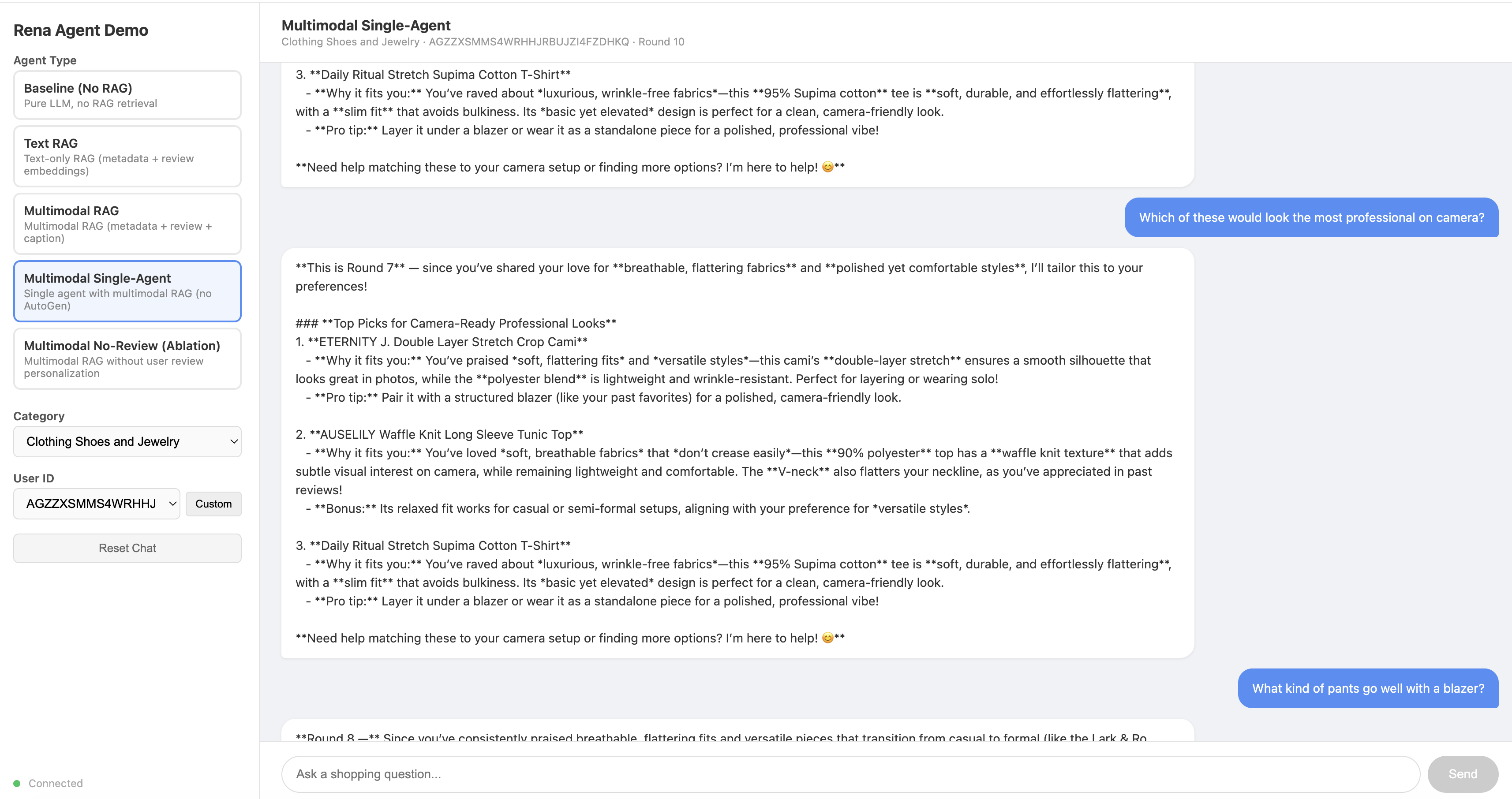}
    \caption{MultimodalRag Single Agent User Interaction 6}
\end{figure*}
\begin{figure*}
    \centering
    \includegraphics[width=\linewidth]{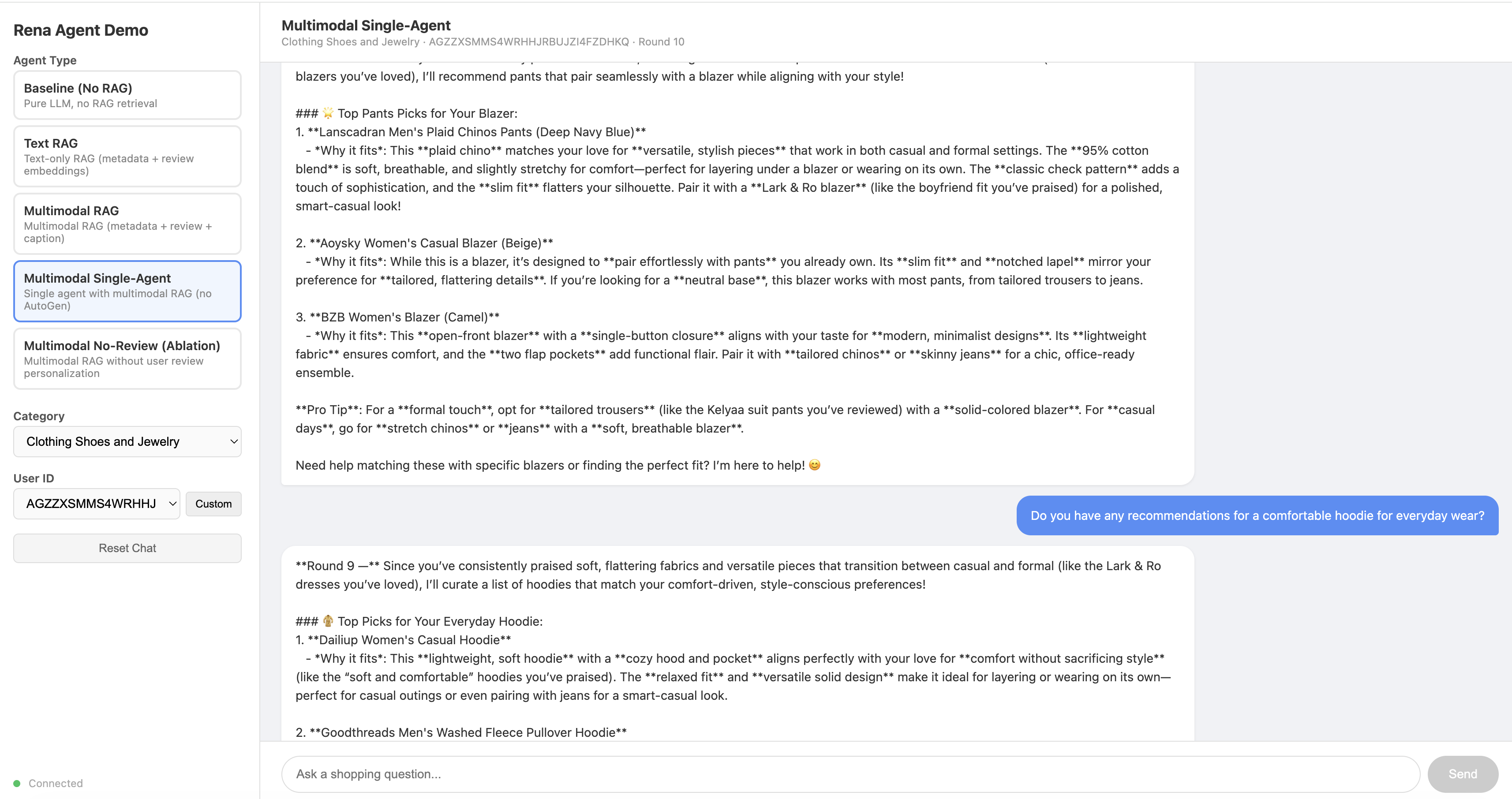}
    \caption{MultimodalRag Single Agent User Interaction 7}
\end{figure*}
\begin{figure*}
    \centering
    \includegraphics[width=\linewidth]{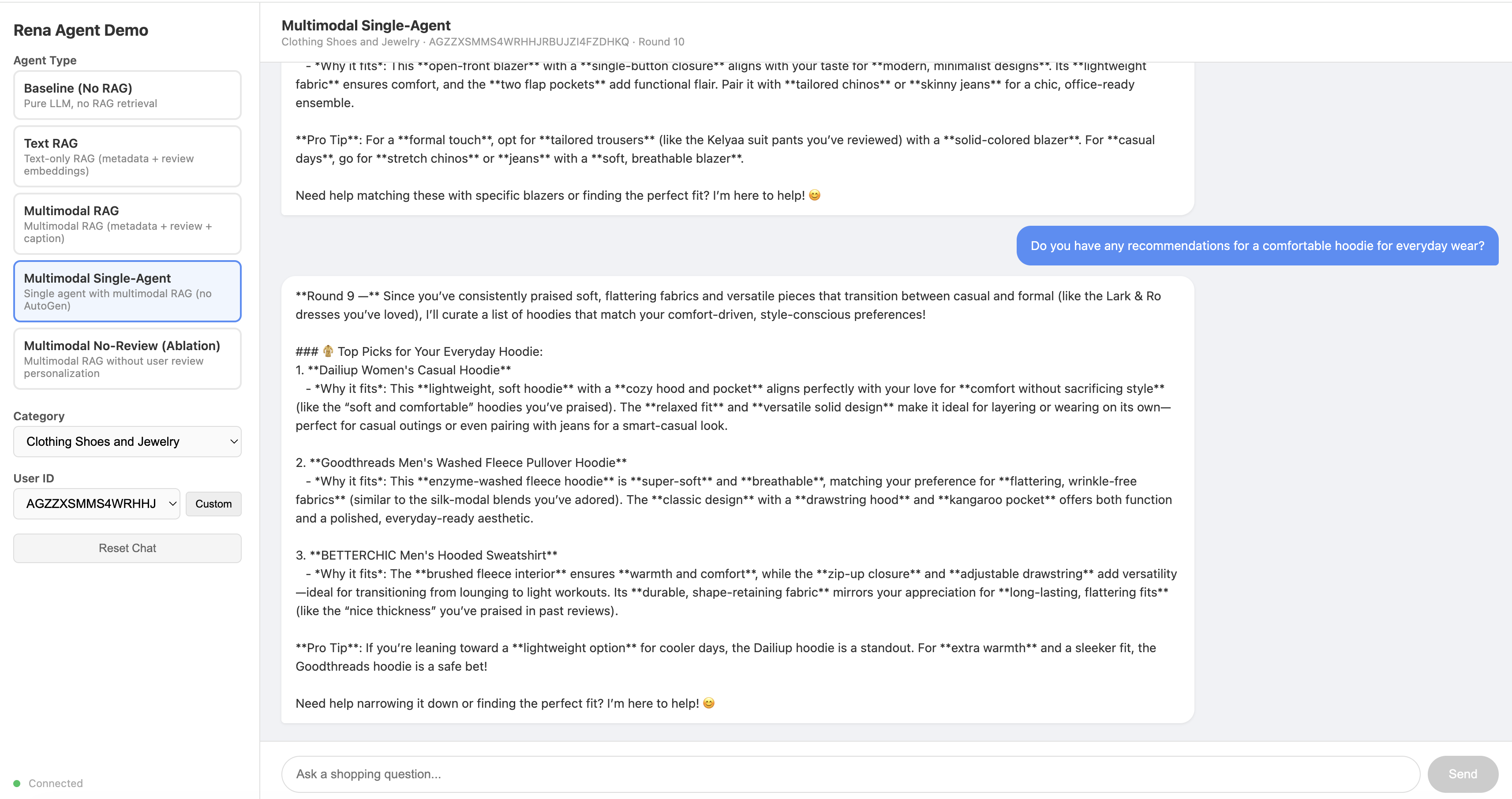}
    \caption{MultimodalRag Single Agent User Interaction 8}
\end{figure*}

\begin{figure*}
    \centering
    \includegraphics[width=\linewidth]{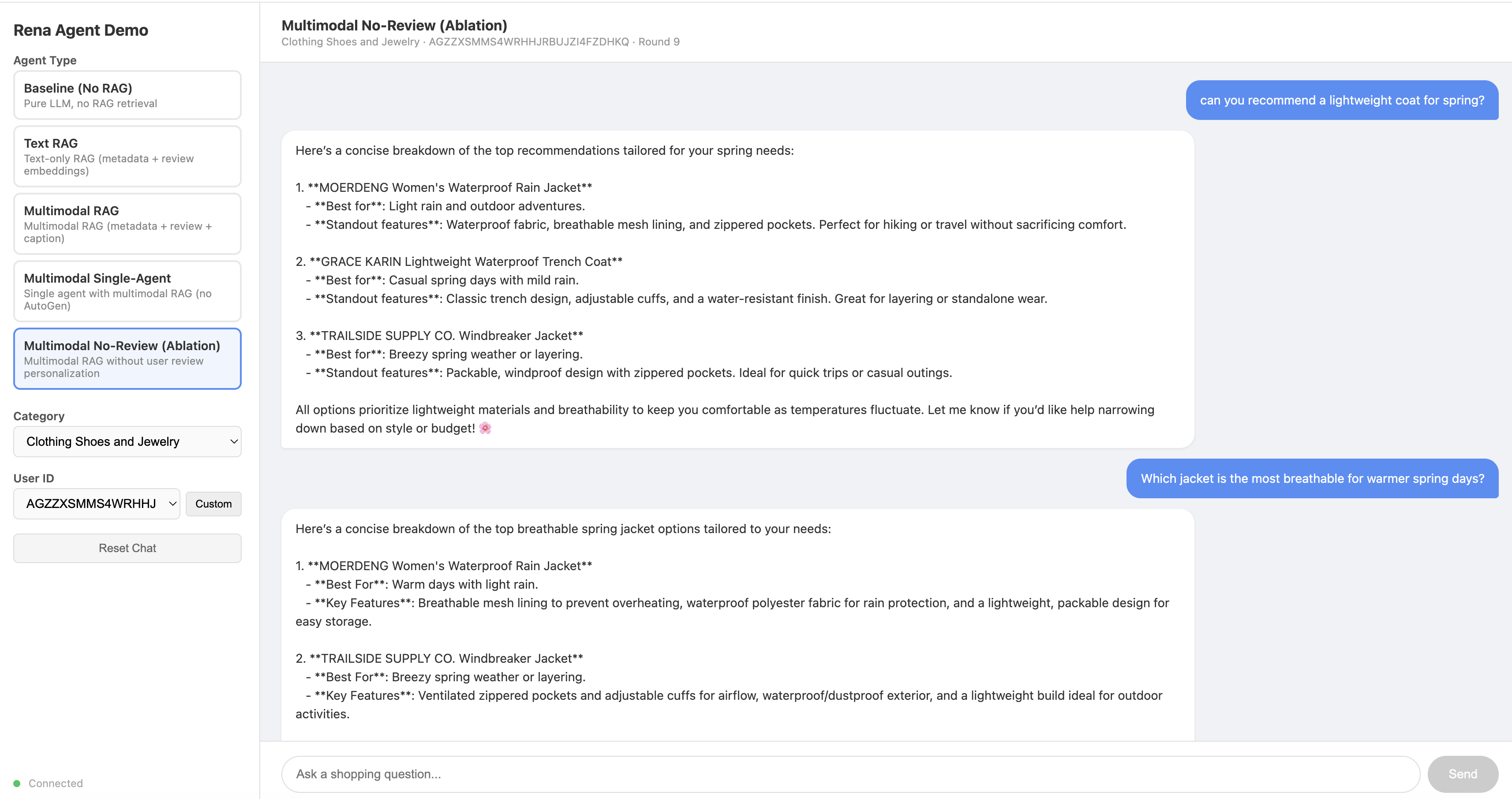}
    \caption{\textbf{MultimodalRag no Review} User Interaction 1}
\end{figure*}
\begin{figure*}
    \centering
    \includegraphics[width=\linewidth]{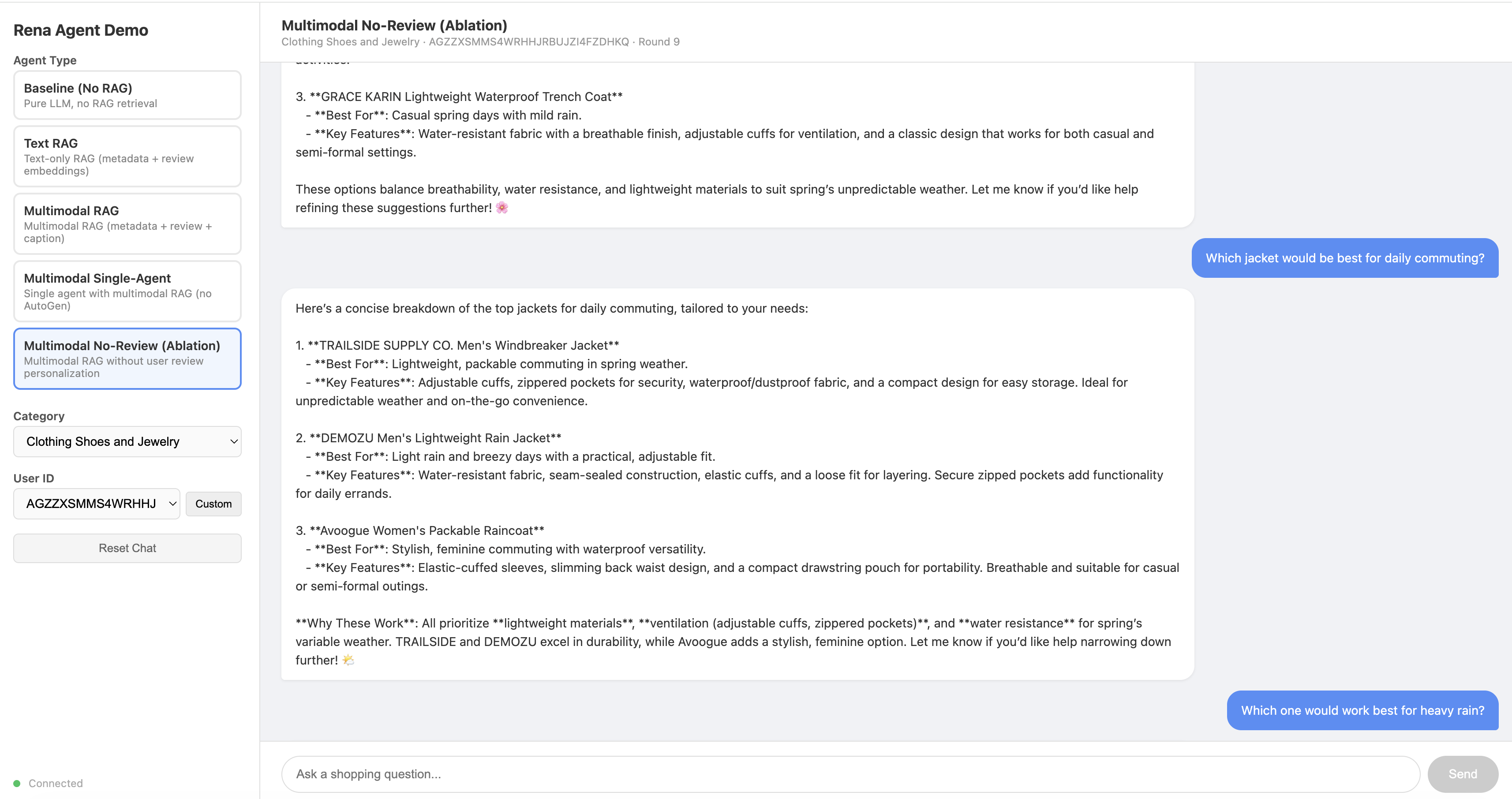}
    \caption{MultimodalRag no Review User Interaction 2}
\end{figure*}
\begin{figure*}
    \centering
    \includegraphics[width=\linewidth]{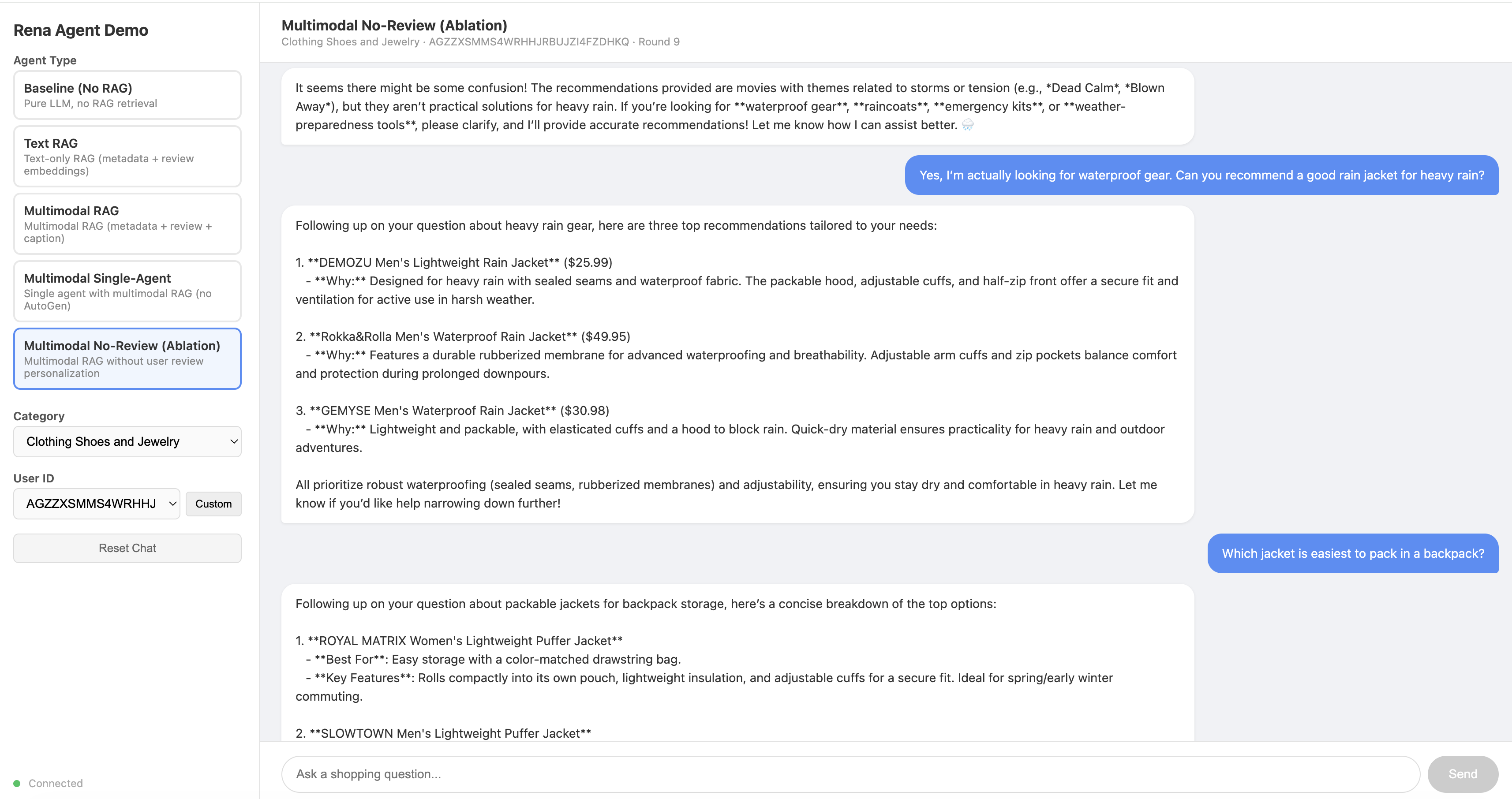}
    \caption{MultimodalRag no Review User Interaction 3}
\end{figure*}
\begin{figure*}
    \centering
    \includegraphics[width=\linewidth]{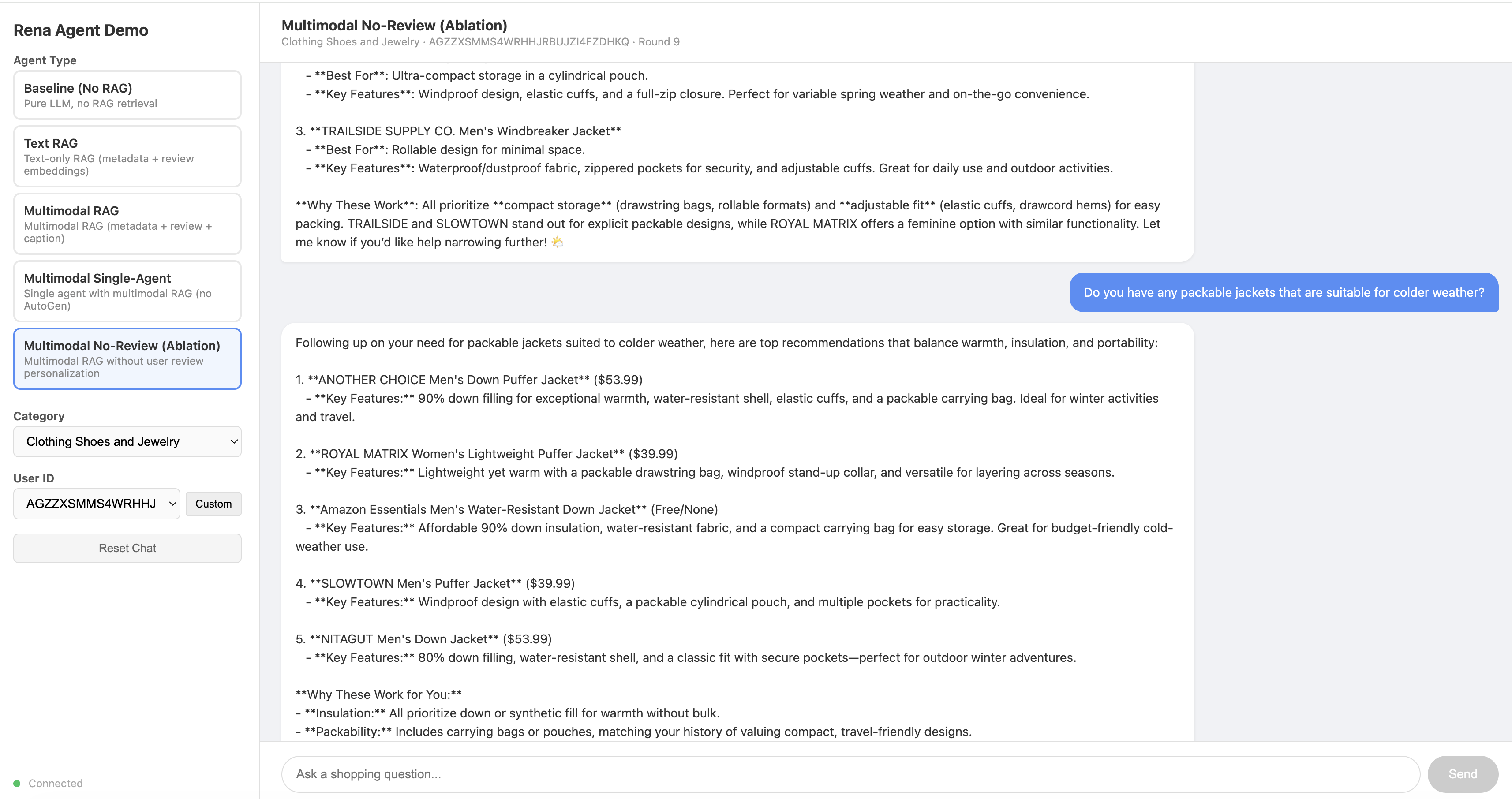}
    \caption{MultimodalRag no Review User Interaction 4}
\end{figure*}
\begin{figure*}
    \centering
    \includegraphics[width=\linewidth]{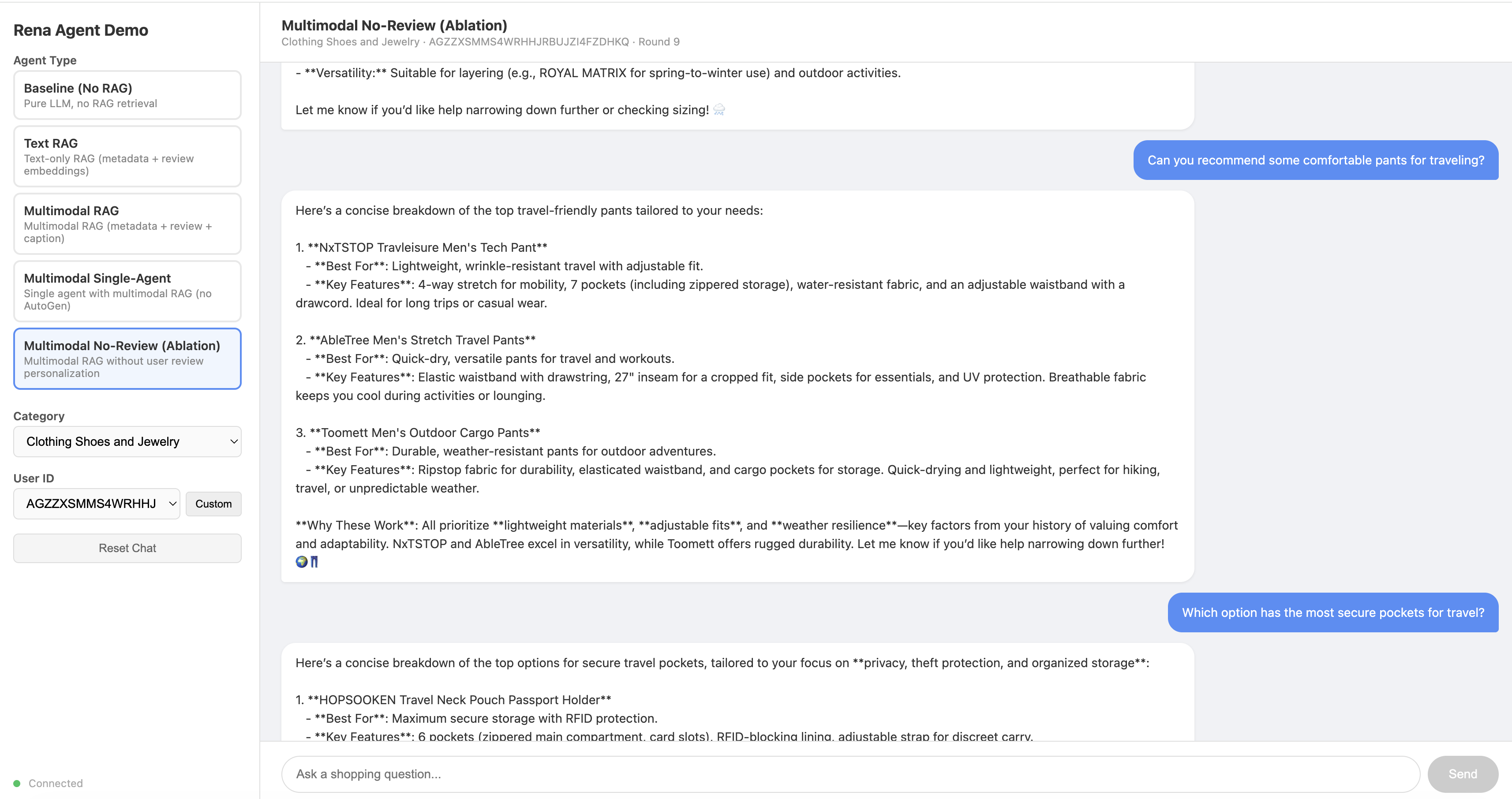}
    \caption{MultimodalRag no Review User Interaction 5}
\end{figure*}
\begin{figure*}
    \centering
    \includegraphics[width=\linewidth]{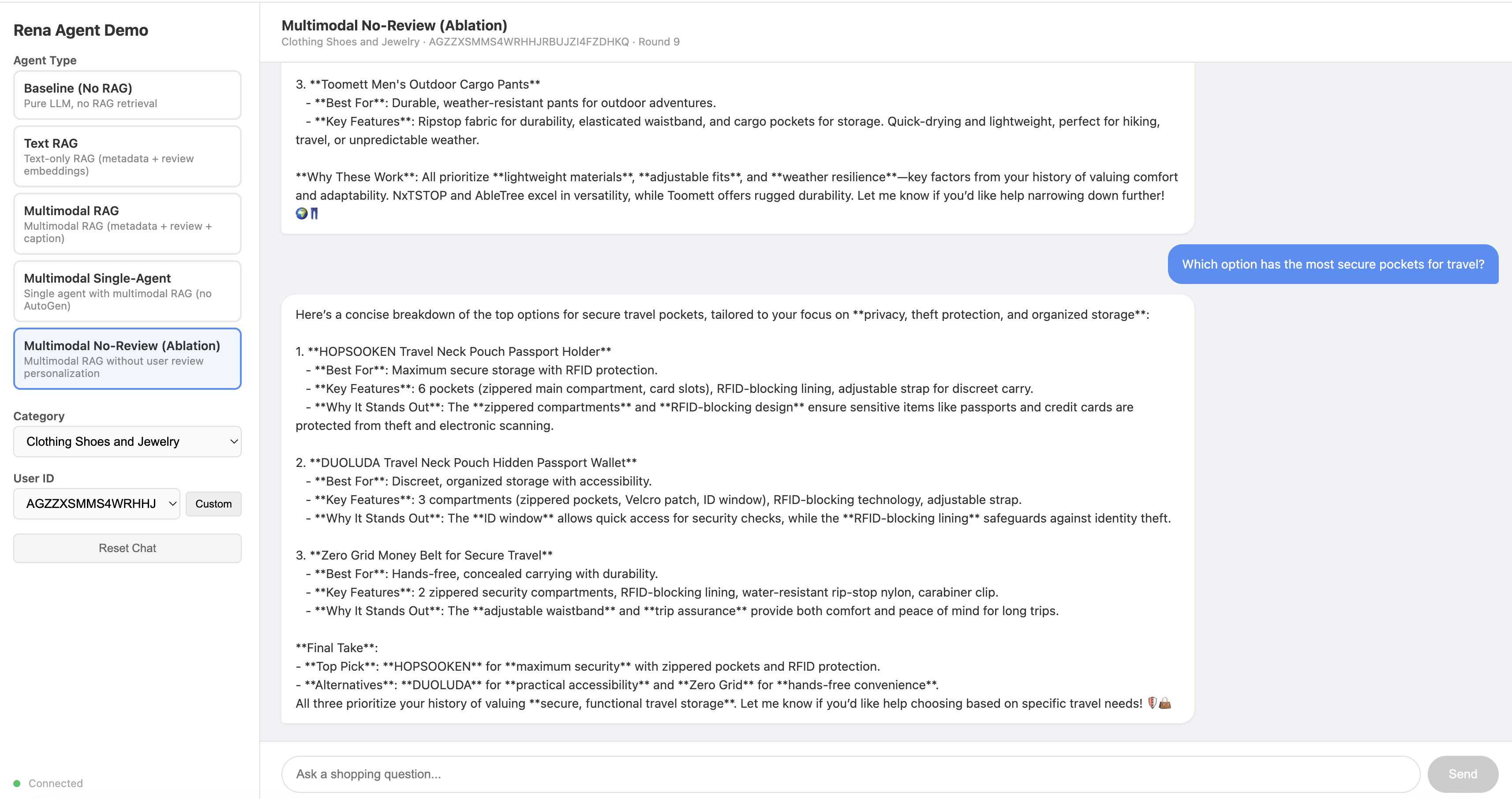}
    \caption{MultimodalRag no Review User Interaction 6}
\end{figure*}
\begin{figure*}
    \centering
    \includegraphics[width=\linewidth]{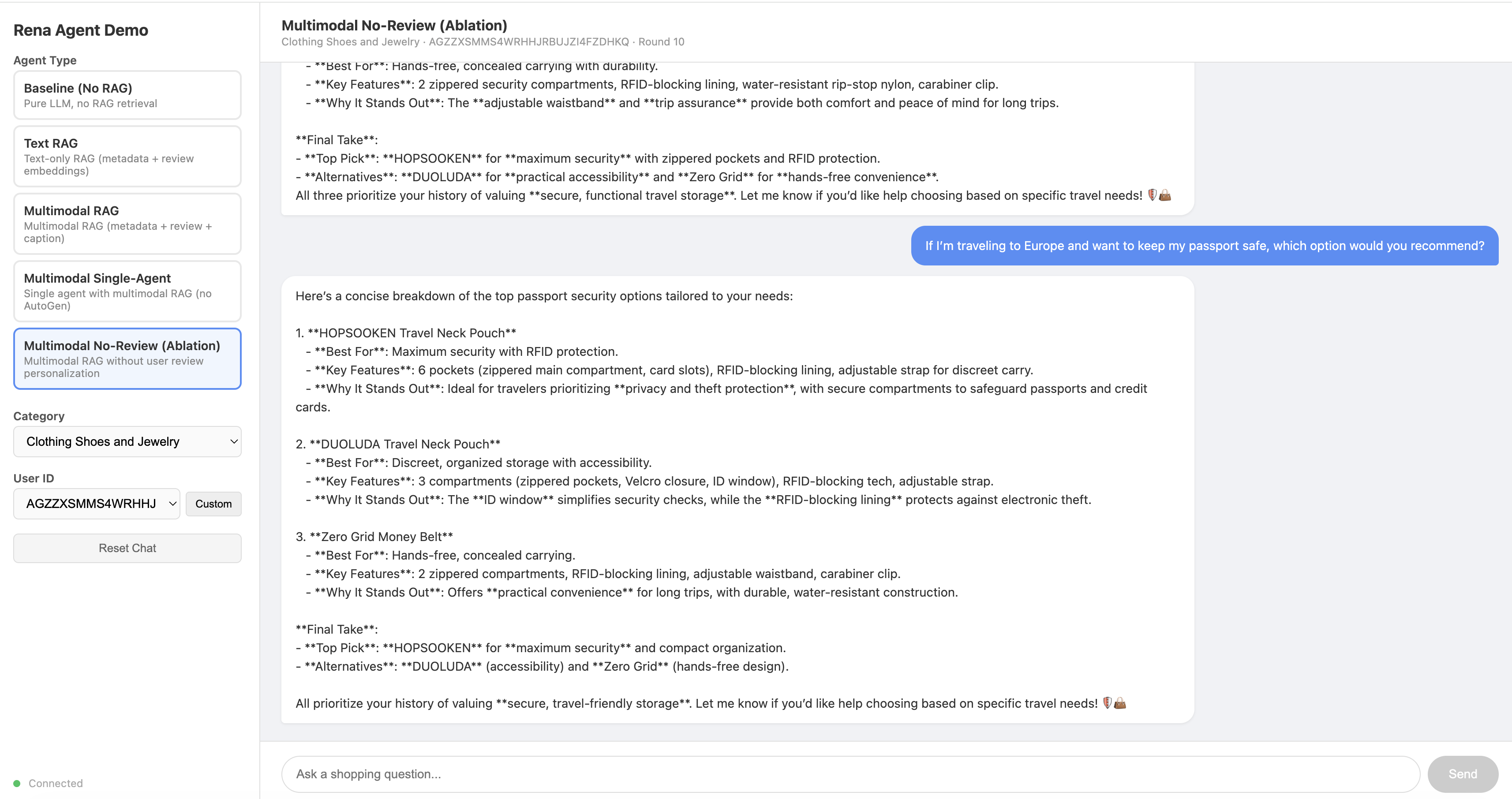}
    \caption{MultimodalRag no Review User Interaction 7}
\end{figure*}

\begin{figure*}
    \centering
    \includegraphics[width=\linewidth]{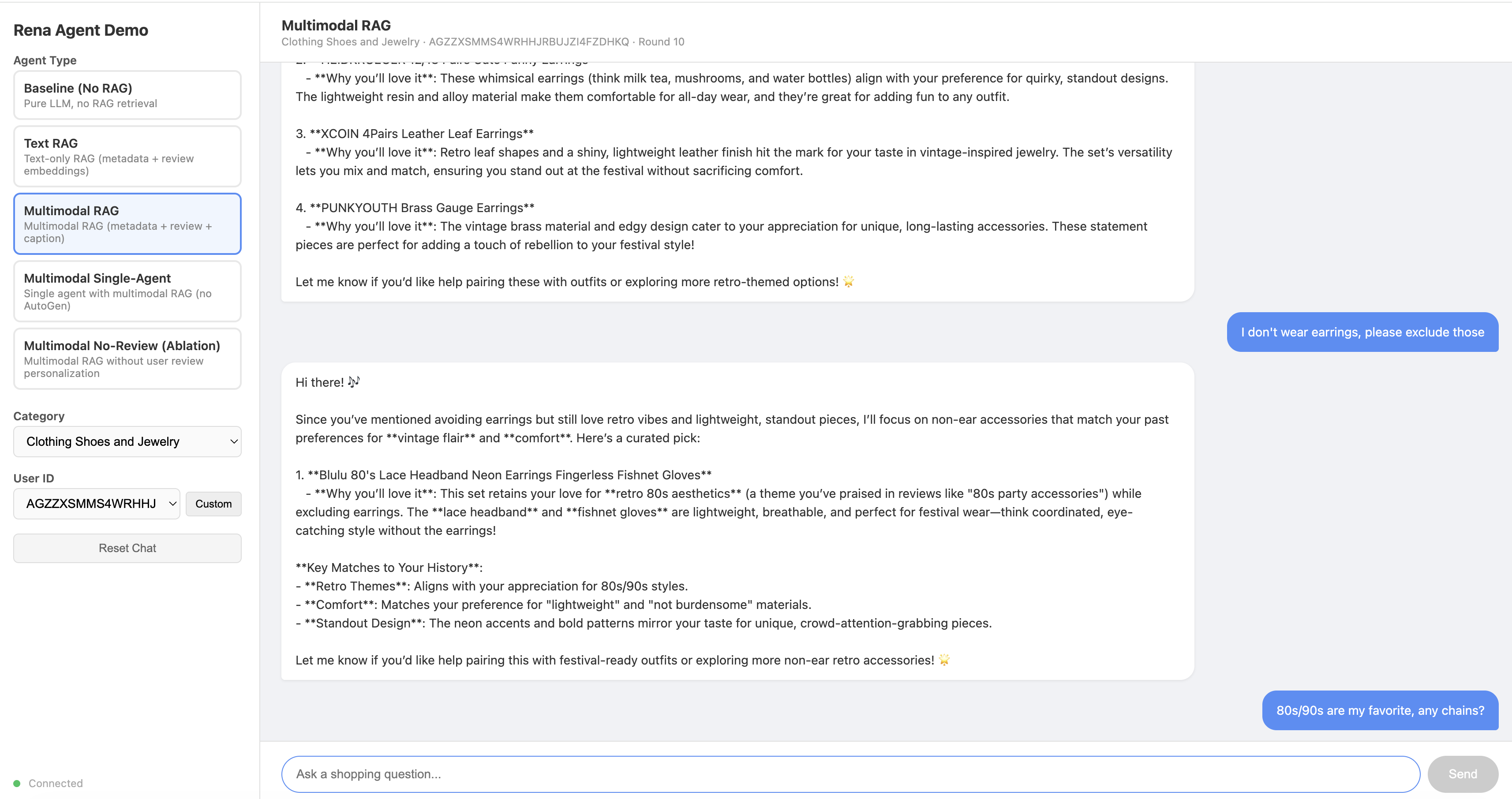}
    \caption{\textbf{MultimodalRag Full} User Interaction 1}
\end{figure*}
\begin{figure*}
    \centering
    \includegraphics[width=\linewidth]{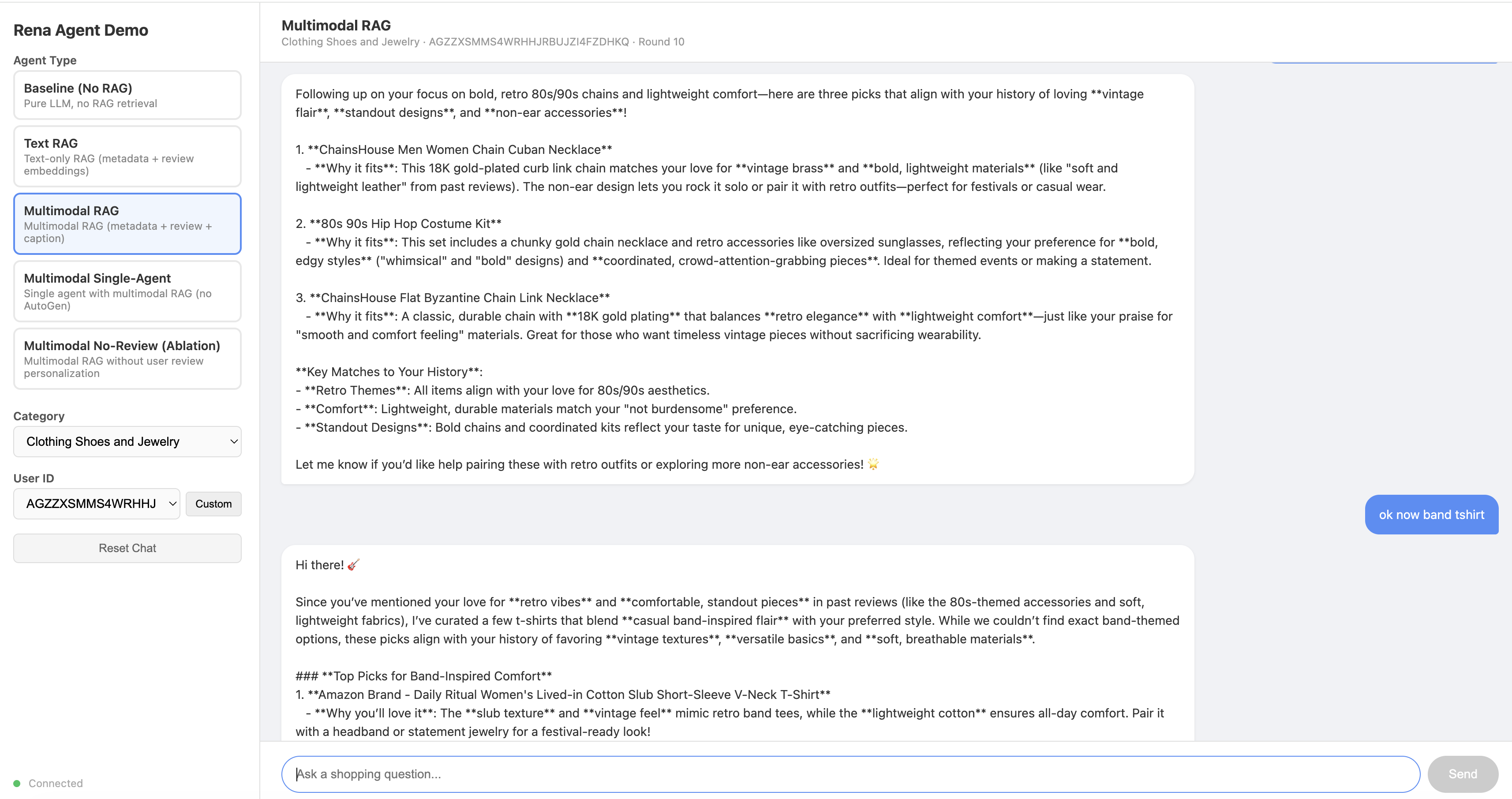}
    \caption{MultimodalRag Full User Interaction 2}
\end{figure*}
\begin{figure*}
    \centering
    \includegraphics[width=\linewidth]{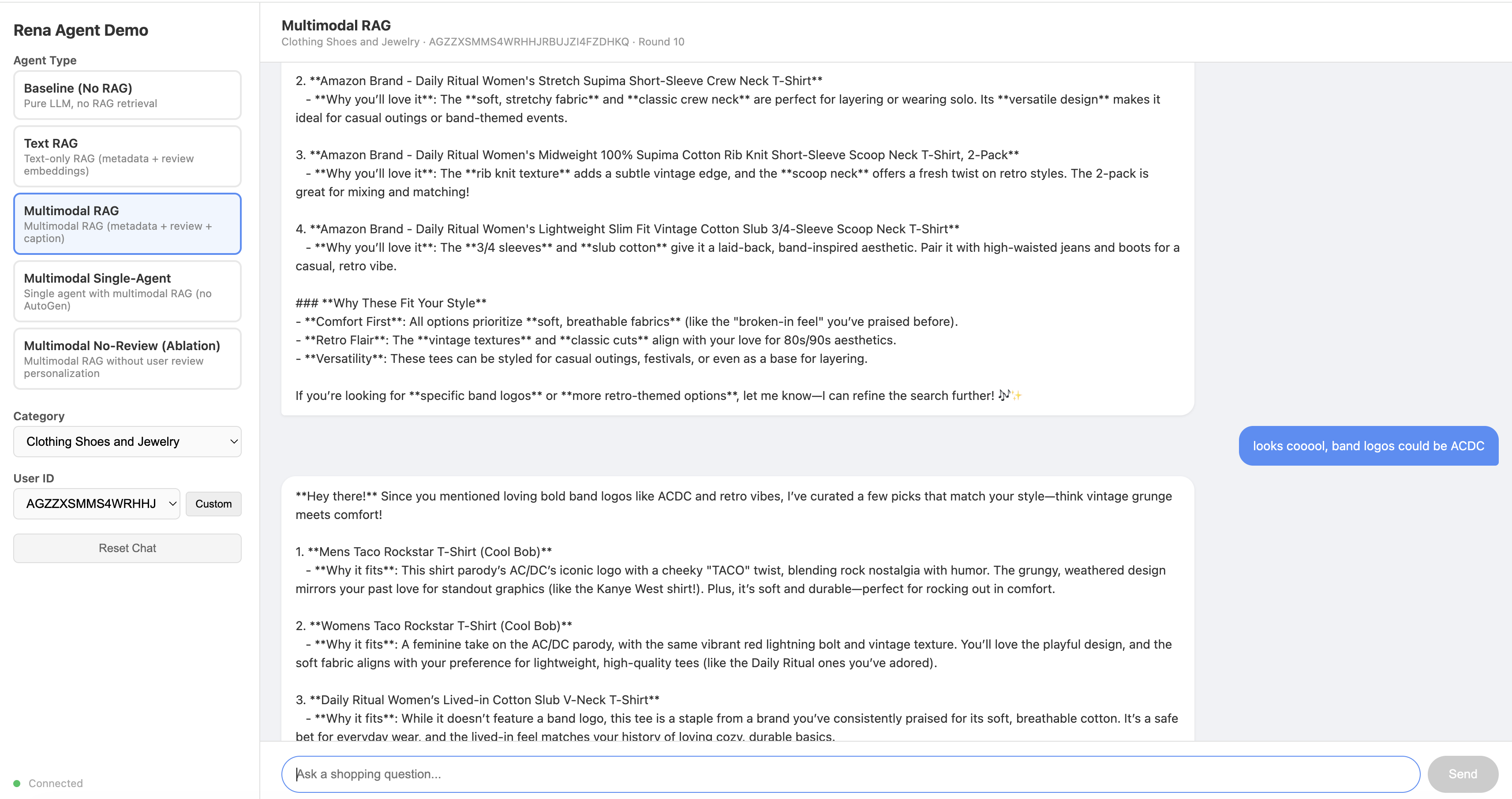}
    \caption{MultimodalRag Full User Interaction 3}
\end{figure*}
\begin{figure*}
    \centering
    \includegraphics[width=\linewidth]{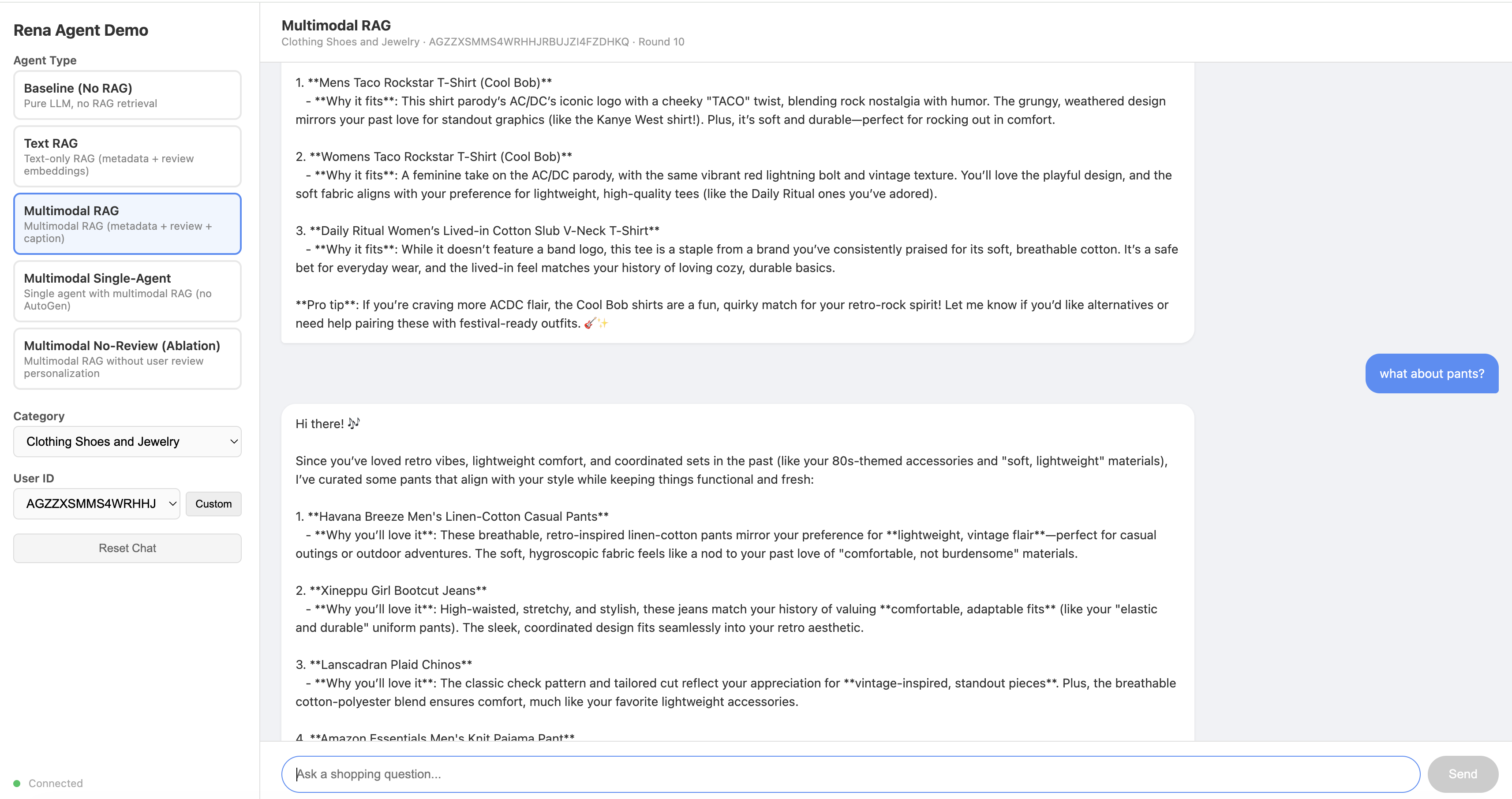}
    \caption{MultimodalRag Full User Interaction 4}
\end{figure*}
\begin{figure*}
    \centering
    \includegraphics[width=\linewidth]{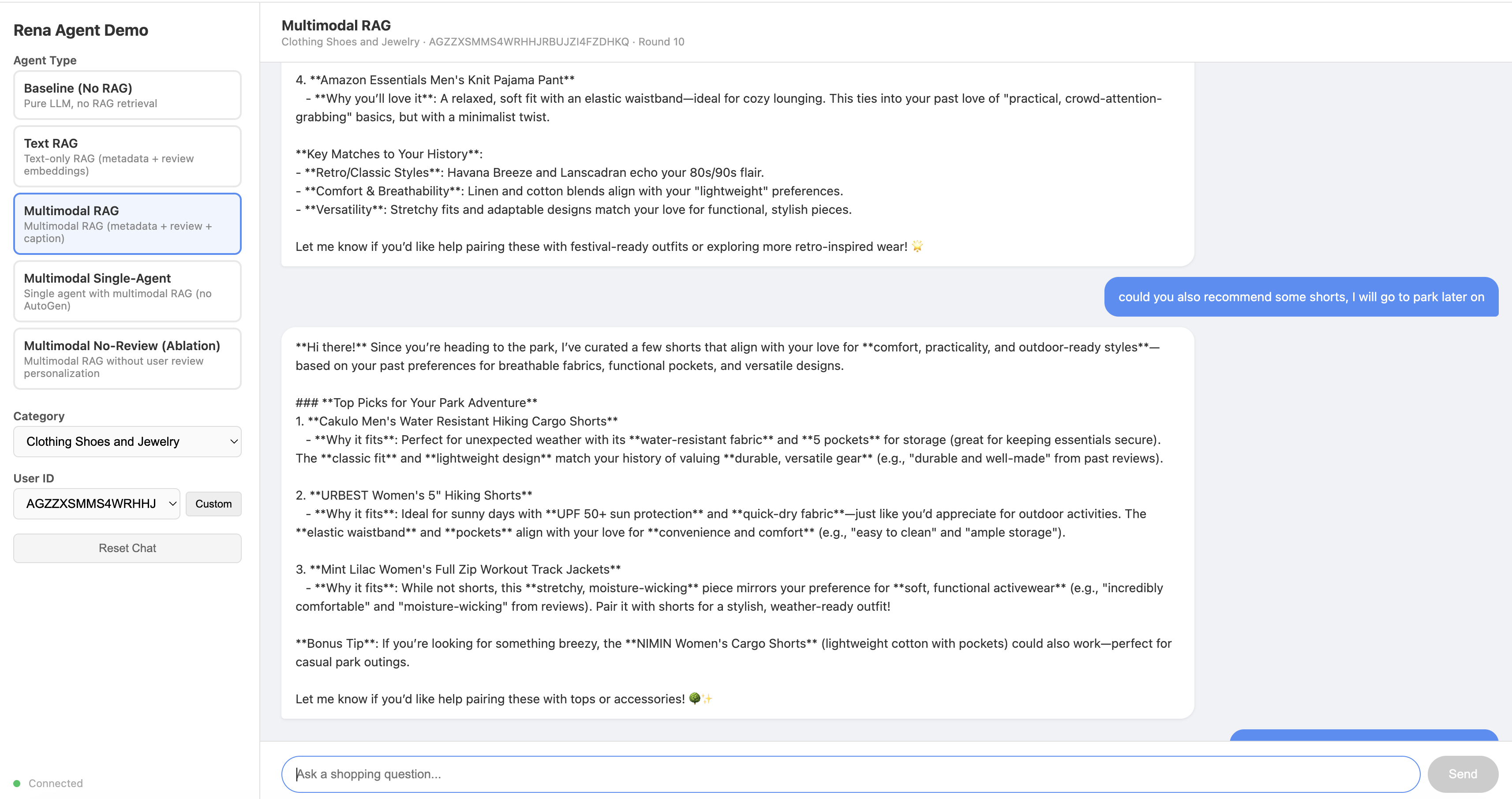}
    \caption{MultimodalRag Full User Interaction 5}
\end{figure*}
\begin{figure*}
    \centering
    \includegraphics[width=\linewidth]{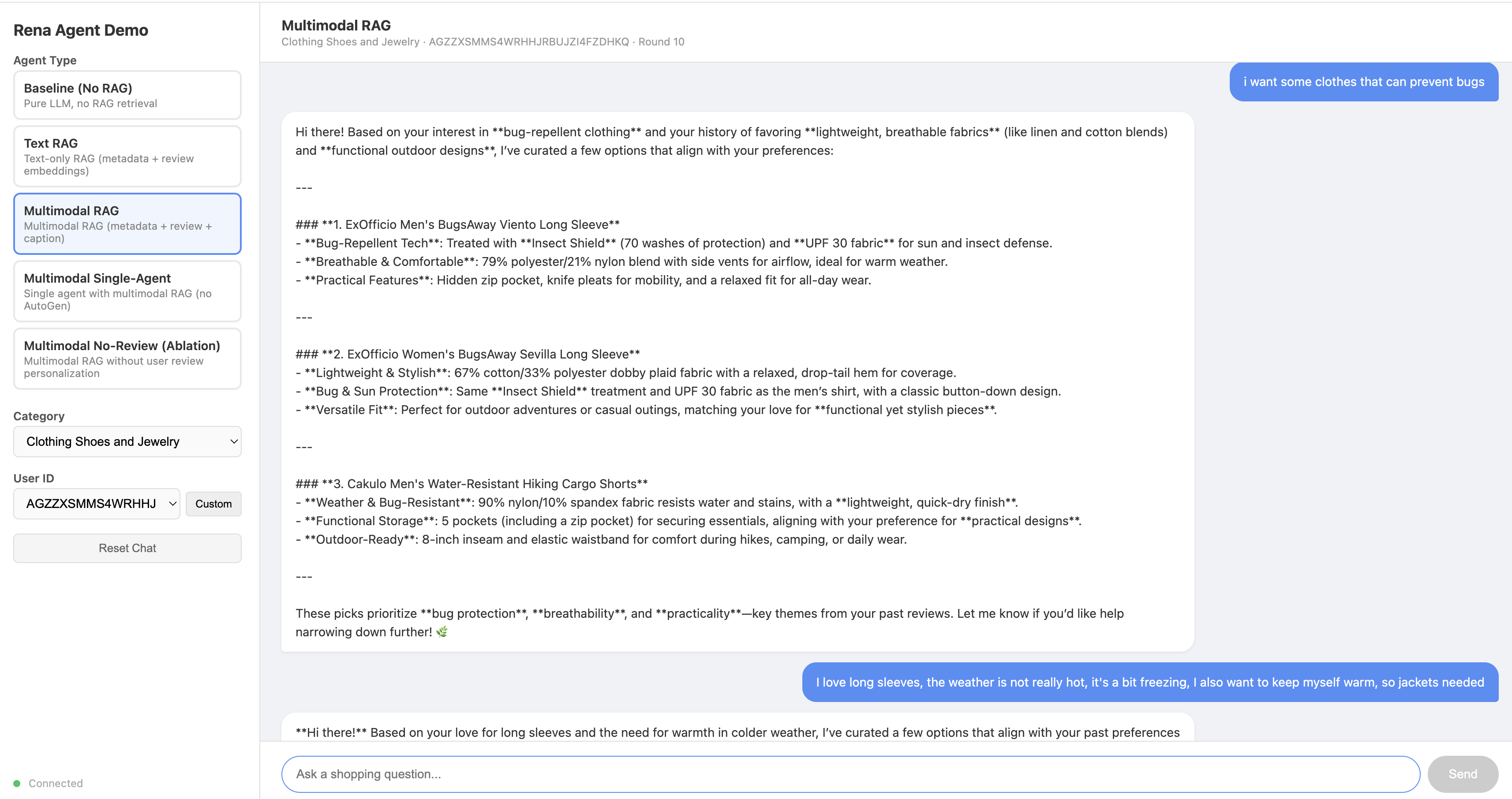}
    \caption{MultimodalRag Full User Interaction 6}
\end{figure*}
\begin{figure*}
    \centering
    \includegraphics[width=\linewidth]{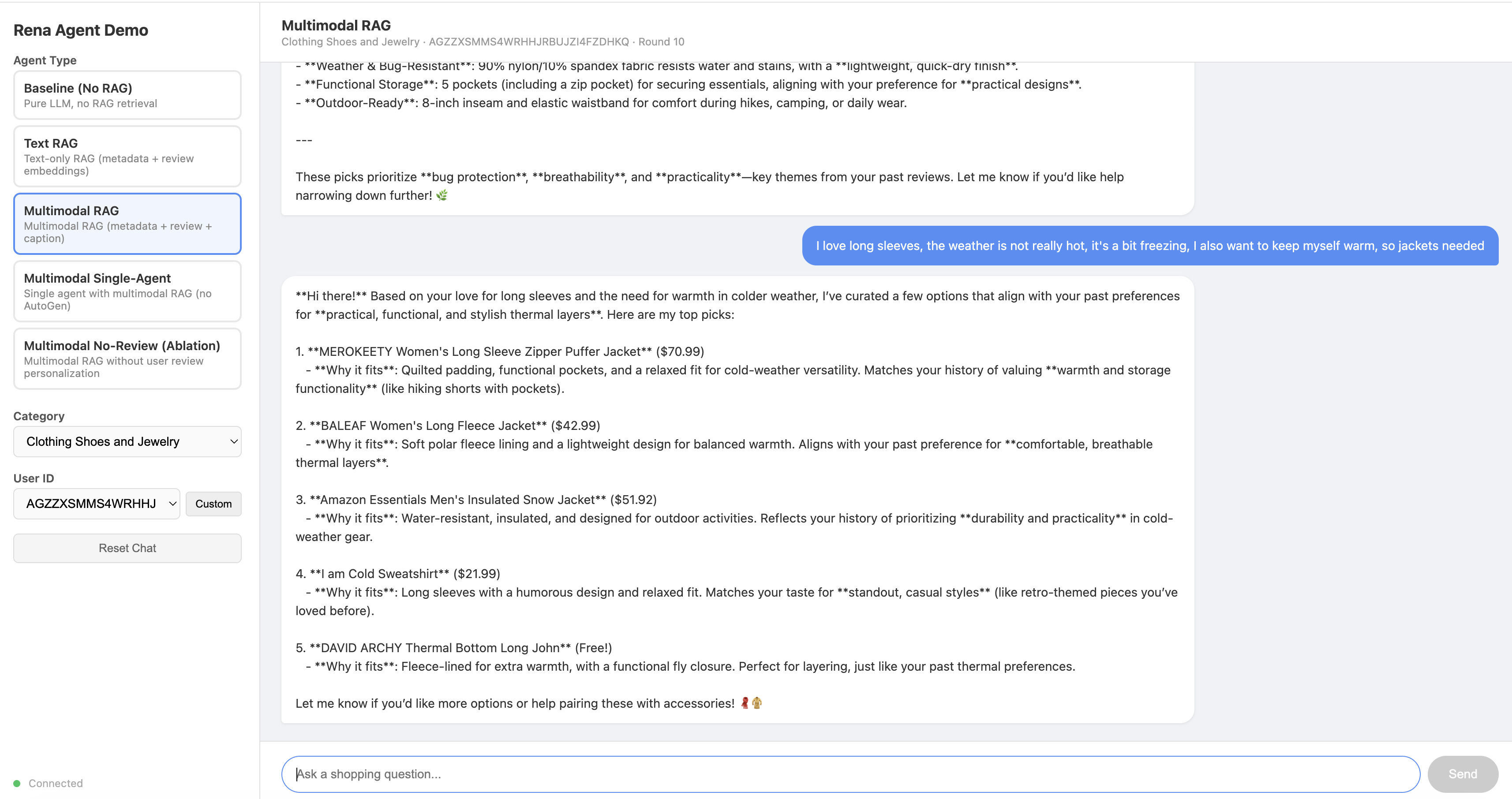}
    \caption{MultimodalRag Full User Interaction 7}
\end{figure*}

%% file: latex/appendix/user_questionnaire.tex
\subsection{User Study Questionnaire}
\label{sec:user_questionnaire}
Five volunteers were recruited through the authors' personal contacts. They used the systems through a web interface and completed the questionnaire below. The same questions were repeated for each system. All volunteers participated without payment. Representative interface screenshots are shown in Appendix~\ref{sec:demo}.

\paragraph{Questions for each system}
The first five questions used a five-point scale; the sixth collected written comments. We reproduce the wording from the feedback form below.
\begin{enumerate}[leftmargin=*,itemsep=3pt]
\item Overall, how satisfied are you with this system? (1 = Very Dissatisfied, 5 = Very Satisfied)
\item The assistant remembered and respected my preferences throughout the conversation.
\item The assistant connected information from earlier parts of the conversation when making recommendations.
\item The conversation flowed naturally and helped me find what I was looking for.
\item The assistant's tone was consistent, friendly, and appropriate throughout.
\item Please share any additional comments about this system.
\end{enumerate}

\begin{samepage}
\paragraph{Questions after trying all systems}
\begin{enumerate}[leftmargin=*,itemsep=3pt]
\item Which system did you prefer most overall?
\item Which system felt most personalized to you?
\item Is there anything else you'd like to share about your overall experience?
\end{enumerate}
\end{samepage}

%% file: latex/appendix/artifact_licensing.tex
\subsection{Data and Code Licensing}
\label{sec:artifact_licensing}
We use Amazon Reviews 2023, which McAuley Lab makes available primarily for research. The maintainers state that they cannot assign a license to the dataset or determine its terms of use.\footnote{\url{https://huggingface.co/datasets/McAuley-Lab/Amazon-Reviews-2023/discussions/1}} Our original code is released under the MIT License. This license does not cover third-party code or data.